  \documentclass[aps,prd,reprint,superscriptaddress,nofootinbib,floatfix,longbibliography]{revtex4-2}

  \usepackage[T1]{fontenc}
  \usepackage[utf8]{inputenc}
  \usepackage[english]{babel}

  \usepackage[dvipsnames]{xcolor}

  \usepackage{amsmath,amsbsy,amssymb,amsfonts,latexsym,cancel}
  \usepackage{newtxtext}
  \usepackage{bm}
  \usepackage{mathrsfs}
  \usepackage{euscript}

  \usepackage{array}
  \usepackage{dcolumn}
  \usepackage{longtable}
  \usepackage{rotating}
  \usepackage{booktabs}
  \usepackage[caption=false]{subfig}
  \usepackage{placeins} 
  \usepackage{multirow}
  \usepackage{enumerate}

  \usepackage{silence}
  \definecolor{DarkGreen}{RGB}{108,134,36}

  \usepackage{orcidlink}
  \usepackage{hyperref}
  \hypersetup{
    colorlinks=true,
    linkcolor=DarkGreen,
    citecolor=DarkGreen,
    urlcolor=DarkGreen
  }

  \newcommand{\CONUS}{\textsc{Conus}$+$}
  \newcommand{\CEvNS}{CE$\nu\mathcal N$S}
  
  \newcommand{\SM}{\mathrm{SM}}

\begin{document}

  \renewcommand{\sectionautorefname}{Section}
  \renewcommand{\subsectionautorefname}{Section}
  \renewcommand{\subsubsectionautorefname}{Section}

  \title{Count-Model Validation and Vector NSI Inference with Public \texorpdfstring{\CONUS{}}{CONUS+} Reactor Spectra}

  \author{Ch.~M.~Benavides~\orcidlink{0009-0006-9765-1169}}
  \email{chmbenavides@gmail.com}
  \affiliation{Departamento de Física, Universidad de Nariño, Pasto, Colombia}

  \author{C.~S.~Muñoz-López~\orcidlink{0009-0004-0189-3266}}
  \email{cristiansantiagocsml123@gmail.com}
  \affiliation{Departamento de Física, Universidad de Nariño, Pasto, Colombia}

  \author{B.C.~Cañas~\orcidlink{0000-0002-7334-5304}}
  \email{blanca.canas@unipamplona.edu.co}
  \affiliation{Departamento de Física, Universidad de Pamplona, Pamplona, Colombia}

  \author{E.~Rojas~\orcidlink{0000-0002-7412-1345}}
  \email{rojas@udenar.edu.co}
  \affiliation{Departamento de Física, Universidad de Nariño, Pasto, Colombia}

  \date{\today}

  \begin{abstract}
    We assess how count-model validation and the reconstruction of public reactor spectra affect statistical inference for coherent elastic neutrino-nucleus scattering (\CEvNS{}) and vector neutrino-quark nonstandard interactions (NSI) in \CONUS{}. Approximate counts in $64$ bins per reactor phase are tested at fixed nominal means. Poisson fluctuations are compatible with the ON spectrum (deviance $p=0.833$), but not with the OFF spectrum ($p=0.0300$). An effective negative-binomial (NegBin) OFF model with $\hat\kappa^{\mathrm{OFF}}=150^{+220}_{-70}$ improves compatibility ($p=0.469$; dispersion uncertainty from a Monte Carlo calibrated $68.27\%$ interval), supported by a bootstrap dispersion test ($p=0.0291$). The selected likelihood gives a conditional signal significance of $4.84\sigma$ and a strength $\hat s=1.17\pm0.25$ (nominal $1\sigma$) compatible with the Standard Model (SM; $p=0.481$). With a common Gaussian signal-normalization constraint in both NSI analyses, the count likelihood reduces the total length of the nominal $1\sigma$ diagonal allowed sets by $11.4\%$ and widens flavor-changing intervals by about $8.7\%$ relative to the separate $19$-bin excess-spectrum $\Delta\chi^2$ analysis. Nominal $90\%$ two-dimensional regions overlap by $81.0$-$97.1\%$ on the sampled grids, without changing the analytic degeneracies. Reconstruction sensitivity tests preserve the model selection and SM signal-strength compatibility, but shift the conditional significance to $4.30\sigma$-$5.39\sigma$ and the regional overlaps to $67.4$-$97.5\%$. These differences reflect distinct data reductions and statistical assumptions, not the isolated effect of the likelihood choice. The conditional significance is not a replacement for the collaboration result.
  \end{abstract}

  \maketitle


\section{Introduction}
\label{sec:introduction}

  Coherent elastic neutrino-nucleus scattering (\CEvNS{}) probes the neutral-current interaction through the coherent nuclear weak charge~\cite{Freedman:1973yd}. Following its first observation by COHERENT~\cite{COHERENT:2017ipa}, the \CONUS{} collaboration reported the first direct observation with reactor antineutrinos, at a significance of $3.7\sigma$~\cite{Ackermann:2025obx}. Germanium detectors provide access to low nuclear recoil energies, making this process sensitive to vector neutrino-quark nonstandard interactions (NSI)~\cite{CONUS:2024lnu,DeRomeri:2025csu}. These interactions modify the coherent weak charge and can interfere with the Standard Model (SM) amplitude when the final neutrino flavor is unchanged~\cite{Barranco:2005yy,DeRomeri:2025csu}.

  The first \CONUS{} data have been used to constrain NSI and other low-energy interactions~\cite{DeRomeri:2025csu,AtzoriCorona:2025ygn,Chattaraj:2025fvx}. Our earlier study considered vector NSI, light $Z'$ mediators, and elastic antineutrino-electron scattering (E$\nu$eS), including the disconnected solutions and two-dimensional degeneracies associated with the effective weak charge~\cite{Benavides:2026jdz}. Here we address a complementary statistical question: which count distributions adequately describe the reconstructed public spectra, and how do the resulting constraints compare with those from an excess-spectrum analysis?

  The published Nature article~\cite{Ackermann:2025obx} provides two spectral representations of the same first-run measurement (published online July 30, 2025, DOI \href{https://doi.org/10.1038/s41586-025-09322-2}{10.1038/s41586-025-09322-2}). Fig.~3a gives a background-subtracted excess in $19$ bins between $160$ and $350\,\mathrm{eVee}$, while Extended Data Fig.~4a-b gives reactor-ON and reactor-OFF rates in $64$ bins per phase between $160$ and $800\,\mathrm{eVee}$. The energy scale is expressed in electron-equivalent electronvolts ($\mathrm{eVee}$), calibrated to the ionization response to electron recoils. We digitize these panels and reconstruct approximate ON/OFF counts using the effective exposures. In the absence of event-level data, the full covariance, and the experimental background nuisance model, the two representations are analyzed separately rather than combined as independent measurements.

  For binned Poisson data, likelihood-based statistics provide a framework for parameter estimation and goodness-of-fit testing that accounts for the discrete nature of the observations~\cite{Baker:1983tu,ParticleDataGroup:2024cfk}. Our analysis first assesses Poisson and negative-binomial (NegBin) fluctuations around fixed SM signal and background templates using Monte Carlo goodness-of-fit tests, residual diagnostics, bin-deletion checks, and model comparison. We then use the selected count law for two distinct inferential tasks: testing the presence and nominal SM strength of the signal, and constraining vector NSI with a profiled background normalization and an effective Gaussian signal-normalization constraint. The likelihood-based inference follows the profile-likelihood framework commonly employed in particle physics~\cite{Cowan:2010js}. The latter normalization constraint is also used in the quadratic analysis of the background-subtracted excess, following the phenomenological treatment adopted in Refs.~\cite{DeRomeri:2025csu,Benavides:2026jdz}.

  The comparison quantifies changes in one-parameter allowed sets and two-parameter regions while separating their weak-charge geometry from their numerical extent. Because the two analyses differ in energy range, data representation, fluctuation model, and background treatment, their differences cannot be attributed to the choice of likelihood alone. We additionally test the sensitivity of model selection, signal inference, and NSI constraints to specified alternative count reconstructions. The aim is to establish which conclusions are stable under these public-data approximations, rather than to reproduce the full experimental likelihood.

  The paper is organized as follows. \autoref{sec:data-physics} describes the reconstruction of the public \CONUS{} spectra and the theoretical prediction of \CEvNS{} events. \autoref{sec:methodology} introduces the statistical framework for count-model validation, model selection, and parameter inference. \autoref{sec:results} presents the results of the count-model validation and selection, the statistical significance of the \CEvNS{} signal and its compatibility with the SM prediction, and the vector NSI confidence regions and their geometric comparison. Finally, \autoref{sec:discussion} discusses the robustness and limitations of the analysis, and \autoref{sec:conclusions} summarizes the main findings.

\section{Public data and \texorpdfstring{\CEvNS{}}{CEvNS} physical prediction}
\label{sec:data-physics}

  The SM and vector-NSI predictions for the first \CONUS{} measurement follow from the reactor antineutrino flux, coherent nuclear cross section, and detector response. Together with the two published spectral representations, these inputs determine the count means and the excess-spectrum prediction used in the statistical analyses.

\subsection{The \texorpdfstring{\CONUS{}}{CONUS+} experiment and public spectra}
\label{sec:experiment}

  The first \CONUS{} observation used high-purity germanium detectors located $20.7\,\mathrm{m}$ from the $3.6\,\mathrm{GW}$ Leibstadt reactor in Switzerland. The present analysis includes detectors C2, C3, and C5; C4 was excluded because of the rate instabilities reported by the collaboration~\cite{Ackermann:2025obx}. Their thresholds, exposures, and the reactor inputs are summarized in \autoref{tab:conus-parameters}.

  \begin{table}[htbp]
    \centering
    \caption{Experimental specifications of the first \CONUS{} \CEvNS{} dataset used to construct the public-data prediction~\cite{Ackermann:2025obx}. Detector entries are ordered as C2/C3/C5. The fission fractions are ordered as $^{235}\mathrm{U}$/$^{238}\mathrm{U}$/$^{239}\mathrm{Pu}$/$^{241}\mathrm{Pu}$. The measured and predicted event totals provide a global normalization reference but do not replace the spectral information used in the analysis.}
    \label{tab:conus-parameters}
    \begin{tabular}{lc}
      \toprule \toprule
      Parameter & Value \\
      \midrule
      Reactor-detector baseline & $20.7\,\mathrm{m}$ \\
      Ionization thresholds C2/C3/C5 & $180/160/170\,\mathrm{eVee}$ \\
      Active masses C2/C3/C5 & $0.95/0.94/0.94\,\mathrm{kg}$ \\
      Accepted $\mathrm{ON}$ live time C2/C3/C5 & $117/110/119\,\mathrm{days}$ \\
      Total $\mathrm{ON/OFF}$ exposure & $327/60\,\mathrm{kg\cdot day}$ \\
      Antineutrino flux & $1.5\times10^{13}\,\mathrm{cm}^{-2}\mathrm{s}^{-1}$ \\
      Reactor thermal power & $3.6\,\mathrm{GW}$ \\
      Fission fractions & $53/8/32/7\,\%$ \\
      Measured \CEvNS{} excess & $395\pm106$ events \\
      SM expectation & $347\pm59$ events \\
      Data/SM prediction & $1.14\pm0.36$ \\
      SM prediction uncertainty & $17\,\%$ \\
      \bottomrule \bottomrule
    \end{tabular}
  \end{table}

  The $347\pm59$ events in \autoref{tab:conus-parameters} are the collaboration's global SM expectation obtained with its complete three-detector analysis. They are not the normalization used directly in our reconstructed likelihood. Our independently calculated SM spectrum, evaluated bin by bin with the public exposures, detector thresholds, and response model specified below, gives
  \begin{equation}
    \sum_{i=1}^{64}R_i^{\mathrm{SM}}=320.1501\ \mathrm{events}.
    \label{eq:sm-template-total}
  \end{equation}
  This total is $26.8499$ events, or $7.74\%$, below the collaboration's central expectation and differs from it by only $0.46$ times the quoted $59$-event uncertainty. As in our preceding analysis~\cite{Benavides:2026jdz}, the residual offset may reflect the approximations of our independent calculation: the coherent-limit choice $F_W\boldsymbol{(}|\mathbf{q}|^2\boldsymbol{)}\simeq1$, the replacement of the natural-germanium isotope sum by one abundance-weighted effective nucleus, unit detection efficiency within the accepted range, effective time-integrated reactor inputs, and the adopted numerical integration and interpolation procedures. The two values are therefore normalization-compatible but are not interchangeable: $347\pm59$ documents the official global prediction, whereas \eqref{eq:sm-template-total} is the fixed spectral template produced by our approximated calculation and used consistently in the public-data likelihood.

\begin{figure*}[htbp]
    \centering
    \IfFileExists{Pictures/Data/conus_plus_extended_and_excess_counts.pdf}{%
      \includegraphics[width=0.88\textwidth]{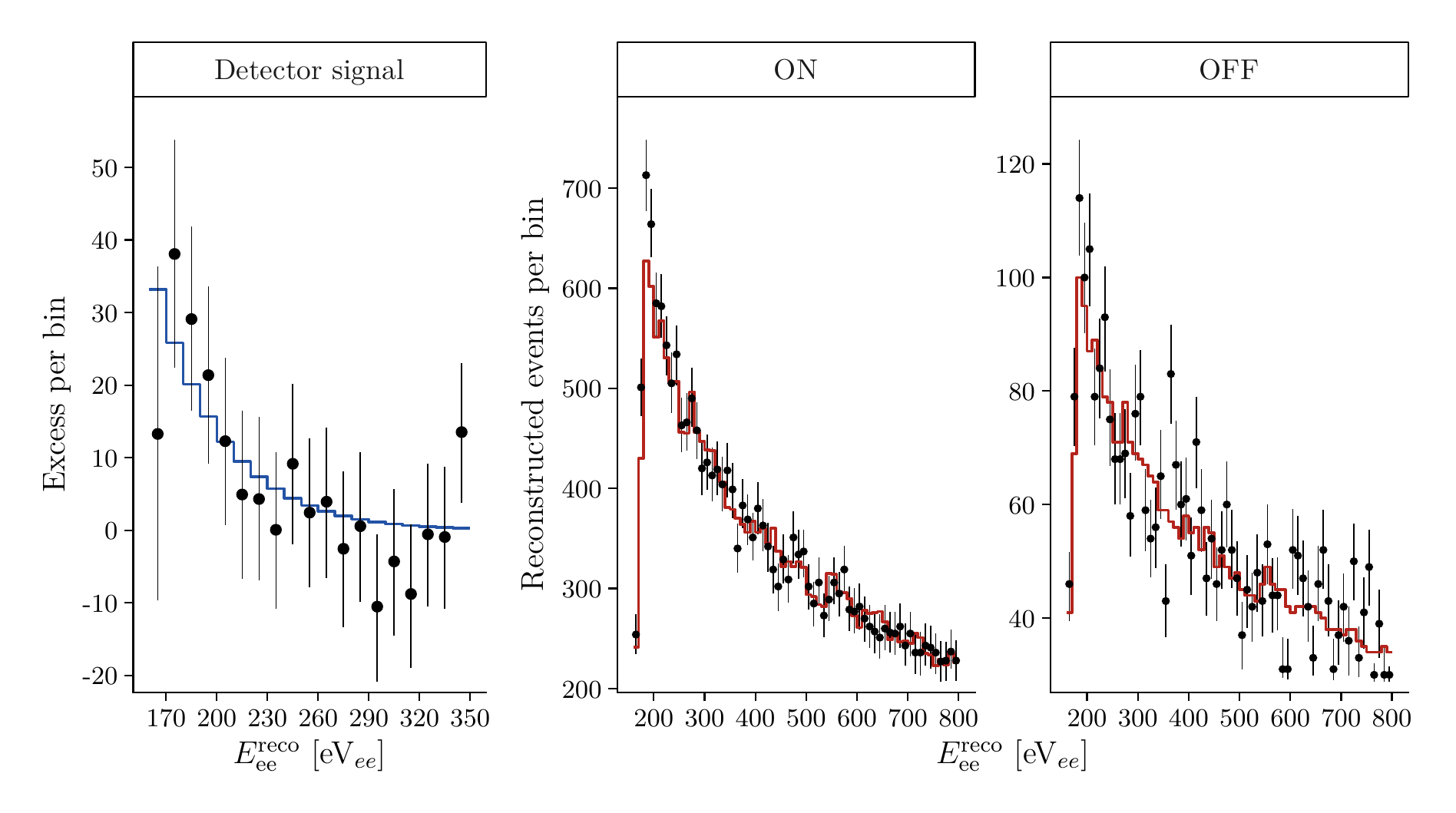}
    }{%
      \fbox{\parbox[c][5cm][c]{0.75\textwidth}{\centering Figura pendiente:\\ \texttt{conus\_plus\_extended\_and\_excess\_counts.pdf}}}%
    }
    \caption{Digitized and adapted public \CONUS{} spectra from Ref.~\cite{Ackermann:2025obx}, and their statistical roles. In all three panels, the black points are experimental data published by \CONUS{}, and the vertical bars are the corresponding uncertainties reported by the collaboration. Left: the points are the $19$ background-subtracted signal bins digitized from Fig.~3a, whereas the blue line is our theoretical \CEvNS{} prediction within the SM; the displayed $19$-bin portion is used to construct $\Delta\chi^2$. Middle and right: the points are the reconstructed reactor-ON and reactor-OFF counts adapted from the normalized rates published in Extended Data Fig.~4a-b. The red lines are the corresponding \CONUS{} background models, converted from published rates to expected counts. The $\mathrm{ON/OFF}$ data and background models are used to assess the count distributions and construct $\Delta q$.}
    \label{fig:public-on-off}
  \end{figure*}

  The spectra summarized in \autoref{fig:public-on-off} were digitized from two distinct public data products~\cite{Ackermann:2025obx}. Fig.~3a contains $19$ background-subtracted excess bins, each $10\,\mathrm{eVee}$ wide, over $160$-$350\,\mathrm{eVee}$; their digitized central values and uncertainties enter the quadratic $\chi^2$ analysis. Extended Data Fig.~4a-b contains $\mathrm{ON}$ and $\mathrm{OFF}$ rates in $64$ bins per reactor phase over $160$-$800\,\mathrm{eVee}$. Counts reconstructed from the published rates are used to validate the count distributions and construct the $\mathrm{ON/OFF}$ likelihood. The products are analyzed separately because their published graphical representations do not provide the correlations required to combine or transform them into the official experimental likelihood.

  The $\mathrm{ON/OFF}$ spectra are published as rates per $\mathrm{kg\cdot day}$ per $10\,\mathrm{eVee}$ bin, rather than as integer counts. For reactor phase $\phi\in\{\mathrm{ON},\mathrm{OFF}\}$, the digitized data rate $\widetilde y_i^\phi$ and background rate $\widetilde b_i^\phi$ are converted using the effective mass-time exposure $\mathcal E_i^\phi$:
  \begin{equation}
    \widetilde n_i^\phi=\widetilde y_i^\phi\mathcal E_i^\phi,
    \qquad B_i^\phi=\widetilde b_i^\phi\mathcal E_i^\phi.
    \label{eq:public-rate-reconstruction}
  \end{equation}
  The effective mass-time exposure depends on which detectors are above the threshold~\cite{Ackermann:2025obx}. The $\mathrm{ON/OFF}$ pairs $(\mathcal E_i^{\mathrm{ON}},\mathcal E_i^{\mathrm{OFF}})$ adopted in our reconstruction are $(103.3,19.9)$, $(215.6,39.9)$, and $(326.8,59.7)\,\mathrm{kg\cdot day}$ for $i=1$, $i=2$, and $i\geq3$, respectively, corresponding to C3, C3+C5, and C2+C3+C5. The plotted rates are already normalized to $10\,\mathrm{eVee}$ bins. Unlike $\mathcal E_i^\phi$, the detector factor $\mathcal E_d$ in \eqref{eq:nuclear-differential-rate} also includes the number of target nuclei.

  The reconstruction rule selects the nonnegative integer whose rate is closest to $\widetilde y_i^\phi$ within the graphical tolerance $\delta_{\mathrm{dig}}=0.02\,\mathrm{events}\,\mathrm{kg}^{-1}\mathrm{d}^{-1}(10\,\mathrm{eVee})^{-1}$. If the set is empty, the rule uses the nearest nonnegative integer to $\widetilde n_i^\phi$, with ties resolved upward. The rounding difference, when the original continuous reading is available, and the half-width of the graphical tolerance in count units are
  \begin{equation}
    \delta_i^\phi=n_i^\phi-\widetilde n_i^\phi,
    \qquad
    \delta_{\mathrm{count},i}^\phi=\mathcal E_i^\phi\delta_{\mathrm{dig}}.
    \label{eq:rounding-and-count-tolerance}
  \end{equation}
  The archived central rates reproduce all stored integer counts to numerical precision. For example, the largest absolute differences after multiplying the archived rates by the exposures are below $1.5\times10^{-12}$ events in ON and $3.0\times10^{-13}$ events in OFF. These checks establish internal consistency. The continuous readings before rounding could not be recovered, so the original $\delta_i^\phi$ and its size relative to the model standard deviation cannot be measured from the archive. The background means $B_i^\phi$ remain continuous.

  Using the archived rates as centers, the graphical half-widths in \eqref{eq:rounding-and-count-tolerance} are $2.066$, $4.312$, and $6.536$ events for ON, and $0.398$, $0.798$, and $1.194$ events for OFF, in order of increasing exposure. There are multiple compatible integers in all $64$ ON bins and in $62$ OFF bins. The first two OFF bins have one candidate each. For Poisson ON and NegBin OFF, the half-width is at most $0.438$ and $0.185$ model standard deviations, respectively. The half-width defines the count perturbations examined in \autoref{sec:discussion}. The displayed experimental error bars and the stochastic variance of the count model retain their separate roles.

\subsection{Standard Model signal}
\label{sec:sm-signal}

  The SM cross section provides the reference prediction used both for count-model validation and for the subsequent inference of vector NSI parameters. In natural units, $\hbar=c=1$, the SM differential cross section for an electron antineutrino with energy $E_{\bar\nu_e}$ and a nuclear recoil $T_{\mathcal N}$ is~\cite{Freedman:1973yd,DeRomeri:2025csu}
  \begin{equation}
    \begin{aligned}
      \left.\frac{d\sigma_{\bar\nu_e\mathcal{N}}}{dT_{\mathcal{N}}}
      \right|_{\mathrm{SM}}
      ={}& \frac{G_F^2m_{\mathcal{N}}}{\pi}
      (Q_V^{\mathrm{SM}})^2
      F_W^2\boldsymbol{(}|\mathbf{q}|^2\boldsymbol{)} \\
      &\times\left[1-\frac{m_{\mathcal{N}}T_{\mathcal{N}}}
      {2E_{\bar\nu_e}^2}-\frac{T_{\mathcal{N}}}{E_{\bar\nu_e}}\right],
    \end{aligned}
    \label{eq:cevns-sm}
  \end{equation}
  where $G_F$ is the Fermi constant, $m_{\mathcal{N}}$ is the nuclear mass, $|\mathbf{q}|\simeq\sqrt{2m_{\mathcal{N}}T_{\mathcal{N}}}$ is the momentum-transfer magnitude, and $F_W\boldsymbol{(}|\mathbf{q}|^2\boldsymbol{)}$ is the weak nuclear form factor that describes the spatial distribution of weak charge and the loss of coherence at finite momentum transfer. For reactor antineutrinos, $|\mathbf{q}|R_{\mathcal{N}}\ll1$, where $R_{\mathcal{N}}$ is the nuclear radius; consequently, the numerical calculation uses the coherent-limit approximation $F_W\boldsymbol{(}|\mathbf{q}|^2\boldsymbol{)}\simeq1$~\cite{Benavides:2026jdz}. The SM weak charge is~\cite{Freedman:1973yd,DeRomeri:2025csu}
  \begin{equation}
    \begin{aligned}
      Q_V^{\mathrm{SM}}&=Zg_V^p+Ng_V^n,\\
      g_V^p&=\frac{1-4\sin^2\theta_W}{2},
      &g_V^n&=-\frac{1}{2}.
    \end{aligned}
    \label{eq:weak-charge}
  \end{equation}
  Here, $Z$ and $N$ are the proton and neutron numbers, $g_V^p$ and $g_V^n$ are the proton and neutron vector couplings, and $\theta_W$ is the weak mixing angle. We use $\sin^2\theta_W=0.23857$ in the low-energy regime~\cite{ParticleDataGroup:2024cfk}.

\subsection{Nonstandard interactions}
\label{sec:bsm-framework}

  Vector neutrino-quark NSI modify the coherent weak charge and hence the \CEvNS{} rate. These interactions provide a model-independent parametrization of physics beyond the SM through effective four-fermion operators~\cite{Ohlsson:2012kf,Farzan:2017xzy}. The corresponding effective interaction is~\cite{Papoulias:2015vxa}
  \begin{equation}
    \mathcal{L}_{\mathrm{NSI}}=
    -2\sqrt{2}G_F\sum_{f,f',q,C}
    \varepsilon_{ff'}^{qC}
    (\bar\nu_f\gamma^\mu P_L\nu_{f'})
    (\bar q\gamma_\mu P_Cq).
    \label{eq:nsi-lagrangian}
  \end{equation}
  Here $f,f'\in\{e,\mu,\tau\}$ denote neutrino flavors, $q\in\{u,d\}$ denotes the quark, and $C\in\{L,R\}$ denotes the chirality. The projectors are $P_L=(1-\gamma^5)/2$ and $P_R=(1+\gamma^5)/2$. The dimensionless coefficients $\varepsilon_{ff'}^{qC}$ parametrize the new interaction relative to $G_F$, and the coherent vector combination is $\varepsilon_{ff'}^{qV}=\varepsilon_{ff'}^{qL}+\varepsilon_{ff'}^{qR}$~\cite{Barranco:2005yy,DeRomeri:2025csu}. The analysis assumes real NSI coefficients; for complex flavor-changing coefficients, the corresponding terms in \eqref{eq:nsi-effective-charge} require absolute squares.

  For the incident reactor flavor $f=e$, we define the coherent quark weights as $C_u=2Z+N$ and $C_d=Z+2N$. The complete SM+NSI squared effective charge is~\cite{Barranco:2005yy,Canas:2019fjw,DeRomeri:2025csu}
  \begin{equation}
    \begin{aligned}
      \mathcal Q_{\mathrm{eff}}^2
      \boldsymbol{(}\boldsymbol{\varepsilon}\boldsymbol{)}={}&
      \left[Q_V^{\mathrm{SM}}+C_u\varepsilon_{ee}^{uV}
      +C_d\varepsilon_{ee}^{dV}\right]^2\\
      &+\sum_{f'=\mu,\tau}
      \left[C_u\varepsilon_{ef'}^{uV}
      +C_d\varepsilon_{ef'}^{dV}\right]^2.
    \end{aligned}
    \label{eq:nsi-effective-charge}
  \end{equation}
  Diagonal NSI interfere with the SM amplitude, whereas contributions involving different unobserved final-state neutrino flavors add incoherently. The SM+NSI cross section follows from \eqref{eq:cevns-sm} by replacing $(Q_V^{\mathrm{SM}})^2$ with $\mathcal Q_{\mathrm{eff}}^2\boldsymbol{(}\boldsymbol{\varepsilon}\boldsymbol{)}$. For $\boldsymbol{\varepsilon}=\mathbf{0}$, the SM cross section is recovered. 

\subsection{Reactor flux, detector response, and binned prediction}
\label{sec:signal-prediction}

  The event prediction is obtained by folding the \CEvNS{} cross section with the reactor flux and detector response and integrating over each reconstructed-energy bin, following the formalism described in Refs.~\cite{DeRomeri:2025csu,Benavides:2026jdz}. For detectors C2, C3, and C5, the total prediction is
  \begin{equation}
    R_i\boldsymbol{(}\boldsymbol{\varepsilon}\boldsymbol{)}
    =\sum_{d\in\{\mathrm{C2,C3,C5}\}}
    \int_{E_{i}^{\mathrm{low}}}^{E_{i}^{\mathrm{high}}}
    \frac{dR_d\boldsymbol{(}\boldsymbol{\varepsilon}\boldsymbol{)}}
    {dE_{\mathrm{ee}}^{\mathrm{reco}}}
    \,dE_{\mathrm{ee}}^{\mathrm{reco}},
    \label{eq:nuclear-bin-prediction}
  \end{equation}
  where $d$ labels the three included detectors and $E_i^{\mathrm{low}}$ and $E_i^{\mathrm{high}}$ are the reconstructed-energy bin edges. The detector contribution is
  \begin{equation}
    \begin{aligned}
      \frac{dR_d}{dE_{\mathrm{ee}}^{\mathrm{reco}}}={}&
      \mathcal E_d\int_{T_{\mathcal N}^{\min}}^{T_{\mathcal N}^{\max}}
      dT_{\mathcal N}\,\mathcal F_d\boldsymbol{(}T_{\mathcal N}\boldsymbol{)}
      \mathcal G_d\boldsymbol{(}E_{\mathrm{ee}}^{\mathrm{reco}},T_{\mathcal N}\boldsymbol{)}\\
      &\times\int_{E_{\bar\nu_e}^{\min}\boldsymbol{(}T_{\mathcal N}\boldsymbol{)}}^{10\,\mathrm{MeV}}
      dE_{\bar\nu_e}\,
      \frac{d\Phi_{\bar\nu_e}}{dE_{\bar\nu_e}}
      \frac{d\sigma_{\bar\nu_e\mathcal N}
      \boldsymbol{(}\boldsymbol{\varepsilon}\boldsymbol{)}}{dT_{\mathcal N}}.
    \end{aligned}
    \label{eq:nuclear-differential-rate}
  \end{equation}
  The factor $\mathcal E_d$ contains the number of target nuclei and the live time, $\mathcal F_d$ is the detection efficiency, and $\mathcal G_d$ is the detector response. The detection efficiency is taken as unity within the accepted energy range, consistent with the near-unity trigger efficiency reported above the experimental thresholds~\cite{Ackermann:2025obx}. Natural germanium is approximated by a single abundance-weighted effective nucleus ($Z=32$, $N_{\mathrm{eff}}=40.7053$) with a mean nuclear mass. Isotope-specific recoil spectra are not summed. Detector thresholds determine the accepted reconstructed-energy bins.

  The response is modeled as a Gaussian centered on the mean ionization energy $E_{\mathrm{er}}\boldsymbol{(}T_{\mathcal N}\boldsymbol{)}$~\cite{Lindner:2024eng,Chattaraj:2025fvx},
  \begin{equation}
    \mathcal G_d\boldsymbol{(}E_{\mathrm{ee}}^{\mathrm{reco}},T_{\mathcal N}\boldsymbol{)}
    =\frac{\exp\left[-\dfrac{\left(E_{\mathrm{ee}}^{\mathrm{reco}}-E_{\mathrm{er}}\boldsymbol{(}T_{\mathcal N}\boldsymbol{)}\right)^2}{2\sigma_{\mathrm{res}}^2\boldsymbol{(}T_{\mathcal N}\boldsymbol{)}}\right]}
    {\sqrt{2\pi}\,\sigma_{\mathrm{res}}\boldsymbol{(}T_{\mathcal N}\boldsymbol{)}},
    \label{eq:nuclear-response}
  \end{equation}
  where the energy resolution is parameterized as~\cite{Ackermann:2025obx,CONUS:2024lnu,Benavides:2026jdz}
  \begin{equation}
    \begin{aligned}
      \sigma_{\mathrm{res}}\boldsymbol{(}T_{\mathcal N}\boldsymbol{)}
      &={}\sqrt{\sigma_0^2+F_{\mathrm{Fano}}\eta
      E_{\mathrm{er}}\boldsymbol{(}T_{\mathcal N}\boldsymbol{)}},\\
      \sigma_0&=20.38\,\mathrm{eVee},
      &F_{\mathrm{Fano}}&=0.1096,\\
      \eta&=2.96\,\mathrm{eVee}.
    \end{aligned}
    \label{eq:energy-resolution}
  \end{equation}
  Here $\sigma_0$ is the baseline energy resolution, $F_{\mathrm{Fano}}$ is the Fano factor, and $\eta$ is the mean energy required to produce an electron-hole pair in germanium. The ionization energy is $E_{\mathrm{er}}\boldsymbol{(}T_{\mathcal N}\boldsymbol{)}=Q_F\boldsymbol{(}T_{\mathcal N}\boldsymbol{)}T_{\mathcal N}$, where the Lindhard model gives~\cite{Rink:2022rsx,Bonhomme:2022lcz}
  \begin{equation}
    \begin{aligned}
      Q_F\boldsymbol{(}T_{\mathcal N}\boldsymbol{)}
      &={}\frac{k g\boldsymbol{(}\epsilon\boldsymbol{)}}
      {1+k g\boldsymbol{(}\epsilon\boldsymbol{)}},
      &k&=0.162,\\
      g\boldsymbol{(}\epsilon\boldsymbol{)}
      &={}3\epsilon^{0.15}+0.7\epsilon^{0.6}+\epsilon,
      &\epsilon&=11.5Z^{-7/3}\frac{T_{\mathcal N}}{\mathrm{keV}}.
    \end{aligned}
    \label{eq:lindhard-quenching}
  \end{equation}

  The recoil integration begins at $T_{\mathcal N}^{\min}=2.72\times10^{-5}\,\mathrm{MeV}$, well below the observable threshold. Its upper limit and the minimum antineutrino energy follow from two-body kinematics~\cite{Benavides:2026jdz},
  \begin{equation}
    \begin{aligned}
      T_{\mathcal N}^{\max}
      &={}\frac{2(E_{\bar\nu_e}^{\max})^2}
      {m_{\mathcal N}+2E_{\bar\nu_e}^{\max}},\\
      E_{\bar\nu_e}^{\min}\boldsymbol{(}T_{\mathcal N}\boldsymbol{)}
      &={}\frac{T_{\mathcal N}}{2}
      +\sqrt{\left(\frac{T_{\mathcal N}}{2}\right)^2
      +\frac{m_{\mathcal N}T_{\mathcal N}}{2}}.
    \end{aligned}
    \label{eq:kinematic-limits}
  \end{equation}
  where $E_{\bar\nu_e}^{\max}=10\,\mathrm{MeV}$.

  The reactor flux combines the four dominant fissile isotopes $l\in\{^{235}\mathrm{U},^{238}\mathrm{U},^{239}\mathrm{Pu},^{241}\mathrm{Pu}\}$ and is given by~\cite{Baldoncini:2014vda,Zhang:2023zif,Benavides:2026jdz}
  \begin{equation}
    \frac{d\Phi_{\bar\nu_e}}{dE_{\bar\nu_e}}
    =\frac{P_{\mathrm{th}}}{4\pi L^2\sum_l f_lE_l^{\mathrm{fis}}}
    \sum_l f_l\lambda_l\boldsymbol{(}E_{\bar\nu_e}\boldsymbol{)},
    \label{eq:reactor-flux}
  \end{equation}
  where $P_{\mathrm{th}}$ is the reactor thermal power, $L$ is the reactor--detector baseline, $f_l$ are the fission fractions, $E_l^{\mathrm{fis}}$ are the mean energies released per fission, and $\lambda_l\boldsymbol{(}E_{\bar\nu_e}\boldsymbol{)}$ is the antineutrino spectrum per fission. The central values of $E_l^{\mathrm{fis}}$ reported in Table~II of Ref.~\cite{Baldoncini:2014vda} are adopted. For the time-integrated \CONUS{} dataset, $P_{\mathrm{th}}$ and $f_l$ are treated as effective constants, whose values are listed in \autoref{tab:conus-parameters}.

  For antineutrino energies above $2\,\mathrm{MeV}$, each isotopic spectrum is described by the phenomenological parametrization~\cite{Mueller:2011nm}
  \begin{equation}
    \lambda_l\boldsymbol{(}E_{\bar\nu_e}\boldsymbol{)}
    =\exp\left[
      \sum_{k=1}^{6}a_{lk}E_{\bar\nu_e}^{k-1}
    \right],
    \label{eq:isotope-spectrum}
  \end{equation}
  where the antineutrino energy $E_{\bar\nu_e}$ is expressed in MeV. The central values of the coefficients $a_{lk}$ reported in Refs.~\cite{Mueller:2011nm,Baldoncini:2014vda} are adopted, and their correlations are not propagated in the fixed flux prediction. For antineutrino energies below $2\,\mathrm{MeV}$, the spectrum of Ref.~\cite{Kopeikin:2012zz} is used.

  The predicted counts $R_i\boldsymbol{(}\boldsymbol{\varepsilon}\boldsymbol{)}$ enter the $\mathrm{ON}$ mean in \eqref{eq:expected-counts}. When all NSI couplings vanish, they reduce to the SM template, $R_i\boldsymbol{(}\mathbf{0}\boldsymbol{)}=R_i^{\mathrm{SM}}$.

\section{Statistical methodology}
\label{sec:methodology}

  The statistical analysis proceeds in two stages. First, the $64$ reconstructed bins in each reactor phase are used to test Poisson and negative-binomial fluctuations at fixed SM and background means. The diagnostics in \autoref{sec:validation-methods} and the model comparison in \autoref{sec:model-comparison-methods} guide the choice of count law. Second, normalization parameters and NSI couplings are introduced for inference in \autoref{sec:inference-methods}. A quadratic analysis of the separate $19$-bin excess spectrum provides a comparison, not an independent measurement.

\subsection{Candidate count models}
\label{sec:count-models}

  We first connect the physical predictions to count random variables, then specify the probability laws used to assess fluctuations around the nominal means.

\subsubsection{Random variables and expected counts}
\label{sec:count-random-variables}

  For bin $i=1,\ldots,64$ and reactor phase $\phi\in\{\mathrm{ON},\mathrm{OFF}\}$, let $N_i^\phi$ denote the nonnegative integer-valued random variable representing the count. Its realization $n_i^\phi$ is the integer reconstructed from the published spectrum (\autoref{sec:experiment}). While $N_i^\phi$ describes possible counts in repeated experiments under the model, $n_i^\phi$ is fixed by the reconstructed data. The expectation is $\mu_i^\phi=\mathbb{E}\boldsymbol{[}N_i^\phi\boldsymbol{]}$~\cite{ParticleDataGroup:2024cfk}. Following the signal and background decomposition employed by the \CONUS{} collaboration~\cite{Ackermann:2025obx}, before introducing normalization parameters, the nominal conditional means are
  \begin{equation}
    \begin{aligned}
      \mathbb{E}\boldsymbol{[}N_i^{\mathrm{ON}}\boldsymbol{]}
      &=\mu_{i,0}^{\mathrm{ON}}
      \boldsymbol{(}\boldsymbol{\varepsilon}\boldsymbol{)}\\
      &=B_i^{\mathrm{ON}}+R_i
      \boldsymbol{(}\boldsymbol{\varepsilon}\boldsymbol{)},\\
      \mathbb{E}\boldsymbol{[}N_i^{\mathrm{OFF}}\boldsymbol{]}
      &=\mu_{i,0}^{\mathrm{OFF}}=B_i^{\mathrm{OFF}}.
    \end{aligned}
    \label{eq:expected-counts}
  \end{equation}
  Here the expectations are evaluated at fixed physical parameters, $B_i^\phi$ is the continuous background expectation, and $R_i\boldsymbol{(}\boldsymbol{\varepsilon}\boldsymbol{)}$ is the predicted signal. Unlike the realized counts, the means need not be integers. Counts in different bins and phases are assumed conditionally independent, as in the factorized likelihood framework commonly adopted for binned counting experiments~\cite{Baker:1983tu,Cowan:2010js}. This assumption is part of the public-data model, since the full covariance and nuisance-parameter description is unavailable. Digitization tolerance is not included as an extra variance component.

\subsubsection{Poisson and negative-binomial count models}
\label{sec:count-distributions}

  The validation compares two conditional distributions for $N_i^\phi$. Throughout this stage, the means are fixed at the SM values $\mu_i^{\mathrm{ON}}=\mu_{i,0}^{\mathrm{ON}}\boldsymbol{(}\mathbf{0}\boldsymbol{)}$ and $\mu_i^{\mathrm{OFF}}=\mu_{i,0}^{\mathrm{OFF}}$ from \eqref{eq:expected-counts}; no background or signal normalization is fitted. The Poisson candidate is~\cite{ParticleDataGroup:2024cfk,Baker:1983tu}
  \begin{equation}
    N_i^\phi\sim\operatorname{Poisson}
    \boldsymbol{(}\mu_i^\phi\boldsymbol{)},
    \qquad
    \operatorname{Var}\boldsymbol{(}N_i^\phi\boldsymbol{)}=\mu_i^\phi,
    \label{eq:poisson-model}
  \end{equation}
  whereas the NegBin candidate is parameterized by its mean and a positive dispersion parameter $\kappa^\phi$~\cite{Bohme:2025nvu},
  \begin{equation}
    N_i^\phi\sim\operatorname{NegBin}
    \boldsymbol{(}\mu_i^\phi,\kappa^\phi\boldsymbol{)},
    \qquad
    \operatorname{Var}\boldsymbol{(}N_i^\phi\boldsymbol{)}
    =\mu_i^\phi+\frac{\boldsymbol{(}\mu_i^\phi\boldsymbol{)}^2}{\kappa^\phi}.
    \label{eq:negbin-model}
  \end{equation}
  The Poisson distribution imposes equality between the conditional mean and variance (equidispersion), whereas the NegBin distribution accommodates overdispersion through the additional quadratic term in \eqref{eq:negbin-model}. This distinction allows us to assess whether the reconstructed spectra require fluctuations beyond those described by the Poisson model.

  Denoting the Poisson and NegBin distributions by the subscripts $\mathrm{P}$ and $\mathrm{NB}$, respectively, the probability mass function $p_M$ of model $M$ assigns the probability $\mathbb{P}_M\boldsymbol{(}N_i^\phi=n_i^\phi\boldsymbol{)}$ to the reconstructed realization $n_i^\phi$ at the specified model parameters. The corresponding expressions are~\cite{Baker:1983tu,Bohme:2025nvu}
  \begin{equation}
    \begin{aligned}
      p_{\mathrm{P}}\boldsymbol{(}n_i^\phi\mid\mu_i^\phi\boldsymbol{)}
      &=\frac{e^{-\mu_i^\phi}(\mu_i^\phi)^{n_i^\phi}}{n_i^\phi!},\\
      p_{\mathrm{NB}}\boldsymbol{(}n_i^\phi\mid\mu_i^\phi,\kappa^\phi\boldsymbol{)}
      &=\frac{\Gamma\boldsymbol{(}n_i^\phi+\kappa^\phi\boldsymbol{)}}
      {\Gamma\boldsymbol{(}\kappa^\phi\boldsymbol{)}\,n_i^\phi!}\\
      &\quad\times\left(\frac{\kappa^\phi}{\kappa^\phi+\mu_i^\phi}\right)^{\kappa^\phi}
      \left(\frac{\mu_i^\phi}{\kappa^\phi+\mu_i^\phi}\right)^{n_i^\phi},
    \end{aligned}
    \label{eq:count-pmfs}
  \end{equation}
  The mean is $\mu_i^\phi>0$, the NegBin dispersion parameter is $\kappa^\phi>0$, and $\Gamma$ is the gamma function. At fixed $\mu_i^\phi$, taking the limit $\kappa^\phi\to\infty$ in \eqref{eq:count-pmfs} recovers the Poisson probability mass function. Consequently, the Poisson distribution is a limiting case of the adopted NegBin model.

  Before examining the discrepancies, we specify how the parameters of each candidate model are estimated. For phase $\phi$, model $M$, and fitted parameter vector $\boldsymbol{\vartheta}_{\phi,M}$, conditional independence gives the following log-likelihood and its maximum~\cite{ParticleDataGroup:2024cfk,Cowan:2010js}:
  \begin{equation}
    \begin{aligned}
      \ell_{\phi,M}\boldsymbol{(}\boldsymbol{\vartheta}_{\phi,M}\boldsymbol{)}&=\sum_i\log p_M\boldsymbol{(}n_i^\phi\mid\mu_i^\phi,\boldsymbol{\vartheta}_{\phi,M}\boldsymbol{)},\\
      \hat\ell_{\phi,M}&=\max_{\boldsymbol{\vartheta}_{\phi,M}}\ell_{\phi,M}\boldsymbol{(}\boldsymbol{\vartheta}_{\phi,M}\boldsymbol{)}.
    \end{aligned}
    \label{eq:phase-log-likelihood}
  \end{equation}
  At fixed nominal means, the Poisson model has no freely fitted parameter, whereas the NegBin model estimates only $\kappa^\phi$. We denote the resulting dispersion estimate by $\hat\kappa^\phi$. The subsequent diagnostics use these fixed-mean fits, without fitting a signal or background normalization.

\subsection{Count-model validation}
\label{sec:validation-methods}

  Validation addresses three complementary questions: are the discrepancies consistent with the assumed fluctuations, are the goodness-of-fit decisions stable under individual bin deletions, and does the residual distribution agree with the model? Each diagnostic probes the same observations from a different perspective. The use of complementary goodness-of-fit statistics is motivated by their different sensitivities to departures from the assumed probability model~\cite{ParticleDataGroup:2024cfk,Beaujean:2011zza}.

\subsubsection{Global discrepancies and local residuals}
\label{sec:global-discrepancy-methods}

  The expected variance generally changes from bin to bin because it depends on the count mean, as shown in \eqref{eq:poisson-model} and \eqref{eq:negbin-model}. This mean-dependent variance (heteroscedasticity across bins) implies that the same absolute difference between the observed and predicted counts may have different statistical relevance in different bins. We therefore construct signed local residuals that account for the expected fluctuations, and combine their squared contributions to assess the global discrepancy.

  For each count model $M\in\{\mathrm{P},\mathrm{NB}\}$, let $V_{i,M}^{\phi}\equiv\operatorname{Var}\boldsymbol{(}N_i^\phi\boldsymbol{)}$ denote the model variance, evaluated at $\hat\kappa^\phi$ for NegBin. The Pearson residual is the difference between the observed and expected counts expressed in units of the model standard deviation~\cite{Dunn:1996rqr}:
  \begin{equation}
    r_{P,i}^{\phi,M}=
    \frac{n_i^\phi-\mu_i^\phi}{\sqrt{V_{i,M}^\phi}}.
    \label{eq:pearson-residual}
  \end{equation}
  The sign of \eqref{eq:pearson-residual} distinguishes a count excess from a deficit. When the model is correct and its parameters are known, the random counterpart (with $N_i^\phi$ in place of $n_i^\phi$) has expectation zero and unit variance, but is not necessarily normally distributed~\cite{Dunn:1996rqr}. These moment properties need not hold exactly when the dispersion is estimated from the same data.

  An alternative discrepancy measure uses the likelihood in \eqref{eq:phase-log-likelihood}. Its reference is a saturated model, in which each bin has an independently adjustable mean, following the likelihood-ratio construction for binned data~\cite{Baker:1983tu,ParticleDataGroup:2024cfk}. For either count law, the maximizing mean is $\hat\mu_{i,\mathrm{sat}}^\phi=n_i^\phi$, including the limiting zero-count case. For NegBin, the fitted dispersion is held fixed at the same value in both likelihoods. Evaluating at the observed counts gives
  \begin{equation}
    \begin{aligned}
      \hat L_{\phi,M}&=\prod_{i=1}^{64}
      p_M\boldsymbol{(}n_i^\phi\mid\mu_i^\phi,\hat\kappa^\phi\boldsymbol{)},\\
      L_{\phi,M}^{\mathrm{sat}}&=\prod_{i=1}^{64}
      p_M\boldsymbol{(}n_i^\phi\mid n_i^\phi,\hat\kappa^\phi\boldsymbol{)},
    \end{aligned}
    \label{eq:saturated-likelihood}
  \end{equation}
  where the dispersion argument is omitted for Poisson, as in \eqref{eq:count-pmfs}. In the saturated likelihood, the first $n_i^\phi$ is the observed count and the second is its fitted mean. This likelihood supplies a reference for measuring discrepancies, not a competing physical prediction.

  The contribution of bin $i$ to the deviance is twice the log likelihood ratio between its saturated and candidate probabilities~\cite{Baker:1983tu}. The signed square root of this contribution defines the deviance residual~\cite{Dunn:1996rqr}:
  \begin{equation}
    \begin{aligned}
      d_{i,M}^\phi&=2\log\left[
      \frac{p_M\boldsymbol{(}n_i^\phi\mid n_i^\phi,\hat\kappa^\phi\boldsymbol{)}}
      {p_M\boldsymbol{(}n_i^\phi\mid\mu_i^\phi,\hat\kappa^\phi\boldsymbol{)}}\right],\\
      r_{D,i}^{\phi,M}&=\operatorname{sign}\boldsymbol{(}n_i^\phi-\mu_i^\phi\boldsymbol{)}
      \sqrt{d_{i,M}^\phi}.
    \end{aligned}
    \label{eq:deviance-residual}
  \end{equation}
  In \eqref{eq:deviance-residual}, the numerator is the saturated probability mass and the denominator is evaluated at the candidate mean. Their ratio is at least unity because the saturated mean maximizes the probability at fixed dispersion. Both residuals vanish when the count equals its predicted mean. Their squared contributions define the global Pearson and deviance statistics, respectively~\cite{Baker:1983tu,Dunn:1996rqr,ParticleDataGroup:2024cfk}:
  \begin{equation}
    \begin{aligned}
      X_{\phi,M}^2&=\sum_i(r_{P,i}^{\phi,M})^2
      =\sum_i\frac{(n_i^\phi-\mu_i^\phi)^2}{V_{i,M}^\phi},\\
      D_{\phi,M}&=\sum_i(r_{D,i}^{\phi,M})^2\\
      &=-2\log\left(\frac{\hat L_{\phi,M}}{L_{\phi,M}^{\mathrm{sat}}}\right)
      =\sum_i d_{i,M}^\phi.
    \end{aligned}
    \label{eq:global-discrepancies}
  \end{equation}
  The statistics in \eqref{eq:global-discrepancies} quantify the aggregate discrepancy, while the signed residuals locate and distinguish excesses from deficits across energy. The interpretation of a goodness-of-fit statistic depends on its sampling distribution under the assumed probability model~\cite{Beaujean:2011zza}. To assess whether a global discrepancy is unusually large under the candidate model, we approximate its sampling distribution by parametric Monte Carlo rather than imposing an asymptotic chi-squared reference~\cite{ParticleDataGroup:2024cfk,Grosso:2023scl}. The broader use of resampling-based calibration to determine goodness-of-fit probabilities in physical applications is discussed in Ref.~\cite{Christ:2024nxz}.

  For each phase and candidate, a pseudoexperiment is an independent realization of the entire $64$-bin spectrum. Each simulated count $n_{i,j}^\phi$ is a realization of a count random variable $N_{i,j}^\phi$ drawn at the same fixed nominal mean used for the observed bin. For NegBin, the generating dispersion is the estimate $\hat\kappa^\phi$ obtained from the observed spectrum. The generation step is summarized by
  \begin{equation}
    N_{i,j}^\phi\sim
    \begin{cases}
      \operatorname{Poisson}\boldsymbol{(}\mu_i^\phi\boldsymbol{)},& M=\mathrm{P},\\
      \operatorname{NegBin}\boldsymbol{(}\mu_i^\phi,\hat\kappa^\phi\boldsymbol{)},& M=\mathrm{NB},
    \end{cases}
    \label{eq:gof-pseudoexperiments}
  \end{equation}
  Here $j=1,\ldots,N_{\mathrm{MC}}$ indexes pseudoexperiments, with independent draws across bins and pseudoexperiments. Since estimating model parameters from the observed data can alter the sampling distribution of goodness-of-fit statistics and their associated $p$-values~\cite{Beaujean:2011zza}, we consider two complementary calibration procedures. The reference goodness-of-fit calibration holds the observed $\hat\kappa^\phi$ fixed when evaluating both discrepancies in each NegBin replica. A separate calibration refits $\kappa^\phi$ in each of the same replicas and uses that replica's estimate in the Pearson variance and in both the saturated and fixed-mean likelihoods for the deviance. The former procedure evaluates discrepancies relative to the fitted count law treated as fixed, whereas the latter reproduces the dispersion-estimation step within each pseudoexperiment. Both NegBin calibrations use the plug-in generating dispersion and do not constitute exact tests for every possible value of $\kappa^\phi$. For each calibration, the upper-tail $p$-value is the fraction of simulated discrepancies at least as large as the observed one. We use $N_{\mathrm{MC}}=5000$ and seed $1234$. A value below the nominal threshold $0.05$ indicates an unusual discrepancy conditional on the adopted count model and mean templates.

  To control the reporting precision of simulated tail probabilities, let $K$ denote the number of exceedances among $N_{\mathrm{MC}}$ independent pseudoexperiments and let $p_\star$ be their generating-model tail probability at the observed statistic. Conditional on the observed statistic and the simulation model, $K\sim\operatorname{Binomial}\boldsymbol{(}N_{\mathrm{MC}},p_\star\boldsymbol{)}$. With $\tilde p=K/N_{\mathrm{MC}}$, the estimated Monte Carlo standard errors of the ordinary and the plus-one estimators are
  \begin{equation}
    \begin{aligned}
      \widehat{\mathrm{SE}}_{\mathrm{MC}}\boldsymbol{(}\tilde p\boldsymbol{)}
      &=\sqrt{\frac{\tilde p(1-\tilde p)}{N_{\mathrm{MC}}}},\\
      \widehat{\mathrm{SE}}_{\mathrm{MC}}\boldsymbol{(}
      \frac{K+1}{N_{\mathrm{MC}}+1}\boldsymbol{)}
      &=\frac{\sqrt{N_{\mathrm{MC}}\tilde p(1-\tilde p)}}
      {N_{\mathrm{MC}}+1}.
    \end{aligned}
    \label{eq:mc-reporting-precision}
  \end{equation}
  Equation~\eqref{eq:mc-reporting-precision} follows from the binomial variance of $K$ and quantifies only the finite-simulation error, not experimental uncertainty. The dispersion and signal/background scale estimates and their statistical uncertainties are rounded following the Particle Data Group (PDG) conventions~\cite{ParticleDataGroup:2024cfk}. For Monte Carlo $p$-values we adopt the same rounding principle using their simulation standard errors, without displaying an uncertainty for every probability. Ranges use a common decimal position, with additional digits retained when needed to distinguish a probability from a decision threshold. All computations and decisions use unrounded values.

  The relationship between these two measures can be made explicit for Poisson counts. Expanding the binwise Poisson deviance~\cite{Baker:1983tu} about $n_i^\phi=\mu_i^\phi$ for $|n_i^\phi-\mu_i^\phi|/\mu_i^\phi\ll1$ gives
  \begin{equation}
    \begin{aligned}
      &2\left[n_i^\phi\log\left(\frac{n_i^\phi}{\mu_i^\phi}\right)
      -(n_i^\phi-\mu_i^\phi)\right]\\
      &\quad=\frac{(n_i^\phi-\mu_i^\phi)^2}{\mu_i^\phi}
      +\mathcal O\left(\frac{|n_i^\phi-\mu_i^\phi|^3}{(\mu_i^\phi)^2}\right).
    \end{aligned}
    \label{eq:deviance-pearson-limit}
  \end{equation}
  Here the Poisson deviance uses the convention $0\log 0=0$. Equation~\eqref{eq:deviance-pearson-limit} shows that the familiar quadratic Pearson contribution is the leading approximation to the Poisson deviance for small relative discrepancies; it does not require replacing the count likelihood by a Gaussian model. The exact deviance retains its likelihood interpretation outside this local approximation. We next examine the stability of the global conclusions in \autoref{sec:deletion-methods}.

\subsubsection{Robustness against dominant bins}
\label{sec:deletion-methods}

  A global goodness-of-fit decision may be sensitive to a small number of energy bins. Single-case deletion diagnostics provide a framework for assessing the influence of individual observations on statistical conclusions by examining changes in discrepancy measures and their associated $p$-values~\cite{Williams:1987deletion,Pena:2006validation}.

  To assess this dependence, each bin is deleted in turn, the dispersion is refitted for the NegBin model, and the goodness-of-fit test is recalibrated for the reduced spectrum using the Monte Carlo procedure described in \autoref{sec:global-discrepancy-methods}. This recalibration accounts for the reduced number of observations and, for NegBin, the updated dispersion estimate~\cite{Beaujean:2011zza,Grosso:2023scl}.

  Changes in the nominal $5\%$ compatibility decision identify bins that affect the global conclusion. Such changes indicate sensitivity to individual observations rather than evidence that the corresponding counts are necessarily inconsistent with the model. This diagnostic does not prescribe permanent removal of any observation from the analysis.

\subsubsection{Distributional consistency of residuals}
\label{sec:quantile-methods}

  After examining discrepancy magnitudes and the stability of goodness-of-fit decisions, we ask whether the residuals follow the distribution expected under the count law. Pearson residuals are standardized but retain a distribution that depends on the underlying model. Randomized quantile residuals map the probabilistic position of each count onto a common standard-normal scale, facilitating checks of its center, spread, and tails~\cite{Dunn:1996rqr}.

  For bin $i$, phase $\phi$, and model $M$, the cumulative distribution function at an integer argument $m\geq0$ is
  \begin{equation}
    \begin{aligned}
      F_{i,M}^\phi\boldsymbol{(}m\boldsymbol{)}
      &=\mathbb P_M\boldsymbol{(}N_i^\phi\leq m\boldsymbol{)}\\
      &=\sum_{k=0}^{m}
      p_M\boldsymbol{(}k\mid\mu_i^\phi,\kappa^\phi\boldsymbol{)},
    \end{aligned}
    \label{eq:count-cdf}
  \end{equation}
  where $F_{i,M}^\phi\boldsymbol{(}-1\boldsymbol{)}=0$, and the dispersion argument is omitted for Poisson. Initially, the distribution parameters in \eqref{eq:count-cdf} are regarded as known. For a continuous random variable with its correct continuous cumulative distribution function, evaluation of that function at the random variable gives a uniform variable on $(0,1)$ (the probability integral transform). Applying the inverse standard-normal cumulative distribution function $\Phi^{-1}$ then gives a standard-normal variable~\cite{Dunn:1996rqr}. For a discrete count, however, direct evaluation of $F_{i,M}^\phi\boldsymbol{(}N_i^\phi\boldsymbol{)}$ yields only the upper endpoints of the jumps of the cumulative distribution function and is not continuously uniform.

  To fill these intervals in probability space, let $A_i^\phi\sim\operatorname{Uniform}\boldsymbol{(}0,1\boldsymbol{)}$ be an auxiliary random variable independent of the count, with independent draws across bins. Following the randomization procedure introduced in Ref.~\cite{Dunn:1996rqr}, the randomized probability integral transform is
  \begin{equation}
    \begin{aligned}
      U_{i,M}^\phi={}&F_{i,M}^\phi\boldsymbol{(}N_i^\phi-1\boldsymbol{)}\\
      &+A_i^\phi\big[F_{i,M}^\phi\boldsymbol{(}N_i^\phi\boldsymbol{)}\\
      &\hspace{3.3em}-F_{i,M}^\phi\boldsymbol{(}N_i^\phi-1\boldsymbol{)}\big].
    \end{aligned}
    \label{eq:randomized-pit}
  \end{equation}
  The randomization in \eqref{eq:randomized-pit} acts on the probability interval associated with a count, not on the experimental count itself. For any realization $N_i^\phi=n_i^\phi$, the interval between the two cumulative probabilities has length
  \begin{equation}
    \begin{aligned}
      F_{i,M}^\phi\boldsymbol{(}n_i^\phi\boldsymbol{)}
      &-F_{i,M}^\phi\boldsymbol{(}n_i^\phi-1\boldsymbol{)}\\
      &=p_M\boldsymbol{(}n_i^\phi\mid\mu_i^\phi,\kappa^\phi\boldsymbol{)}.
    \end{aligned}
    \label{eq:cdf-jump}
  \end{equation}
  Conditional on the count, $U_{i,M}^\phi$ is uniform over its probability interval, with density equal to the reciprocal of the interval length. The probability of selecting that interval is the same length, by \eqref{eq:cdf-jump}. Their product gives the marginal density within the interval:
  \begin{equation}
    \begin{aligned}
      f_{U_{i,M}^\phi}\boldsymbol{(}u\boldsymbol{)}
      &=\frac{p_M\boldsymbol{(}n_i^\phi\mid\mu_i^\phi,\kappa^\phi\boldsymbol{)}}
      {p_M\boldsymbol{(}n_i^\phi\mid\mu_i^\phi,\kappa^\phi\boldsymbol{)}}=1,
    \end{aligned}
    \label{eq:randomized-pit-density}
  \end{equation}
  for $F_{i,M}^\phi\boldsymbol{(}n_i^\phi-1\boldsymbol{)}<u<F_{i,M}^\phi\boldsymbol{(}n_i^\phi\boldsymbol{)}$ whenever the associated mass is positive. Since these intervals partition $(0,1)$ up to endpoints of probability zero, $U_{i,M}^\phi\sim\operatorname{Uniform}\boldsymbol{(}0,1\boldsymbol{)}$. This explains why uniform randomization is required in the affine interpolation of \eqref{eq:randomized-pit}: an arbitrary auxiliary distribution would generally produce nonuniform density within each interval. It is the transformed variable, rather than the discrete count or its original probability law, that becomes continuous.

  The randomized quantile residual is obtained by applying the inverse standard-normal cumulative distribution function~\cite{Dunn:1996rqr}. For the observed count $n_i^\phi$ and a realization $a_i^\phi$ of the auxiliary variable, it is evaluated as
  \begin{equation}
    \begin{aligned}
      u_{i,M}^\phi={}&F_{i,M}^\phi\boldsymbol{(}n_i^\phi-1\boldsymbol{)}\\
      &+a_i^\phi\big[F_{i,M}^\phi\boldsymbol{(}n_i^\phi\boldsymbol{)}\\
      &\hspace{3.3em}-F_{i,M}^\phi\boldsymbol{(}n_i^\phi-1\boldsymbol{)}\big],\\
      r_{Q,i}^{\phi,M}={}&\Phi^{-1}\boldsymbol{(}u_{i,M}^\phi\boldsymbol{)}.
    \end{aligned}
    \label{eq:quantile-residual}
  \end{equation}
  By \eqref{eq:randomized-pit-density}, the random counterpart $\Phi^{-1}\boldsymbol{(}U_{i,M}^\phi\boldsymbol{)}$ follows the standard-normal distribution exactly when the count model is correct and its parameters are known. Thus, for example, a residual of $+2$ represents a randomized cumulative probability corresponding to two standard deviations above the mean of the standard-normal reference, not necessarily a count two model standard deviations above its mean. In the actual NegBin diagnostic, $F_{i,M}^\phi$ is evaluated at the fitted dispersion $\hat\kappa^\phi$. Exact standard normality is not guaranteed after estimation from the same observations~\cite{Dunn:1996rqr}. More generally, parameter estimation can modify the sampling distribution of goodness-of-fit diagnostics~\cite{Beaujean:2011zza}, which motivates calibration using simulated samples.

  To assess the distribution jointly, we order the residuals in \eqref{eq:quantile-residual} and compare them with the corresponding theoretical normal quantiles in a quantile--quantile (Q--Q) plot. For each phase--model combination, let $r_{(j)}^{\mathrm{obs}}$ denote the $j$th ordered observed quantile residual, with $j=1,\ldots,J$ and $J=64$. For notational simplicity, the phase and model indices are suppressed in the following expressions. The theoretical normal quantiles are evaluated at the midpoint plotting positions,
  \begin{equation}
    q_j=\Phi^{-1}\boldsymbol{(}
    \frac{j-\frac{1}{2}}{J}
    \boldsymbol{)}.
    \label{eq:normal-reference-quantiles}
  \end{equation}
  A worm plot is a detrended Q--Q plot that displays the differences between the ordered residuals and the corresponding normal reference quantiles~\cite{vanBuuren:2001worm}. Its observed ordinates are therefore
  \begin{equation}
    w_j^{\mathrm{obs}}=r_{(j)}^{\mathrm{obs}}-q_j,
    \label{eq:worm-ordinates}
  \end{equation}
  plotted against the quantiles in \eqref{eq:normal-reference-quantiles}. Departures from the horizontal zero reference in \eqref{eq:worm-ordinates} reveal discrepancies in the location, spread, or shape of the residual distribution.

  Simulation envelopes provide a reference against which the observed departures can be evaluated. Unlike pointwise envelopes, which describe the expected variability at individual quantiles, simultaneous envelopes are constructed to control departures over the entire diagnostic curve~\cite{Warton:2023envelopes}. This distinction is important because the ordered residuals are statistically dependent, even when the original counts are independent.

  Following the Monte Carlo calibration principle discussed in \autoref{sec:global-discrepancy-methods} and Ref.~\cite{Grosso:2023scl}, we generate $N_{\mathrm{MC}}=5000$ independent count spectra for each phase--model combination according to \eqref{eq:gof-pseudoexperiments}, with the nominal means held fixed. For NegBin, the counts are generated using the dispersion estimate $\hat\kappa^\phi$ obtained from the observed spectrum. However, the dispersion is refitted by maximum likelihood in each pseudoexperiment, yielding $\hat\kappa^{\phi,(b)}$ for replica $b=1,\ldots,N_{\mathrm{MC}}$. These replica-specific estimates are used to evaluate the corresponding cumulative distribution functions in \eqref{eq:quantile-residual}, thereby reproducing the dispersion-estimation step within the simulated diagnostic~\cite{Beaujean:2011zza,Warton:2023envelopes}.

  Independent auxiliary uniform variables are generated for each simulated spectrum, whereas the observed Poisson and NegBin residuals share the same auxiliary uniform realizations within each reactor phase. To prevent numerical divergences in the inverse normal transformation, the cumulative probabilities in \eqref{eq:quantile-residual} are restricted numerically to $[\epsilon_{\mathrm{mach}},1-\epsilon_{\mathrm{mach}}]$, where $\epsilon_{\mathrm{mach}}$ denotes machine precision. This restriction is a numerical safeguard rather than a modification of the theoretical randomized probability integral transform. The simulation workflow uses a base random seed of $1234$, with the worm-plot calculation continuing the random-number stream of the preceding validation computations.

  Let $r_{(j)}^{(b)}$ denote the $j$th ordered randomized quantile residual in replica $b$. The Monte Carlo center and scale at each ordered position are estimated from the simulated residuals as
  \begin{equation}
    \begin{aligned}
      \bar r_j&=\frac{1}{N_{\mathrm{MC}}}
      \sum_{b=1}^{N_{\mathrm{MC}}}r_{(j)}^{(b)},\\
      s_j&=\max\left\{
      \sqrt{\frac{1}{N_{\mathrm{MC}}-1}
      \sum_{b=1}^{N_{\mathrm{MC}}}
      (r_{(j)}^{(b)}-\bar r_j)^2},
      \sqrt{\epsilon_{\mathrm{mach}}}
      \right\}.
    \end{aligned}
    \label{eq:worm-mc-center-scale}
  \end{equation}
  Here $\bar r_j$ is the simulated mean of the residual at ordered position $j$, and $s_j$ is its sample standard deviation, subject to a numerical lower bound that prevents division by zero. The position-dependent scale in \eqref{eq:worm-mc-center-scale} accounts for the different fluctuations of ordered residuals across the center and tails of their distribution.

  Following the maximum-deviation approach to simultaneous simulation envelopes~\cite{Warton:2023envelopes}, we quantify the observed and simulated global departures through their maximum absolute standardized deviations from the Monte Carlo center:
  \begin{equation}
    \begin{aligned}
      M^{\mathrm{obs}}
      &=\max_{1\leq j\leq J}
      \left|
      \frac{r_{(j)}^{\mathrm{obs}}-\bar r_j}{s_j}
      \right|,\\
      M_b
      &=\max_{1\leq j\leq J}
      \left|
      \frac{r_{(j)}^{(b)}-\bar r_j}{s_j}
      \right|.
    \end{aligned}
    \label{eq:worm-max-deviation}
  \end{equation}
  Both the observed and simulated statistics in \eqref{eq:worm-max-deviation} use the same Monte Carlo center and scale. Each simulated spectrum also contributes to the estimation of these reference quantities through \eqref{eq:worm-mc-center-scale}.

  An important distinction is that the theoretical quantiles $q_j$ determine the graphical reference in \eqref{eq:worm-ordinates}, whereas the Monte Carlo centers $\bar r_j$ determine the statistical reference in \eqref{eq:worm-max-deviation}. The global statistic therefore measures standardized departures from the distribution expected under the adopted count model and estimation procedure, rather than unstandardized departures from the theoretical normal quantiles.

  The simultaneous critical value is obtained as the empirical $90$th percentile of the simulated maximum-deviation statistics:
  \begin{equation}
    c_{0.90}^{\mathrm{MC}}
    =\widehat Q_{0.90}\boldsymbol{(}
    M_1,\ldots,M_{N_{\mathrm{MC}}}
    \boldsymbol{)},
    \label{eq:worm-critical-value}
  \end{equation}
  where $\widehat Q_{0.90}$ denotes the empirical quantile evaluated using the default linear interpolation convention (type 7) implemented in \texttt{R}. Since the statistic in \eqref{eq:worm-max-deviation} takes the maximum over all ordered positions, the critical value in \eqref{eq:worm-critical-value} defines a simultaneous rather than a pointwise reference.

  The lower and upper limits of the simultaneous envelope on the ordered-residual scale are
  \begin{equation}
    \begin{aligned}
      L_j&=\bar r_j-c_{0.90}^{\mathrm{MC}}s_j,\\
      U_j&=\bar r_j+c_{0.90}^{\mathrm{MC}}s_j.
    \end{aligned}
    \label{eq:quantile-simultaneous-envelope}
  \end{equation}
  The same critical value is applied to every ordered position within a given phase--model combination. To represent the envelope on the worm-plot scale, the theoretical quantiles in \eqref{eq:normal-reference-quantiles} are subtracted from both limits:
  \begin{equation}
    \begin{aligned}
      L_j^{\mathrm{worm}}&=L_j-q_j,\\
      U_j^{\mathrm{worm}}&=U_j-q_j.
    \end{aligned}
    \label{eq:worm-simultaneous-envelope}
  \end{equation}
  The transformation in \eqref{eq:worm-simultaneous-envelope} preserves the simultaneous coverage event on the ordered-residual scale. Consequently, the observed worm curve lies entirely within its envelope if and only if $M^{\mathrm{obs}}\leq c_{0.90}^{\mathrm{MC}}$.

  Complementarily, the Monte Carlo upper-tail probability is estimated by the fraction of simulated maximum-deviation statistics at least as large as the observed statistic:
  \begin{equation}
    p_{\mathrm{MC}}=
    \frac{1}{N_{\mathrm{MC}}}
    \sum_{b=1}^{N_{\mathrm{MC}}}
    \mathbf{1}_{\{M_b\geq M^{\mathrm{obs}}\}},
    \label{eq:worm-mc-pvalue}
  \end{equation}
  where $\mathbf{1}$ is the indicator function. The estimator in \eqref{eq:worm-mc-pvalue} includes ties among the exceedances. The nominal significance level is $0.10$, corresponding to the simultaneous $90\%$ envelope. This procedure provides a global diagnostic rather than treating individual departures as independent tests.

  To identify the energy bins associated with departures from the envelope, the original bin indices are retained when ordering the observed residuals, with ties resolved by increasing bin index. Let $i_{(j)}$ denote the original energy-bin index associated with ordered position $j$. The set of bins outside the simultaneous envelope is
  \begin{equation}
    \mathcal B_{\mathrm{out}}=
    \left\{
    i_{(j)}:
    r_{(j)}^{\mathrm{obs}}<L_j
    \ \text{or}\
    r_{(j)}^{\mathrm{obs}}>U_j
    \right\}.
    \label{eq:worm-outside-bins}
  \end{equation}
  Equation~\eqref{eq:worm-outside-bins} identifies the individual contributions to the observed departures without interpreting them as independently significant observations. Equality with an envelope boundary is classified as remaining inside the envelope.

  The simultaneous envelopes and upper-tail probabilities are finite-simulation approximations. In particular, the NegBin calibration uses a plug-in generating dispersion, and the same simulated residuals are used to estimate the Monte Carlo center, scale, and reference distribution of the maximum-deviation statistic. Accordingly, the nominal coverage and significance levels should not be interpreted as exact finite-sample guarantees. The resulting diagnostic curves, global statistics, Monte Carlo probabilities, and energy bins outside the envelopes are presented in \autoref{sec:quantile-results}.

\subsection{Model comparison and selection}
\label{sec:model-comparison-methods}

  Once the absolute adequacy and residual behavior of the candidates have been examined, model selection addresses a different question: whether the improvement afforded by an additional dispersion parameter justifies its inclusion. We use a direct test of OFF-phase overdispersion and a relative comparison of all joint phase assignments. The two assessments complement, rather than replace, the validation diagnostics.

\subsubsection{Overdispersion and the Poisson limit}
\label{sec:overdispersion-methods}

  The Poisson distribution is recovered in the limit of infinite NegBin dispersion parameter, as discussed in \autoref{sec:count-distributions}. To examine how strongly the reconstructed counts constrain dispersion, we first compare the likelihood at each trial value of $\kappa^\phi$ with its maximum in the same reactor phase. At fixed nominal means, the NegBin likelihood is $L_{\phi,\mathrm{NB}}\boldsymbol{(}\kappa^\phi\boldsymbol{)}=\exp\boldsymbol{(}\ell_{\phi,\mathrm{NB}}\boldsymbol{(}\kappa^\phi\boldsymbol{)}\boldsymbol{)}$, with the log likelihood defined in \eqref{eq:phase-log-likelihood}. Following the likelihood-ratio framework for parameter inference~\cite{Cowan:2010js,ParticleDataGroup:2024cfk}, its relative likelihood and corresponding likelihood-ratio statistic are
  \begin{equation}
    \begin{aligned}
      \rho_\kappa^\phi\boldsymbol{(}\kappa^\phi\boldsymbol{)}
      &=\frac{L_{\phi,\mathrm{NB}}\boldsymbol{(}\kappa^\phi\boldsymbol{)}}
      {L_{\phi,\mathrm{NB}}\boldsymbol{(}\hat\kappa^\phi\boldsymbol{)}},\\
      \Delta q_\kappa^\phi\boldsymbol{(}\kappa^\phi\boldsymbol{)}
      &=-2\log\rho_\kappa^\phi\boldsymbol{(}\kappa^\phi\boldsymbol{)}\\
      &=2\left[\hat\ell_{\phi,\mathrm{NB}}
      -\ell_{\phi,\mathrm{NB}}\boldsymbol{(}\kappa^\phi\boldsymbol{)}\right].
    \end{aligned}
    \label{eq:dispersion-relative-likelihood}
  \end{equation}
  The two quantities in \eqref{eq:dispersion-relative-likelihood} describe the same likelihood curve: the maximum corresponds to $\rho_\kappa^\phi=1$ and $\Delta q_\kappa^\phi=0$, whereas larger values of $\Delta q_\kappa^\phi$ indicate a greater loss of likelihood relative to the best fit. The horizontal thresholds $\Delta q_\kappa^\phi=1,4,9$ correspond to the nominal two-sided $1\sigma$, $2\sigma$, and $3\sigma$ levels for one parameter under the regular Wilks approximation~\cite{Wilks:1938dza,Cowan:2010js}. We retain these references for both phases and calibrate the OFF confidence sets by Monte Carlo in \autoref{sec:dispersion-coverage-methods}, since the Poisson boundary and finite sample may affect their coverage. The following test first assesses specifically whether the OFF counts require a departure from Poisson sampling.

  The excess variation of the OFF counts can be tested directly by comparing the nested Poisson and NegBin probability laws. Define $\tau=1/\kappa^{\mathrm{OFF}}$, so that $\tau=0$ represents the Poisson limit. The null hypothesis is $H_0:\tau=0$, whereas the alternative is $H_1:\tau>0$. The likelihood-ratio statistic measures the improvement obtained by allowing additional dispersion~\cite{Wilks:1938dza,Self:1987liang,Cowan:2010js}:
  \begin{equation}
    T_{\mathrm{disp}}=2\left[\hat\ell_{\mathrm{OFF},\mathrm{NB}}-\hat\ell_{\mathrm{OFF},\mathrm{P}}\right].
    \label{eq:dispersion-lrt}
  \end{equation}
  Here, $\hat\ell_{\mathrm{OFF},\mathrm{NB}}$ and $\hat\ell_{\mathrm{OFF},\mathrm{P}}$ are the maximized OFF log likelihoods in \eqref{eq:phase-log-likelihood}. Since the fixed-mean NegBin likelihood converges to the Poisson likelihood as $\kappa^{\mathrm{OFF}}\to\infty$, the statistic in \eqref{eq:dispersion-lrt} is also the limiting value of $\Delta q_\kappa^{\mathrm{OFF}}$ in \eqref{eq:dispersion-relative-likelihood}. This connects the dispersion test directly with the likelihood curves. Even when Poisson is the correct model, fitting an additional parameter may improve the likelihood through sampling fluctuations alone. We must therefore assess whether the observed improvement is unusually large under $H_0$. Since $\tau=0$ lies on the boundary of the parameter space, the regularity conditions for the usual Wilks chi-squared approximation are not satisfied~\cite{Self:1987liang,Cowan:2010js}. We instead calibrate the statistic by a parametric bootstrap under the null hypothesis, following the general principle of determining reference distributions through simulated experiments~\cite{Beaujean:2011zza,ParticleDataGroup:2024cfk}.

  Unlike the goodness-of-fit simulations in \autoref{sec:global-discrepancy-methods}, which generate pseudoexperiments from each candidate, this bootstrap generates all pseudoexperiments from the OFF Poisson model. In each of the $N_{\mathrm{boot}}=10\,000$ replicas, the $64$ bin counts are drawn independently as
  \begin{equation}
    N_{i,j}^{\mathrm{OFF}}\sim\operatorname{Poisson}
    \boldsymbol{(}\mu_i^{\mathrm{OFF}}\boldsymbol{)},
    \quad i=1,\ldots,64,
    \label{eq:dispersion-bootstrap-sampling}
  \end{equation}
  where $\mu_i^{\mathrm{OFF}}=B_i^{\mathrm{OFF}}$ remains fixed, and $j=1,\ldots,N_{\mathrm{boot}}$ indexes the replicas. The corresponding realizations $n_{i,j}^{\mathrm{OFF}}$ constitute the generated spectra. Both candidate likelihoods are evaluated for each generated spectrum: Poisson has no free parameter at the nominal means, whereas NegBin is fitted by maximizing its likelihood over the dispersion parameter, using the same numerical procedure as for the observed counts. This refitting step reproduces the effect of parameter estimation on the sampling distribution of the test statistic~\cite{Beaujean:2011zza}. Denoting the resulting log likelihoods by $\hat\ell_{\mathrm{OFF},M}^{(j)}$, we obtain the simulated statistic
  \begin{equation}
    T_j=2\left[\hat\ell_{\mathrm{OFF},\mathrm{NB}}^{(j)}
    -\hat\ell_{\mathrm{OFF},\mathrm{P}}^{(j)}\right].
    \label{eq:dispersion-bootstrap-statistic}
  \end{equation}
  The ensemble of these statistics approximates the reference distribution of \eqref{eq:dispersion-lrt} under Poisson sampling. Let $T_{\mathrm{obs}}$ be the statistic evaluated on the reconstructed data. Following the Monte Carlo rank-based approach discussed in Ref.~\cite{Phipson:2010pvalues}, the upper-tail probability is estimated as
  \begin{equation}
    p_{\mathrm{MC}}=\frac{1+\sum_{j=1}^{N_{\mathrm{boot}}}\mathbb I\boldsymbol{(}T_j\geq T_{\mathrm{obs}}\boldsymbol{)}}{N_{\mathrm{boot}}+1},
    \label{eq:bootstrap-p-value}
  \end{equation}
  where $\mathbb I\boldsymbol{(}\cdot\boldsymbol{)}$ equals one when its condition is true and zero otherwise. The unit added to the numerator and denominator in \eqref{eq:bootstrap-p-value} accounts for the observed statistic in the Monte Carlo rank and avoids a zero probability estimate in a finite simulation~\cite{Phipson:2010pvalues}. The decision rule is to reject $H_0$ in favor of additional dispersion if $p_{\mathrm{MC}}<\alpha_{\mathrm{test}}$, with nominal significance level $\alpha_{\mathrm{test}}=0.05$. The bootstrap tail probability is not the probability that either model is true~\cite{Beaujean:2011zza,ParticleDataGroup:2024cfk}. It is also distinct from the relative model-selection rule developed in \autoref{sec:aic-methods}.

\subsubsection{Monte Carlo calibration of dispersion intervals}
\label{sec:dispersion-coverage-methods}

  The bootstrap test in \eqref{eq:bootstrap-p-value} addresses only the Poisson null. To estimate uncertainty in the OFF dispersion itself, we invert likelihood-ratio tests over trial dispersions, including the exact Poisson endpoint. This procedure follows the frequentist construction of confidence sets based on likelihood-ratio ordering introduced by Feldman and Cousins~\cite{Feldman:1997qc,ParticleDataGroup:2024cfk}.

  At each generating value $\kappa_0$ we generate independent OFF spectra $\mathbf{N}$ with the fixed nominal background means and compute $\Delta q_\kappa^{\mathrm{OFF}}\boldsymbol{(}\kappa_0;\mathbf{N}\boldsymbol{)}$ as in \eqref{eq:dispersion-relative-likelihood}, maximizing the likelihood separately for each spectrum. The maximization compares the finite-dispersion solution with the exact Poisson endpoint and the generating point. The latter comparison guards against numerical losses in the maximization.

  For a critical function $c_j$ at confidence level $\gamma_j=\mathbb P\boldsymbol{(}\chi_1^2\leq j^2\boldsymbol{)}$, $j=1,2,3$, the frequentist coverage at $\kappa_0$ is the probability that the confidence set constructed from a repeated experiment contains the generating parameter~\cite{Feldman:1997qc,ParticleDataGroup:2024cfk}. In terms of the likelihood-ratio statistic, this probability is
  \begin{equation}
    C_j\boldsymbol{(}\kappa_0\boldsymbol{)}=
    \mathbb P_{\kappa_0}\boldsymbol{[}
    \Delta q_\kappa^{\mathrm{OFF}}\boldsymbol{(}\kappa_0;\mathbf{N}\boldsymbol{)}
    \leq c_j\boldsymbol{(}\kappa_0\boldsymbol{)}\boldsymbol{]}.
    \label{eq:dispersion-coverage}
  \end{equation}
  The nominal procedure uses $c_j=j^2$, corresponding to the asymptotic reference discussed in \autoref{sec:overdispersion-methods} and based on Wilks' theorem~\cite{Wilks:1938dza}. However, the finite sample and the presence of the Poisson boundary may lead to departures from the nominal coverage.

  To account for these effects, the critical values are estimated separately at each generating parameter from the simulated likelihood-ratio distribution, following the conventional Monte Carlo implementation of the Feldman--Cousins construction~\cite{Feldman:1997qc,Berns:2023dmi}. Let $Q_{(r)}\boldsymbol{(}\kappa_0\boldsymbol{)}$ denote the $r$th ordered simulated likelihood-ratio statistic among $N_{\mathrm{cal}}$ replicas at $\kappa_0$. We estimate the point-dependent critical value by
  \begin{equation}
    \hat c_j\boldsymbol{(}\kappa_0\boldsymbol{)}=
    Q_{(r_j)}\boldsymbol{(}\kappa_0\boldsymbol{)},\qquad
    r_j=\left\lceil(N_{\mathrm{cal}}+1)\gamma_j\right\rceil.
    \label{eq:dispersion-calibrated-cut}
  \end{equation}
  The ceiling function selects an integer rank compatible with the finite-simulation correction associated with Monte Carlo rank-based tests~\cite{Phipson:2010pvalues}. Ties at the critical value in \eqref{eq:dispersion-calibrated-cut} are accepted. This construction accounts for the finite number of calibration replicas and avoids relying exclusively on an asymptotic reference distribution.

  Since the critical values are evaluated at a finite number of generating points, an interpolation procedure is required to construct the confidence sets over the parameter domain~\cite{Berns:2023dmi}. With $\tau=1/\kappa^{\mathrm{OFF}}$, including $\tau=0$ for Poisson, we interpolate the critical values linearly in $\tau$. Denoting this interpolated function by $\bar c_j$, the calibrated set for the observed spectrum is obtained by retaining all parameter values whose likelihood-ratio statistic does not exceed the corresponding critical value:
  \begin{equation}
    \mathcal C_j^{\mathrm{MC}}=\left\{
    \kappa\in[10,\infty]:
    \Delta q_\kappa^{\mathrm{OFF}}\boldsymbol{(}\kappa\boldsymbol{)}
    \leq\bar c_j\boldsymbol{(}1/\kappa\boldsymbol{)}\right\}.
    \label{eq:dispersion-calibrated-set}
  \end{equation}
  The notation $[10,\infty]$ in \eqref{eq:dispersion-calibrated-set} includes the Poisson endpoint explicitly. Unlike the nominal procedure, which uses the same critical value throughout the parameter domain, the calibrated procedure allows the acceptance threshold to depend on the generating dispersion.

  We use $42$ generating points (including Poisson and refinements near the crossings), with $N_{\mathrm{cal}}=40\,000$ replicas per point. A separate sample of $N_{\mathrm{test}}=15\,000$ replicas per point assesses the coverage of both procedures through
  \begin{equation}
    \widehat C_j\boldsymbol{(}\kappa_0\boldsymbol{)}=
    \frac{1}{N_{\mathrm{test}}}\sum_{m=1}^{N_{\mathrm{test}}}
    \mathbb I\boldsymbol{[}Q_m^{\mathrm{test}}\leq
    c_j\boldsymbol{(}\kappa_0\boldsymbol{)}\boldsymbol{]},
    \label{eq:dispersion-coverage-estimate}
  \end{equation}
  where $Q_m^{\mathrm{test}}$ is the likelihood-ratio statistic at the generating value in the independent verification sample and $c_j$ is either the nominal or the calibrated critical value. The use of an independent verification sample separates the estimation of the critical values from the assessment of their achieved coverage. The simulation standard error follows from \eqref{eq:mc-reporting-precision} with $N_{\mathrm{MC}}=N_{\mathrm{test}}$, conditional on the chosen cut.

  Calibration is approximate between grid points and conditional on the adopted OFF count law and fixed background means. Moreover, the empirical coverage is subject to finite-simulation uncertainty. The ON dispersion is not calibrated by these OFF simulations.

\subsubsection{Akaike information criterion}
\label{sec:aic-methods}

  Whereas the preceding test addresses the OFF-phase dispersion parameter, the Akaike information criterion (AIC)~\cite{Akaike:1974vps,Liddle:2007fy} provides a relative comparison of all candidate model assignments, using the maximized log-likelihood in \eqref{eq:phase-log-likelihood}. It is defined as
  \begin{equation}
    \operatorname{AIC}_{\phi,M}=-2\hat\ell_{\phi,M}+2K_{\phi,M},
    \label{eq:aic}
  \end{equation}
  where $K_{\phi,M}=\dim\boldsymbol{(}\boldsymbol{\vartheta}_{\phi,M}\boldsymbol{)}$ is the number of freely estimated parameters: $K_{\phi,\mathrm{P}}=0$ and $K_{\phi,\mathrm{NB}}=1$ in this validation. The first term in \eqref{eq:aic} rewards a larger maximized likelihood, whereas the second penalizes model complexity. Thus, adding a parameter is favored only if the improvement in fit compensates for its penalty~\cite{Akaike:1974vps}. The NegBin candidate retains its one fitted dispersion parameter in this comparison even if the fitted value approaches the Poisson limit. Conditional independence makes the phase log likelihoods, parameter counts, and AIC values additive. Let $m=(M,M')$ denote one of the four joint ON/OFF assignments in the candidate set $\mathcal M$. The joint AIC and its minimization rule are
  \begin{equation}
    \begin{aligned}
      \operatorname{AIC}_m&=\operatorname{AIC}_{\mathrm{ON},M}
      +\operatorname{AIC}_{\mathrm{OFF},M'},\\
      m^\star&\in\underset{m\in\mathcal M}{\operatorname{arg\,min}}\,
      \operatorname{AIC}_m.
    \end{aligned}
    \label{eq:aic-selection-rule}
  \end{equation}
  For example, within a given phase, NegBin has a lower AIC than Poisson if
  \begin{equation}
    \begin{aligned}
      2\left[\hat\ell_{\phi,\mathrm{NB}}-\hat\ell_{\phi,\mathrm{P}}\right]
      &>2\left[K_{\phi,\mathrm{NB}}-K_{\phi,\mathrm{P}}\right]\\
      &=2.
    \end{aligned}
    \label{eq:aic-dispersion-rule}
  \end{equation}
  Equality gives the same AIC for both models. Equation~\eqref{eq:aic-dispersion-rule} is a relative selection rule, not a hypothesis test: the penalty of two per additional parameter is not a critical value calibrated at a specified significance level. This distinction explains how the likelihood improvement can enter both the dispersion test and the AIC comparison while yielding different interpretations.

  To quantify the relative support of the four joint assignments, we define the AIC differences and Akaike weights following Ref.~\cite{Liddle:2007fy}:
  \begin{equation}
    \Delta_m=\operatorname{AIC}_m-\min_r\operatorname{AIC}_r,\qquad
    w_m=\frac{\exp\boldsymbol{(}-\Delta_m/2\boldsymbol{)}}
    {\sum_r\exp\boldsymbol{(}-\Delta_r/2\boldsymbol{)}}.
    \label{eq:akaike-weight}
  \end{equation}
  The weights $w_m$ sum to one and quantify relative AIC support among the four assignments; they are neither goodness-of-fit $p$-values nor, in the present analysis, posterior model probabilities. We use the conventional parameter-count penalty as a relative score, with caution at the Poisson boundary. A low AIC does not establish absolute adequacy, so the final choice also draws on global discrepancies, deletion robustness, the residual distribution, and the dispersion test. These diagnostics are complementary, not independent votes.

\subsection{Statistical inference}
\label{sec:inference-methods}

  With a count model selected at fixed nominal means, we introduce the parameters needed for inference. The NSI analysis constrains vector couplings, whereas the signal-strength analysis tests the presence and nominal normalization of an SM-shaped signal. A separate quadratic analysis constrains NSI using the excess spectrum. The mean functions below reflect these distinct questions and data representations, rather than successive additions to a single fit.

\subsubsection{Phase-specific count probabilities}
\label{sec:inference-count-probabilities}

  The NSI and signal-strength fits share the phase-specific count laws established in \autoref{sec:count-models} and the conditional-independence assumption underlying the factorized likelihood for binned counting experiments~\cite{Baker:1983tu,Cowan:2010js}. Their mean functions and the treatment of signal normalization differ according to the inferential objective.

  Let $M$ and $M'$ denote the candidate count models assigned to the $\mathrm{ON}$ and $\mathrm{OFF}$ phases, respectively, with $M,M'\in\{\mathrm P,\mathrm{NB}\}$. Using the Poisson and NegBin probability mass functions defined in \eqref{eq:count-pmfs}~\cite{Baker:1983tu,Bohme:2025nvu}, we introduce the model-indexed probability mass function for reactor phase $\phi$:
  \begin{equation}
    p_M^\phi\boldsymbol{(}n_i^\phi\mid\mu_i^\phi\boldsymbol{)}=
    \begin{cases}
      p_{\mathrm P}\boldsymbol{(}n_i^\phi\mid\mu_i^\phi\boldsymbol{)},
      &M=\mathrm P,\\
      p_{\mathrm{NB}}\boldsymbol{(}n_i^\phi\mid\mu_i^\phi,\kappa^\phi\boldsymbol{)},
      &M=\mathrm{NB}.
    \end{cases}
    \label{eq:model-indexed-pmf}
  \end{equation}
  For any NegBin component, the dispersion parameter $\kappa^\phi$ is held fixed at its maximum-likelihood estimate $\hat\kappa^\phi$ obtained during validation. Unlike the validation stage, the mean $\mu_i^\phi$ now depends on the parameters introduced for inference. Consequently, \eqref{eq:model-indexed-pmf} provides a common probabilistic representation for constructing the NSI and signal-strength likelihoods, conditional on the adopted count laws and their fitted dispersion parameters.

\subsubsection{Profile-likelihood inference for vector NSI}
\label{sec:nsi-profile-methods}

  To constrain vector NSI, the parameter of interest is $\boldsymbol{\varepsilon}$, which modifies the physical signal through \eqref{eq:nuclear-bin-prediction}. Following the likelihood-based treatment of systematic uncertainties in particle physics~\cite{Cowan:2010js,ParticleDataGroup:2024cfk}, we distinguish uncertainty in the background normalization from systematic uncertainty in the predicted signal normalization, treating both as nuisance parameters.

  A common positive factor $b$ multiplies the nominal background templates in both reactor phases, so that the OFF data constrain the background entering the ON prediction. A separate fractional signal-normalization shift $\alpha$ rescales the CEvNS prediction by $1+\alpha$. The latter is constrained by a Gaussian normalization term with width $\sigma_\alpha=0.169$, accounting for the experimental systematic uncertainties reported by the \CONUS{} collaboration~\cite{Ackermann:2025obx} and following the effective normalization prescription adopted in Refs.~\cite{DeRomeri:2025csu,Benavides:2026jdz}. Unlike an unconstrained signal-strength parameter, this nuisance shift carries a likelihood penalty for departures from zero. The resulting means are
  \begin{equation}
    \begin{aligned}
      \mu_i^{\mathrm{ON}}
      \boldsymbol{(}\boldsymbol{\varepsilon},b,\alpha\boldsymbol{)}
      &=bB_i^{\mathrm{ON}}+(1+\alpha)R_i
      \boldsymbol{(}\boldsymbol{\varepsilon}\boldsymbol{)},\\
      \mu_i^{\mathrm{OFF}}\boldsymbol{(}b\boldsymbol{)}
      &=bB_i^{\mathrm{OFF}}.
    \end{aligned}
    \label{eq:nsi-inference-means}
  \end{equation}
  Assuming conditional independence of the reconstructed counts, the count likelihood is the product of the phase-specific probabilities defined in \eqref{eq:model-indexed-pmf}. Following the constrained-likelihood framework for the treatment of nuisance parameters~\cite{Cowan:2010js,ParticleDataGroup:2024cfk}, the NSI likelihood includes, in addition, the Gaussian normalization constraint (up to a constant independent of the fitted parameters):
  \begin{equation}
    \begin{aligned}
      L_{M,M'}^{\mathrm{counts}}\boldsymbol{(}
      \boldsymbol{\varepsilon},b,\alpha\boldsymbol{)}
      ={}&\prod_{i=1}^{64}p_M^{\mathrm{ON}}
      \boldsymbol{[}n_i^{\mathrm{ON}}\mid
      \mu_i^{\mathrm{ON}}
      \boldsymbol{(}\boldsymbol{\varepsilon},b,\alpha\boldsymbol{)}\boldsymbol{]}\\
      &\times\prod_{i=1}^{64}p_{M'}^{\mathrm{OFF}}
      \boldsymbol{[}n_i^{\mathrm{OFF}}\mid
      \mu_i^{\mathrm{OFF}}\boldsymbol{(}b\boldsymbol{)}\boldsymbol{]},\\[2pt]
      L_{M,M'}^{\mathrm{NSI}}\boldsymbol{(}
      \boldsymbol{\varepsilon},b,\alpha\boldsymbol{)}
      ={}&L_{M,M'}^{\mathrm{counts}}\boldsymbol{(}
      \boldsymbol{\varepsilon},b,\alpha\boldsymbol{)}\\
      &\times\exp\boldsymbol{[}-\frac{1}{2}
      \left(\frac{\alpha}{\sigma_\alpha}\right)^2\boldsymbol{]}.
    \end{aligned}
    \label{eq:joint-likelihood}
  \end{equation}
  The Gaussian constraint in \eqref{eq:joint-likelihood} is an effective treatment of the global signal-normalization uncertainty, not a reconstruction of the full experimental nuisance likelihood. Its contribution to minus twice the log likelihood is the quadratic pull term, as employed in previous analyses of the \CONUS{} data~\cite{DeRomeri:2025csu,Benavides:2026jdz}. Taking the logarithm of \eqref{eq:joint-likelihood} gives
  \begin{equation}
    \begin{aligned}
      q_{M,M'}\boldsymbol{(}\boldsymbol{\varepsilon},b,\alpha\boldsymbol{)}
      &=-2\log L_{M,M'}^{\mathrm{NSI}}
      \boldsymbol{(}\boldsymbol{\varepsilon},b,\alpha\boldsymbol{)}\\
      &=-2\log L_{M,M'}^{\mathrm{counts}}
      \boldsymbol{(}\boldsymbol{\varepsilon},b,\alpha\boldsymbol{)}
      +\left(\frac{\alpha}{\sigma_\alpha}\right)^2.
    \end{aligned}
    \label{eq:nsi-penalized-objective}
  \end{equation}
  The Gaussian contribution in \eqref{eq:nsi-penalized-objective} is treated as a likelihood constraint within the frequentist analysis, rather than as a Bayesian prior for $\alpha$. For any NegBin component, the dispersion parameter is fixed at its estimate obtained during validation, using its full numerical precision. The numerical NSI scans use the nuisance domain $\mathcal D=[0.2,2]\times[-0.95,0.95]$ for $(b,\alpha)$; the selected count likelihood uses a positive numerical floor of $10^{-12}$ for the means.

  To account for the nuisance parameters when evaluating a specified NSI point, we adopt the profile-likelihood approach~\cite{Cowan:2010js,ParticleDataGroup:2024cfk}. For a given pair $(M,M')$, double hats denote conditional maximum-likelihood estimates at fixed $\boldsymbol{\varepsilon}$, whereas single hats denote components of a joint optimum. Let $\mathcal S$ denote the NSI parameter domain of the particular scan, with the other couplings fixed. Since maximizing the likelihood is equivalent to minimizing minus twice its logarithm, we define the conditional and global estimators, respectively, by
  \begin{equation}
    \begin{aligned}
      &\bigl(\hat{\hat b}_{M,M'}\boldsymbol{(}\boldsymbol{\varepsilon}\boldsymbol{)},
      \hat{\hat\alpha}_{M,M'}\boldsymbol{(}\boldsymbol{\varepsilon}\boldsymbol{)}\bigr)\\
      &\qquad\in\underset{(b,\alpha)\in\mathcal D}{\operatorname{argmin}}
      q_{M,M'}\boldsymbol{(}\boldsymbol{\varepsilon},b,\alpha\boldsymbol{)},\\[2pt]
      &\bigl(\hat{\boldsymbol{\varepsilon}}_{M,M'},\hat b_{M,M'},\hat\alpha_{M,M'}\bigr)\\
      &\qquad\in\underset{\boldsymbol{\varepsilon}\in\mathcal S,\,(b,\alpha)\in\mathcal D}{\operatorname{argmin}}
      q_{M,M'}\boldsymbol{(}\boldsymbol{\varepsilon},b,\alpha\boldsymbol{)}.
    \end{aligned}
    \label{eq:nsi-profile-estimators}
  \end{equation}
  The estimators in \eqref{eq:nsi-profile-estimators} define the constrained profile-likelihood-ratio statistic~\cite{Cowan:2010js}:
  \begin{equation}
    \begin{aligned}
      \Delta q_{M,M'}\boldsymbol{(}\boldsymbol{\varepsilon}\boldsymbol{)}
      ={}&-2\log
      \frac{L_{M,M'}^{\mathrm{NSI}}
      \boldsymbol{(}\boldsymbol{\varepsilon},
      \hat{\hat b}_{M,M'},\hat{\hat\alpha}_{M,M'}\boldsymbol{)}}
      {L_{M,M'}^{\mathrm{NSI}}
      \boldsymbol{(}\hat{\boldsymbol{\varepsilon}}_{M,M'},
      \hat b_{M,M'},\hat\alpha_{M,M'}\boldsymbol{)}},
    \end{aligned}
    \label{eq:delta-q}
  \end{equation}
  where the double-hat arguments $\boldsymbol{\varepsilon}$ in the numerator have been suppressed for compactness. Equivalently, \eqref{eq:delta-q} is the difference between the conditional and global minima of \eqref{eq:nsi-penalized-objective}.

  Equation~\eqref{eq:delta-q} measures the likelihood loss at a specified NSI point relative to the global optimum. Degenerate global maximizers give the same denominator. The model pair is selected in \autoref{sec:validation}, and its fitted dispersion is held fixed. Following the asymptotic likelihood-ratio prescription of Wilks~\cite{Wilks:1938dza,Cowan:2010js,ParticleDataGroup:2024cfk}, we use nominal thresholds of $\Delta q=1,4,9$ for one parameter at $1\sigma$, $2\sigma$, and $3\sigma$, respectively, and $\Delta q=4.605$ for two parameters at $90\%$ C.L. These thresholds are used as conventional references for comparing the NSI regions. Boundaries, degeneracies, and weak identification can invalidate the usual chi-squared approximation. In particular, the reduction to a single squared effective charge discussed in \eqref{eq:nsi-statistic-charge-reduction} can lead to nonidentifiability in the two-parameter scans. No dedicated coverage calibration has been performed, and the reported confidence levels must therefore be interpreted as nominal.

\subsubsection{Signal significance and SM compatibility}
\label{sec:signal-profile-methods}

  Unlike the NSI scan in \autoref{sec:nsi-profile-methods}, the signal-strength analysis fixes the spectral shape to the SM prediction and tests its presence and nominal normalization.

  Following the signal-strength parametrization used in likelihood-based tests in particle physics~\cite{Cowan:2010js}, we introduce a nonnegative parameter $s$, with $s=0$ denoting the absence of the modeled signal and $s=1$ the nominal SM prediction:
  \begin{equation}
    \mu_i^{\mathrm{ON}}\boldsymbol{(}s,b\boldsymbol{)}
    =bB_i^{\mathrm{ON}}+sR_i^{\mathrm{SM}},
    \qquad
    \mu_i^{\mathrm{OFF}}\boldsymbol{(}b\boldsymbol{)}
    =bB_i^{\mathrm{OFF}}.
    \label{eq:signal-strength-means}
  \end{equation}
  The signal-strength likelihood uses the count-probability product of \eqref{eq:joint-likelihood} with the means in \eqref{eq:signal-strength-means}, without the Gaussian constraint. This follows the factorized likelihood framework for signal and background components~\cite{Baker:1983tu,Cowan:2010js}. Here $s$ is a parameter of interest, not the constrained NSI nuisance shift $\alpha$.
  \begin{equation}
    \begin{aligned}
      L_{M,M'}^{\mathrm{sig}}\boldsymbol{(}s,b\boldsymbol{)}={}&
      \prod_{i=1}^{64}p_M^{\mathrm{ON}}
      \boldsymbol{[}n_i^{\mathrm{ON}}\mid
      \mu_i^{\mathrm{ON}}\boldsymbol{(}s,b\boldsymbol{)}\boldsymbol{]}\\
      &\times\prod_{i=1}^{64}p_{M'}^{\mathrm{OFF}}
      \boldsymbol{[}n_i^{\mathrm{OFF}}\mid
      \mu_i^{\mathrm{OFF}}\boldsymbol{(}b\boldsymbol{)}\boldsymbol{]}.
    \end{aligned}
    \label{eq:signal-likelihood}
  \end{equation}
  In \eqref{eq:signal-likelihood}, the fitted dispersion is held fixed for any NegBin component. Following the same hat convention, $\hat s_{M,M'}$ and $\hat b_{M,M'}$ now denote the joint maximum-likelihood estimates obtained from this signal-strength likelihood over $s\geq0$ and $b>0$, not from the NSI likelihood. For a fixed reference signal strength $s_{\mathrm{ref}}$, the conditional estimate $\hat{\hat b}_{s_{\mathrm{ref}};M,M'}$ maximizes this likelihood over $b>0$ with $s=s_{\mathrm{ref}}$. Following Refs.~\cite{Cowan:2010js,ParticleDataGroup:2024cfk}, the profile-likelihood-ratio statistic under the model pair $(M,M')$ is defined as
  \begin{equation}
    \Delta q_{\mathrm{ref};M,M'}=-2\log
    \frac{L_{M,M'}^{\mathrm{sig}}
    \boldsymbol{(}s_{\mathrm{ref}},
    \hat{\hat b}_{s_{\mathrm{ref}};M,M'}\boldsymbol{)}}
    {L_{M,M'}^{\mathrm{sig}}
    \boldsymbol{(}\hat s_{M,M'},\hat b_{M,M'}\boldsymbol{)}},
    \label{eq:signal-profile-statistic}
  \end{equation}
  For uncertainty estimation, we also profile each normalization parameter while optimizing the other over its admissible domain~\cite{Cowan:2010js,ParticleDataGroup:2024cfk}. In the selected model, writing the joint maximum as $\hat\ell$, the two profile statistics are
  \begin{equation}
    \begin{aligned}
      \Delta q_b\boldsymbol{(}b\boldsymbol{)}&=2\left[
      \hat\ell-\sup_{s\geq0}\log L_{\mathrm{sig}}\boldsymbol{(}s,b\boldsymbol{)}\right],\\
      \Delta q_s\boldsymbol{(}s\boldsymbol{)}&=2\left[
      \hat\ell-\sup_{b>0}\log L_{\mathrm{sig}}\boldsymbol{(}s,b\boldsymbol{)}\right].
    \end{aligned}
    \label{eq:normalization-interval-profiles}
  \end{equation}
  For $x=b,s$, the lower and upper crossings $x_L,x_U$ of $\Delta q_x=1$ in \eqref{eq:normalization-interval-profiles} define the nominal $68.27\%$ interval around the maximum; the corresponding errors are $\hat x-x_L$ and $x_U-\hat x$. These intervals use the one-parameter asymptotic likelihood-ratio reference~\cite{Wilks:1938dza,Cowan:2010js,ParticleDataGroup:2024cfk} and are conditional on the fitted, fixed OFF dispersion and selected count law.

  For $s_{\mathrm{ref}}=0$, the alternative is $s>0$. Since $s=0$ lies on the boundary of the admissible parameter space, we use the one-sided asymptotic discovery prescription of Ref.~\cite{Cowan:2010js}, rather than the ordinary two-sided Wilks reference. With $s\geq0$, a fit at $\hat s_{M,M'}=0$ gives $\Delta q_{0;M,M'}=0$. For an observed positive signal, the asymptotic significance is $Z_0=\sqrt{\Delta q_{0;M,M'}}$, with $p_0=1-\Phi\boldsymbol{(}Z_0\boldsymbol{)}$.

  For compatibility with the SM, $s_{\mathrm{ref}}=1$ lies in the interior of the signal-strength domain. Using the one-parameter asymptotic likelihood-ratio reference~\cite{Wilks:1938dza,Cowan:2010js,ParticleDataGroup:2024cfk}, the signed displacement is $Z_{\mathrm{SM}}=\operatorname{sign}\boldsymbol{(}\hat s-1\boldsymbol{)}\sqrt{\Delta q_{1;M,M'}}$, giving a two-sided probability $p_{\mathrm{bil}}=2[1-\Phi\boldsymbol{(}|Z_{\mathrm{SM}}|\boldsymbol{)}]$ and, for an upward displacement, a directional probability $p_+=1-\Phi\boldsymbol{(}Z_{\mathrm{SM}}\boldsymbol{)}$. Here $\hat s$ denotes the estimate for the model pair under consideration. These probabilities refer to different alternatives: the detection test assesses the absence of the modeled signal, whereas the SM tests assess its nominal normalization. The reference distributions are asymptotic and conditional on the adopted reconstructed-data likelihood, fixed templates, and fitted dispersion; no dedicated calibration of these signal-strength tests is performed.

\subsubsection{Chi-squared inference for vector NSI}
\label{sec:chi-square-methods}

  The constrained count likelihood is compared with a quadratic analysis of the published excess spectrum, following the statistical approach adopted in Refs.~\cite{DeRomeri:2025csu,Benavides:2026jdz}. Because the background has already been subtracted in this representation, no separate background-scale parameter $b$ is fitted; this does not imply that background uncertainty is absent. Both NSI analyses employ the same effective Gaussian signal-normalization constraint, unlike the conditional signal-strength analysis.

  The quadratic comparison uses the separate 19-bin background-subtracted spectrum reported in Fig.~3a of Ref.~\cite{Ackermann:2025obx}, whose bin centers range from $165$ to $345\,\mathrm{eVee}$. Let $R_j^{\mathrm{exp}}$ be the reported excess in bin $j$, and $\sigma_{j,+}$, $\sigma_{j,-}$ its upper and lower uncertainties, respectively. Following the weighted least-squares framework~\cite{ParticleDataGroup:2024cfk} and the normalization prescription adopted in Refs.~\cite{DeRomeri:2025csu,Benavides:2026jdz}, the statistic is
  \begin{equation}
    \chi^2\boldsymbol{(}\boldsymbol{\varepsilon},\alpha\boldsymbol{)}
    =\sum_{j=1}^{19}
    \frac{\left[R_j^{\mathrm{exp}}-(1+\alpha)
    R_j^{\mathrm{th}}\boldsymbol{(}\boldsymbol{\varepsilon}\boldsymbol{)}\right]^2}
    {\sigma_j^2}
    +\left(\frac{\alpha}{\sigma_\alpha}\right)^2,
    \label{eq:chi-square}
  \end{equation}
  Here $R_j^{\mathrm{th}}\boldsymbol{(}\boldsymbol{\varepsilon}\boldsymbol{)}$ is the predicted excess, $\sigma_j=\boldsymbol{(}\sigma_{j,+}+\sigma_{j,-}\boldsymbol{)}/2$ is the symmetrized uncertainty, following Ref.~\cite{Benavides:2026jdz}, and the signal-normalization shift $\alpha$ has the same constraint width $\sigma_\alpha=0.169$ as in \eqref{eq:joint-likelihood}. The spectral contribution to \eqref{eq:chi-square} uses a diagonal covariance approximation because the full bin-to-bin covariance matrix is unavailable for the reconstructed public spectrum. The common normalization nuisance nevertheless introduces a coherent variation of the predicted signal across all bins. Under independent Gaussian measurement errors with known variances and a Gaussian normalization constraint, the statistic in \eqref{eq:chi-square} is equivalent to minus twice the corresponding log likelihood up to a parameter-independent constant~\cite{ParticleDataGroup:2024cfk}. Here it is adopted as a quadratic approximation to the published excess data.

  Because \eqref{eq:chi-square} is quadratic in $\alpha$, its unconstrained conditional minimizer is obtained analytically by setting its derivative with respect to $\alpha$ equal to zero, as in Ref.~\cite{Benavides:2026jdz}:
  \begin{equation}
    \alpha_{\mathrm{unc}}\boldsymbol{(}\boldsymbol{\varepsilon}\boldsymbol{)}
    =
    \frac{\displaystyle\sum_{j=1}^{19}
    \frac{R_j^{\mathrm{th}}\boldsymbol{(}\boldsymbol{\varepsilon}\boldsymbol{)}
    \left[R_j^{\mathrm{exp}}-
    R_j^{\mathrm{th}}\boldsymbol{(}\boldsymbol{\varepsilon}\boldsymbol{)}\right]}
    {\sigma_j^2}}
    {\displaystyle\sum_{j=1}^{19}
    \frac{\left[R_j^{\mathrm{th}}\boldsymbol{(}\boldsymbol{\varepsilon}\boldsymbol{)}\right]^2}
    {\sigma_j^2}+\sigma_\alpha^{-2}}.
    \label{eq:chi-square-alpha-profile}
  \end{equation}
  For the NSI scans, the minimizer in \eqref{eq:chi-square-alpha-profile} is restricted to the interval $[-0.95,0.95]$, consistently with the normalization domain adopted in \autoref{sec:nsi-profile-methods}. Since the quadratic statistic is strictly convex in $\alpha$, the conditional estimate is the projection of the unconstrained solution onto this interval:
  \begin{equation}
    \begin{aligned}
      \hat{\hat\alpha}\boldsymbol{(}\boldsymbol{\varepsilon}\boldsymbol{)}
      =\min\bigl\{0.95,\max\bigl\{-0.95,
      \alpha_{\mathrm{unc}}\boldsymbol{(}\boldsymbol{\varepsilon}\boldsymbol{)}
      \bigr\}\bigr\}.
    \end{aligned}
    \label{eq:chi-square-alpha-bounded}
  \end{equation}
  The nuisance scale $\alpha$ is the only profiled nuisance parameter in the quadratic statistic. The estimates $\hat{\boldsymbol{\varepsilon}}$ and $\hat\alpha$ denote a joint minimizer of \eqref{eq:chi-square} over the NSI scan and the same normalization interval. Following the profiled chi-squared approach~\cite{ParticleDataGroup:2024cfk,Benavides:2026jdz}, the corresponding difference relative to the global minimum is
  \begin{equation}
    \begin{aligned}
      \Delta\chi^2\boldsymbol{(}\boldsymbol{\varepsilon}\boldsymbol{)}
      ={}&\chi^2\boldsymbol{(}\boldsymbol{\varepsilon},
      \hat{\hat\alpha}\boldsymbol{(}\boldsymbol{\varepsilon}\boldsymbol{)}
      \boldsymbol{)}\\
      &-\chi^2\boldsymbol{(}\hat{\boldsymbol{\varepsilon}},
      \hat\alpha\boldsymbol{)}.
    \end{aligned}
    \label{eq:delta-chi-square}
  \end{equation}
  The estimates in the subtracted term are specific to the quadratic statistic. To ensure a comparison at common nominal confidence levels, we adopt for \eqref{eq:delta-chi-square} the same asymptotic thresholds used for \eqref{eq:delta-q}, following the conventional Wilks prescription~\cite{Wilks:1938dza,ParticleDataGroup:2024cfk,Benavides:2026jdz}. The limitations concerning parameter identification and uncalibrated coverage discussed in \autoref{sec:nsi-profile-methods} also apply to these regions.

  Since the two analyses use different representations of the same measurement, with different energy ranges, fluctuation models, and background treatments, their comparison cannot isolate the effect of the statistical formula. Consequently, differences between the resulting NSI regions quantify the combined effect of these methodological choices rather than exclusively the difference between a count likelihood and a quadratic approximation.

\subsection{Geometric diagnostics for two-dimensional NSI regions}
\label{sec:nsi-geometry}

  The two-dimensional NSI regions exhibit characteristic degeneracies originating from the dependence of the \CEvNS{} cross section on the effective nuclear weak charge~\cite{Barranco:2005yy,Giunti:2019xpr,DeRomeri:2025csu}. To distinguish the geometric properties imposed by the physical model from those resulting from the statistical treatment, we derive the analytic charge-level curves and use their geometric properties as reference quantities for the numerical comparison between $\Delta q$ and $\Delta\chi^2$.

  With one effective germanium nucleus, the NSI dependence of the recoil spectrum is governed by the single squared charge $S\boldsymbol{(}\boldsymbol{\varepsilon}\boldsymbol{)}=\mathcal Q_{\mathrm{eff}}^2\boldsymbol{(}\boldsymbol{\varepsilon}\boldsymbol{)}$, defined in \eqref{eq:nsi-effective-charge}, with $Z=32$ and $N_{\mathrm{eff}}=40.7053$. This dependence explains the degeneracy patterns observed in single-target \CEvNS{} analyses~\cite{Barranco:2005yy,Giunti:2019xpr}. An explicit sum over germanium isotopes would not factorize exactly into a single squared charge. Under the adopted approximation, the profiled statistics in \eqref{eq:delta-q} and \eqref{eq:delta-chi-square} have the form
  \begin{equation}
    \Delta\mathcal{T}\boldsymbol{(}\boldsymbol{\varepsilon}\boldsymbol{)}
    =h_{\Delta\mathcal{T}}
    \boldsymbol{(}S\boldsymbol{(}\boldsymbol{\varepsilon}\boldsymbol{)}\boldsymbol{)},
    \qquad
    \Delta\mathcal{T}\in\{\Delta q_{M,M'},\Delta\chi^2\}.
    \label{eq:nsi-statistic-charge-reduction}
  \end{equation}
  The function $h_{\Delta\mathcal T}$ maps the squared charge into the profiled statistic. Consequently, different NSI configurations yielding the same value of $S$ produce identical predictions and identical values of the corresponding profiled statistic. The statistical treatment determines the accepted charge levels, whereas the charge algebra determines the shapes and orientations of their level curves.

  The cancellation between the up- and down-quark contributions has been extensively discussed in the context of \CEvNS{} degeneracies~\cite{Barranco:2005yy,Giunti:2019xpr,DeRomeri:2022twg}. To characterize this cancellation explicitly, we introduce an orthogonal reparametrization of the NSI couplings within a fixed flavor channel into charge-sensitive and cancellation directions:
  \begin{equation}
    \begin{aligned}
      \varepsilon_{\parallel}^{V}
      &=\frac{C_d\varepsilon^{dV}+C_u\varepsilon^{uV}}
      {\sqrt{C_d^2+C_u^2}},\\
      \varepsilon_{\perp}^{V}
      &=\frac{C_u\varepsilon^{dV}-C_d\varepsilon^{uV}}
      {\sqrt{C_d^2+C_u^2}}.
    \end{aligned}
    \label{eq:nsi-rotated-couplings}
  \end{equation}
  Here the flavor indices are suppressed for compactness, and the nuclear coefficients are $C_u=2Z+N_{\mathrm{eff}}$ and $C_d=Z+2N_{\mathrm{eff}}$. Since the NSI contribution to the charge is $C_d\varepsilon^{dV}+C_u\varepsilon^{uV}=\sqrt{C_d^2+C_u^2}\,\varepsilon_{\parallel}^{V}$, only the parallel combination enters the effective charge. The perpendicular combination therefore defines an exactly degenerate direction within the single-effective-nucleus approximation.

  For the flavor-diagonal plane $x=\varepsilon_{ee}^{dV}$ and $y=\varepsilon_{ee}^{uV}$, the characteristic parallel bands originate from the quadratic dependence of the cross section on the sum of the SM and NSI amplitudes~\cite{Barranco:2005yy,DeRomeri:2025csu,Benavides:2026jdz}. Solving \eqref{eq:nsi-effective-charge} at a fixed nonnegative charge level $S_0$ gives
  \begin{equation}
    \begin{aligned}
      y&=-\frac{C_d}{C_u}x
      +\frac{-Q_V^{\mathrm{SM}}\pm\sqrt{S_0}}{C_u},\\
      m_{\mathrm{theory}}&=-\frac{C_d}{C_u}\simeq-1.0831.
    \end{aligned}
    \label{eq:nsi-linear-bands}
  \end{equation}
  Equation~\eqref{eq:nsi-linear-bands} describes two parallel straight lines for $S_0>0$, corresponding to the two possible signs of the effective weak charge. At $S_0=0$, the two lines coincide. The pure flavor-changing NSI plane has the same cancellation slope because its charge contribution involves the same linear combination of quark couplings, although it does not interfere with the SM amplitude. In that case, the squared charge is $(Q_V^{\mathrm{SM}})^2+(C_dx+C_uy)^2$, and the two branches exist for $S_0\geq(Q_V^{\mathrm{SM}})^2$. The resulting parallel bands are consistent with those obtained in previous \CONUS{} analyses~\cite{DeRomeri:2025csu,Benavides:2026jdz}.

  The mixed flavor-diagonal and flavor-changing planes instead exhibit closed charge-level curves, a consequence of the incoherent addition of the corresponding flavor contributions~\cite{Barranco:2005yy,DeRomeri:2025csu}. To characterize these curves analytically, let $x=\varepsilon_{ee}^{qV}$ denote a diagonal coupling and $y=\varepsilon_{e\ell}^{q'V}$ a flavor-changing coupling, where $\ell\in\{\mu,\tau\}$ and $q,q'\in\{u,d\}$. The two flavor-changing channels are considered separately and yield identical predictions under the adopted assumptions. Setting the remaining NSI couplings to zero, \eqref{eq:nsi-effective-charge} reduces to
  \begin{equation}
    S_{qq'}\boldsymbol{(}x,y\boldsymbol{)}
    =\boldsymbol{(}Q_V^{\mathrm{SM}}+C_qx\boldsymbol{)}^2
    +C_{q'}^2y^2,
    \qquad q,q'\in\{u,d\},
    \label{eq:nsi-closed-charge}
  \end{equation}
  where $C_q$ and $C_{q'}$ select the appropriate nuclear coefficients. Completing the square in \eqref{eq:nsi-closed-charge}, the level set $S_{qq'}\boldsymbol{(}x,y\boldsymbol{)}=S_0$, for $S_0>0$, takes the canonical form
  \begin{equation}
    \frac{(x-c_q)^2}{S_0/C_q^2}
    +\frac{y^2}{S_0/C_{q'}^2}=1,
    \qquad
    c_q=-\frac{Q_V^{\mathrm{SM}}}{C_q}.
    \label{eq:nsi-conic}
  \end{equation}
  Equation~\eqref{eq:nsi-conic} describes a circle when $q=q'$ and an ellipse when $q\neq q'$, centered at $(c_q,0)$. At $S_0=0$, the level set reduces to the center. For positive charge levels, the semiaxes along the coordinate directions are $\sqrt{S_0}/C_q$ and $\sqrt{S_0}/C_{q'}$, respectively. Thus, interchanging the quark assignments in the mixed-quark planes changes the center and rotates the major-axis orientation by $90^\circ$, while preserving the semiaxis ratio.

  Writing the ellipse semiaxes as $a_{\mathrm{maj}}\geq a_{\mathrm{min}}>0$, we define their ratio $\rho_{\mathrm{theory}}$ and eccentricity $e_{\mathrm{theory}}$ as geometric reference quantities. Evaluating these quantities with the nuclear coefficients and SM charge adopted in this work gives
  \begin{equation}
    \begin{aligned}
      c_d&\simeq0.1730,
      &c_u&\simeq0.1874,\\
      \rho_{\mathrm{theory}}
      &=\frac{a_{\mathrm{maj}}}{a_{\mathrm{min}}}
      =\frac{C_d}{C_u}\simeq1.0831,\\
      e_{\mathrm{theory}}
      &=\sqrt{1-\frac{a_{\mathrm{min}}^2}{a_{\mathrm{maj}}^2}}
      =\sqrt{1-\frac{C_u^2}{C_d^2}}
      \simeq0.3842.
    \end{aligned}
    \label{eq:nsi-geometric-invariants}
  \end{equation}
  The quantities in \eqref{eq:nsi-geometric-invariants} are derived from the effective weak charge and are independent of the positive charge level $S_0$. Together with the cancellation slope in \eqref{eq:nsi-linear-bands}, they establish analytic reference values against which the numerical contours can be checked.

  In the effective-nucleus model, the centers, cancellation slopes, orientations, semiaxis ratios, and eccentricities are determined by the weak charge and nuclear coefficients. The widths, radii, semiaxis lengths, and accepted areas instead depend on the charge levels admitted by each statistic. We exploit this separation to compare the two statistical constructions without interpreting their common charge-induced degeneracies as independent statistical findings. The geometric checks and quantitative comparison of the allowed regions are presented in \autoref{sec:nsi-geometric-results}.

\section{Results}
\label{sec:results}

  We first present the validation and selection of the count model for the reconstructed reactor spectra. Using the selected count law, we then assess the significance of the \CEvNS{} signal and its compatibility with the SM prediction. Finally, we compare the vector NSI allowed regions obtained from the reconstructed count likelihood and the separate excess-spectrum statistic, including their geometric properties and regional overlap.

\subsection{Validation and model selection}
\label{sec:validation}

  Following \autoref{sec:validation-methods}, we examine global and local discrepancies, single-bin deletion robustness, and the residual distribution before combining the evidence with the dispersion and AIC comparisons.

\subsubsection{Global discrepancies and local residuals}
\label{sec:global-discrepancy-results}

  We first assess the magnitude of discrepancies under each fixed-mean count model and then examine their distribution across energy.

  \autoref{tab:gof} reports the observed Pearson and deviance statistics and their fixed-dispersion upper-tail probabilities. We also compare the NegBin probabilities with those obtained by refitting the dispersion in each replica.

  \begin{table*}[htbp]
    \centering
    \caption{Fixed-mean count-model fits and Monte Carlo goodness-of-fit tests using $5000$ simulated datasets per phase and candidate. The reported $p$-values are upper-tail probabilities for the Pearson statistic $X^2$ and deviance $D$. The OFF dispersion uncertainty is obtained by Monte Carlo calibration at $68.27\%$ (see \autoref{tab:dispersion-intervals}). A dash denotes the Poisson model, and $\infty$ indicates a NegBin fit reaching the upper numerical search boundary, consistent with the Poisson limit.}
    \label{tab:gof}
    \begin{tabular}{llrrrrr}
      \toprule \toprule
      Phase & Model & $\hat\kappa$ & $X^2$ & $p_{X^2}$ & $D$ & $p_D$ \\
      \midrule
      $\mathrm{ON}$ & Poisson & - & $52.42$ & $0.840$ & $52.72$ & $0.833$ \\
      $\mathrm{ON}$ & NegBin & $\infty$ & $52.42$ & $0.851$ & $52.72$ & $0.846$ \\
      $\mathrm{OFF}$ & Poisson & - & $89.87$ & $0.0202$ & $87.23$ & $0.0300$ \\
      $\mathrm{OFF}$ & NegBin & $150^{+220}_{-70}$ & $66.88$ & $0.387$ & $64.46$ & $0.469$ \\
      \bottomrule \bottomrule
    \end{tabular}
  \end{table*}

  For ON, the Poisson goodness-of-fit tests yield $p_{X^2}=0.840$ and $p_D=0.833$ (\autoref{tab:gof}). The ON NegBin fit approaches the Poisson limit, and the refitted calibration reproduces both ON NegBin probabilities to the reported precision. For OFF Poisson, both probabilities ($0.0202$ and $0.0300$) fall below the nominal $5\%$ level. For OFF NegBin, holding the unrounded fitted OFF dispersion fixed gives probabilities $0.387$ and $0.469$, while refitting each replica gives $0.158$ and $0.443$. Both calibrations support the same decisions at the nominal $5\%$ level.

  Above the detector thresholds, the effective $\mathrm{OFF}$ exposure ($59.7\,\mathrm{kg\cdot day}$) is smaller than the $\mathrm{ON}$ exposure ($326.8\,\mathrm{kg\cdot day}$). This increases relative count fluctuations but is already accounted for by Poisson sampling. The $\mathrm{OFF}$ discrepancy instead measures variation around the fixed reconstructed background template, which can include effects of omitted background uncertainties or heterogeneity; it does not establish an intrinsic non-Poisson counting mechanism.

\begin{figure}[htbp]
    \centering
    \IfFileExists{Pictures/Validation/Residuos_Deviance_ON_OFF_2x2.pdf}{%
    \includegraphics[width=0.98\linewidth]{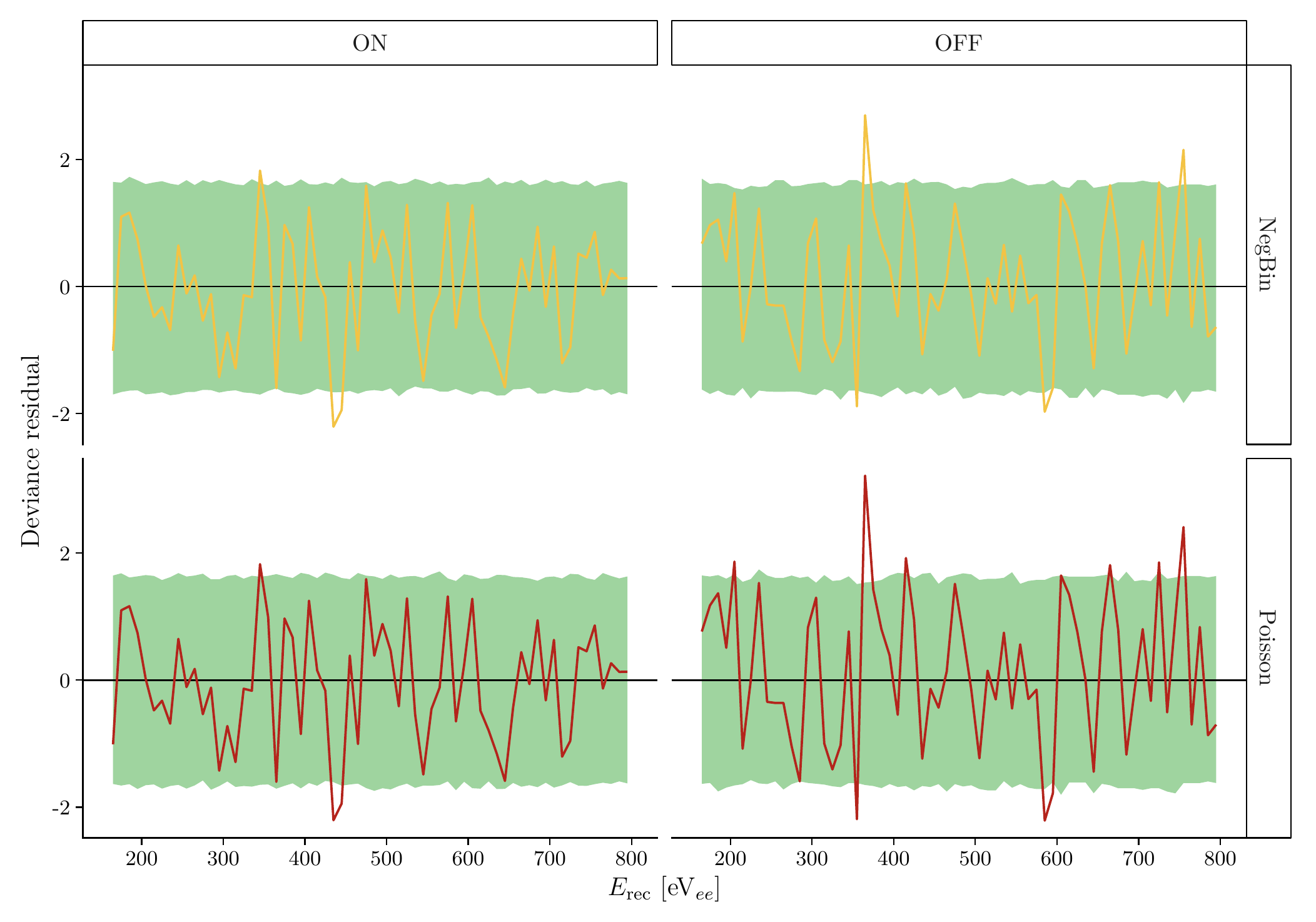}
    }{%
    \fbox{\parbox[c][5cm][c]{0.86\linewidth}{\centering Figura pendiente:\\
    \texttt{Residuos\_Deviance\_ON\_OFF\_2x2.pdf}}}%
    }
    \caption{Signed deviance residuals across reconstructed energy for the $\mathrm{ON}$ and $\mathrm{OFF}$ spectra under Poisson and NegBin models. Shaded areas are pointwise central $90\%$ Monte Carlo envelopes. The $\mathrm{ON}$ results are nearly identical; the NegBin $\mathrm{OFF}$ residuals have smaller excursions.}
    \label{fig:deviance-residuals}
  \end{figure}

  Residuals identify whether the global discrepancies in \autoref{tab:gof} are concentrated in particular energy bins. The signed deviance residuals defined in \eqref{eq:deviance-residual} and displayed in \autoref{fig:deviance-residuals} are compared with pointwise Monte Carlo envelopes, allowing local departures and systematic trends to be distinguished.

  For the $\mathrm{ON}$ spectrum, the residuals fluctuate around zero without a persistent energy trend, and the two models give almost identical patterns. The $\mathrm{OFF}$-Poisson residuals show larger excursions that diminish under the NegBin fit. Since the deviance envelopes are pointwise, individual crossings are not sufficient for a global rejection. The deletion diagnostic in \autoref{sec:robustness} examines whether individual bins affect the global decisions.

\subsubsection{Single-bin deletion robustness}
\label{sec:robustness}

  Following \autoref{sec:deletion-methods}, to evaluate the influence of individual energy bins, each observation is removed in turn, with the NegBin dispersion and Monte Carlo reference recalculated for the remaining $63$ bins. \autoref{tab:deletion-pvalues} reports the ranges of Monte Carlo-calibrated goodness-of-fit \(p\)-values and changes in the nominal $5\%$ compatibility decision.

  \begin{table*}[htbp]
    \centering
    \caption{Range of Monte Carlo goodness-of-fit $p$-values after deleting each bin in turn and repeating the relevant fit. The column $G$ counts deletions that change the compatibility decision at the nominal $5\%$ level and identifies the corresponding bins.}
    \label{tab:deletion-pvalues}
    \begin{tabular}{lllrrl}
      \toprule \toprule
      Phase & Model & Statistic & Minimum $p$ & Maximum $p$ & $G$ and bins \\
      \midrule
      $\mathrm{ON}$ & NegBin & Pearson & $0.821$ & $0.929$ & $0$ \\
      $\mathrm{ON}$ & NegBin & Deviance & $0.814$ & $0.928$ & $0$ \\
      $\mathrm{ON}$ & Poisson & Pearson & $0.818$ & $0.921$ & $0$ \\
      $\mathrm{ON}$ & Poisson & Deviance & $0.810$ & $0.918$ & $0$ \\
      $\mathrm{OFF}$ & NegBin & Pearson & $0.360$ & $0.433$ & $0$ \\
      $\mathrm{OFF}$ & NegBin & Deviance & $0.444$ & $0.503$ & $0$ \\
      $\mathrm{OFF}$ & Poisson & Pearson & $0.0122$ & $0.1018$ & $1$: 21 \\
      $\mathrm{OFF}$ & Poisson & Deviance & $0.0222$ & $0.1190$ & $4$: 20, 21, 43, 60 \\
      \bottomrule \bottomrule
    \end{tabular}
  \end{table*}

  None of the single-bin deletions changes the nominal goodness-of-fit decisions for either $\mathrm{ON}$ model or the $\mathrm{OFF}$ NegBin model. In the $\mathrm{OFF}$ Poisson model, deleting bin $21$, centered at $365\,\mathrm{eVee}$, changes both the Pearson and deviance decisions. The deviance decision also changes when bins $20$, $43$, or $60$ are removed. Bin $21$ produces the largest relative changes in the $\mathrm{OFF}$-Poisson Pearson and deviance statistics, approximately $13.2\%$ and $11.9\%$, respectively; the maxima under the $\mathrm{OFF}$ NegBin model are $4.3\%$ and $3.3\%$, respectively. The fixed-mean $\mathrm{OFF}$-Poisson rejection is therefore influenced by a small number of bins, whereas the adequacy of the $\mathrm{OFF}$ NegBin model is stable under these deletions.

\subsubsection{Distributional consistency of residuals}
\label{sec:quantile-results}

\begin{figure}[htbp]
    \centering
    \IfFileExists{Pictures/Validation/Worm_Residuos_Cuantilicos_ON_OFF.pdf}{%
    \includegraphics[width=0.98\linewidth]{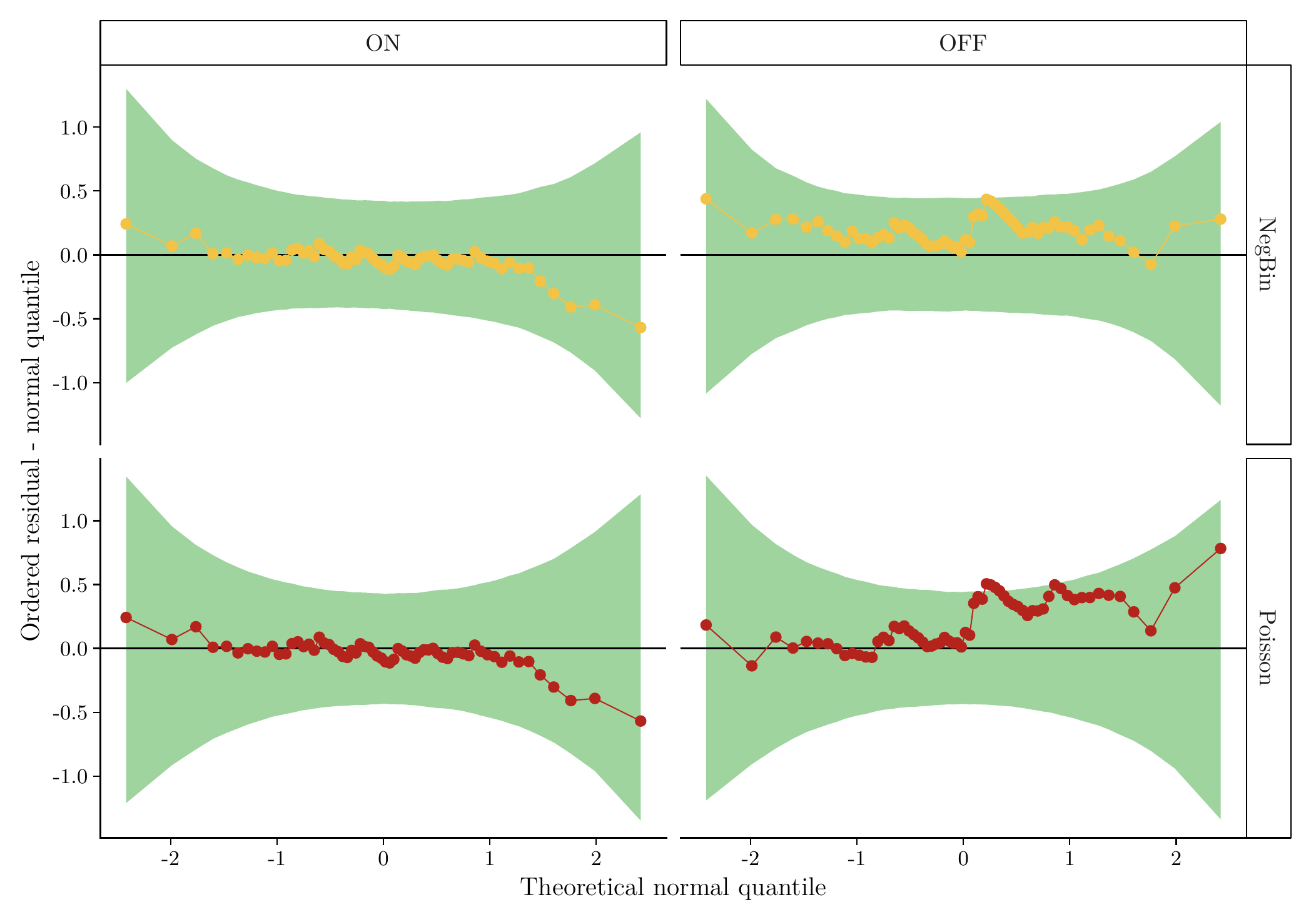}
    }{%
    \fbox{\parbox[c][5cm][c]{0.86\linewidth}{\centering Figura pendiente:\\
    \texttt{Worm\_Residuos\_Cuantilicos\_ON\_OFF.pdf}}}%
    }
    \caption{Worm plots of randomized quantile residuals for the four phase-model combinations, with simultaneous $90\%$ Monte Carlo envelopes. The $\mathrm{OFF}$-Poisson case has positive departures associated with bins $47$, $50$, and $55$; the other model combinations remain inside the corresponding envelopes.}
    \label{fig:worm-plot}
  \end{figure}

  The worm plots in \autoref{fig:worm-plot} complement the preceding discrepancy and deletion diagnostics by comparing ordered randomized quantile residuals with a normal reference, as described in \autoref{sec:quantile-methods}.

  The maximum absolute standardized deviation of each observed residual curve from its Monte Carlo center, defined in \eqref{eq:worm-max-deviation}, is compared with its simulated reference distribution in \autoref{tab:worm-global}. The corresponding critical values and upper-tail probabilities are obtained from \eqref{eq:worm-critical-value} and \eqref{eq:worm-mc-pvalue}, respectively. This provides a simultaneous assessment of distributional compatibility with the adopted count model, including departures in the center and tails of the residual distribution.

  \begin{table}[htbp]
    \centering
    \caption{Maximum absolute standardized deviations $M^{\mathrm{obs}}$ of the ordered randomized quantile residuals from their Monte Carlo centers, defined in \eqref{eq:worm-max-deviation}, together with the simultaneous $90\%$ Monte Carlo thresholds $c_{0.90}^{\mathrm{MC}}$ and the corresponding upper-tail probabilities $p_{\mathrm{MC}}$. The last column identifies the original energy-bin indices outside the simultaneous envelopes.}
    \label{tab:worm-global}
    \begin{tabular}{llrrrl}
      \toprule \toprule
      Phase & Model & $M^{\mathrm{obs}}$ & $c_{0.90}^{\mathrm{MC}}$ & $p_{\mathrm{MC}}$ & Bins \\
      \midrule
      $\mathrm{ON}$ & NegBin & $1.35$ & $2.79$ & $0.9346$ & - \\
      $\mathrm{ON}$ & Poisson & $1.36$ & $2.82$ & $0.904$ & - \\
      $\mathrm{OFF}$ & NegBin & $2.73$ & $2.82$ & $0.122$ & - \\
      $\mathrm{OFF}$ & Poisson & $3.18$ & $2.82$ & $0.0470$ & 47, 50, 55 \\
      \bottomrule \bottomrule
    \end{tabular}
  \end{table}

  The global worm-plot test in \autoref{tab:worm-global} is compatible with both $\mathrm{ON}$ models and $\mathrm{OFF}$ NegBin. $\mathrm{OFF}$ Poisson reaches $M^{\mathrm{obs}}=3.18$ and $p_{\mathrm{MC}}=0.0470$, with the largest positive departures associated with bins $47$, $50$, and $55$. This finding agrees with the goodness-of-fit results, although the diagnostics are correlated because they use the same counts.

\subsubsection{Dispersion likelihoods}
\label{sec:dispersion}

  We now combine the evidence on model adequacy and deletion robustness with the dispersion likelihoods and the relative AIC comparison to select the count law for inference.

\begin{figure}[htbp]
    \centering
    \IfFileExists{Pictures/Validation/Verosimilitud_ON_OFF.pdf}{%
      \includegraphics[width=0.82\linewidth]{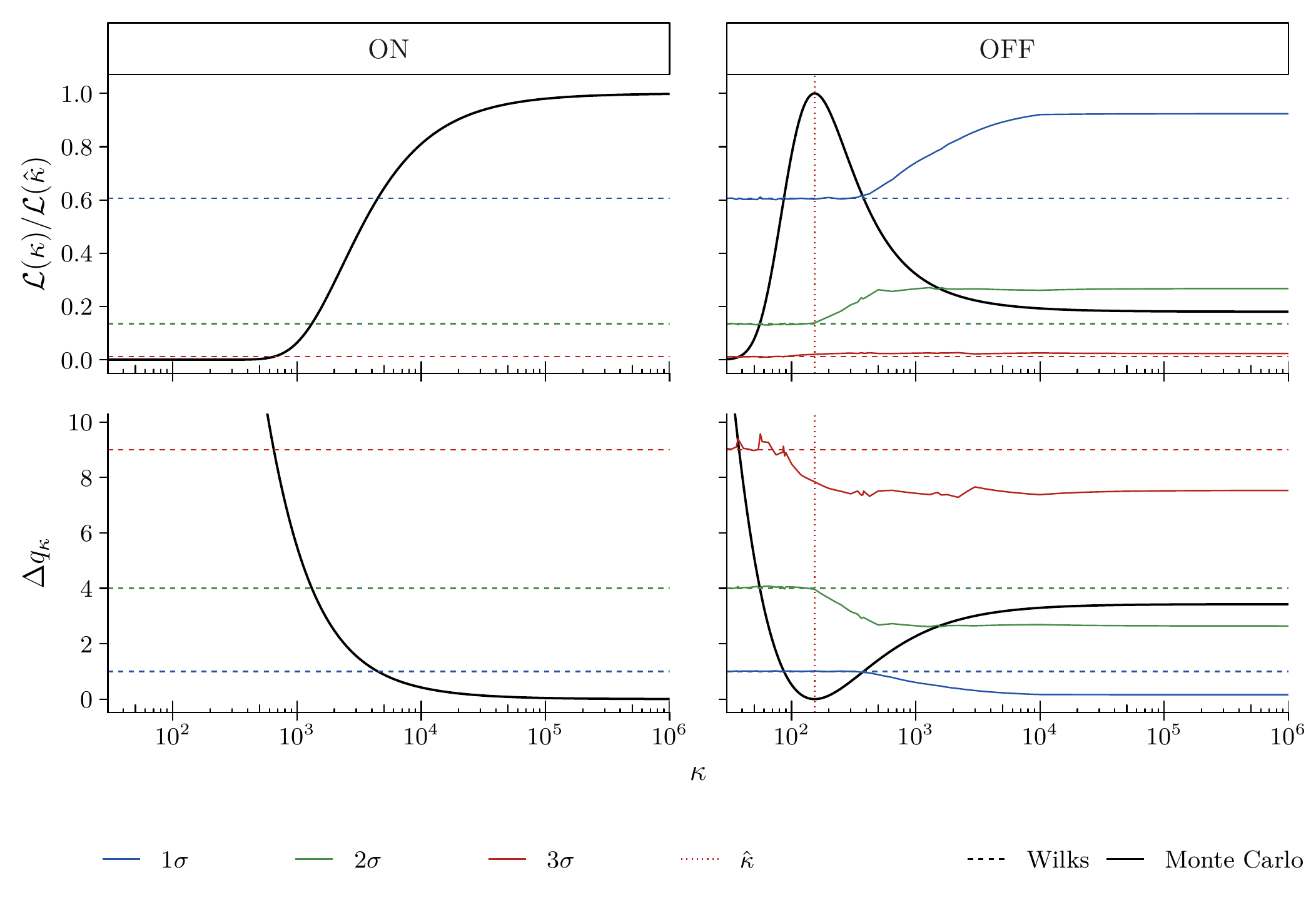}
    }{%
      \fbox{\parbox[c][4.5cm][c]{0.75\linewidth}{\centering Figura pendiente:\\
      \texttt{Verosimilitud\_ON\_OFF.pdf}}}%
    }
    \caption{Negative-binomial dispersion likelihoods at fixed SM and background means, with ON (left) and OFF (right) spectra. The upper panels show the likelihood relative to the Poisson limit for ON and the continuous maximum for OFF. The lower panels show the equivalent statistic $\Delta q_\kappa^\phi$ defined in \eqref{eq:dispersion-relative-likelihood}. Dashed colored lines denote the nominal one-parameter $1\sigma$, $2\sigma$, and $3\sigma$ references ($1$, $4$, and $9$ in the lower panels); solid colored curves in the OFF panels denote the corresponding Monte Carlo critical values, interpolated in inverse dispersion. Upper-panel thresholds are their relative-likelihood transforms. The vertical dotted red line marks the continuous OFF maximum. The ON fit approaches the Poisson limit beyond the displayed range; only the OFF intervals are Monte Carlo calibrated.}
    \label{fig:likelihood-kappa}
  \end{figure}

  \autoref{fig:likelihood-kappa} displays the equivalent representations in \eqref{eq:dispersion-relative-likelihood}: the upper panels show the relative likelihood and the lower panels show $\Delta q_\kappa^\phi$, both on the same logarithmic dispersion axis. The nominal reference levels are shown for both phases, while the OFF panels also include the point-dependent Monte Carlo cuts in \eqref{eq:dispersion-calibrated-cut}, transformed to relative likelihood by $\exp\boldsymbol{(}-\bar c_j/2\boldsymbol{)}$ in the upper panel. Their intersections with the observed curves delimit the sets in \eqref{eq:dispersion-calibrated-set}.

  For ON, the relative likelihood approaches unity as $\kappa^{\mathrm{ON}}$ increases, while $\Delta q_\kappa^{\mathrm{ON}}$ approaches zero. The numerical search reaches its upper bound of $10^8$, and direct evaluation of the Poisson likelihood gives the profile endpoint. We summarize the limiting fit as $\hat\kappa^{\mathrm{ON}}\to\infty$, denoted by $\infty$ in \autoref{tab:gof}. The three nominal levels $1$, $4$, and $9$ cross at approximately $4471$, $1317$, and $651$, respectively, on the low-dispersion side. All three include the Poisson endpoint and leave dispersion unbounded above.

  For OFF, the continuous maximum is included in both plotted profiles, with $\rho_\kappa^{\mathrm{OFF}}=1$ and $\Delta q_\kappa^{\mathrm{OFF}}=0$. Evaluating the Poisson likelihood directly gives $\Delta q_\kappa^{\mathrm{OFF}}\to T_{\mathrm{obs}}=3.43$, as in \eqref{eq:dispersion-lrt}. At the Poisson generating point, the independent verification sample gives nominal coverages of $86.0\%$, $98.2\%$, and $99.88\%$ for the three reference levels. With the calibrated cuts, these become $68.2\%$, $95.6\%$, and $99.77\%$, compared with the target values $68.27\%$, $95.45\%$, and $99.73\%$. The corresponding calibrated coverage standard errors are approximately $0.38$, $0.17$, and $0.039$ percentage points, conditional on the estimated critical values. The overcoverage of the nominal procedure near Poisson motivates using the calibrated OFF sets summarized in \autoref{tab:dispersion-intervals}.

  \begin{table}[htbp]
    \centering
    \caption{Nominal Wilks and Monte Carlo calibrated OFF dispersion confidence sets at fixed background means. Finite endpoints are rounded for presentation; $\infty$ denotes inclusion of the Poisson endpoint. The calibrated sets are obtained by linear interpolation in inverse dispersion on the simulated domain.}
    \label{tab:dispersion-intervals}
    \begin{tabular}{lcc}
      \toprule \toprule
      Level & Nominal & Calibrated \\
      \midrule
      $1\sigma$ & $[86.77,379.85]$ & $[86.99,375.51]$ \\
      $2\sigma$ & $[55.54,\infty]$ & $[55.36,\sim1.5\times10^3]$ \\
      $3\sigma$ & $[37.76,\infty]$ & $[36.93,\infty]$ \\
      \bottomrule \bottomrule
    \end{tabular}
  \end{table}

  The calibrated $1\sigma$ set gives the PDG-rounded estimate $\hat\kappa^{\mathrm{OFF}}=150^{+220}_{-70}$. The $2\sigma$ calibrated set excludes Poisson, unlike the nominal set, while both $3\sigma$ sets include it. Changing from linear to monotone cubic interpolation in inverse dispersion leaves these conclusions unchanged. The upper $2\sigma$ endpoint is especially sensitive to finite-simulation quantile uncertainty: inversion of pointwise $95\%$ order-statistic limits gives upper endpoints of approximately $1.4\times10^3$ and $1.7\times10^3$. These are sensitivity estimates, not simultaneous confidence bounds. The bootstrap below provides the complementary direct test of the Poisson null.

  The observed OFF dispersion likelihood-ratio statistic is $T_{\mathrm{obs}}=3.43$. Among $10\,000$ Poisson bootstrap replicas generated with seed $1234$, $290$ produce a statistic at least as large, giving $p_{\mathrm{MC}}=0.0291$ by \eqref{eq:bootstrap-p-value} (Monte Carlo standard error $0.0017$). This tail probability falls below the nominal $5\%$ level specified in \autoref{sec:overdispersion-methods}, so the Poisson null hypothesis is rejected in favor of additional dispersion under the fixed background template. Together with the goodness-of-fit results, this supports a finite effective OFF dispersion, although the sensitivity to a few bins tempers the evidence against Poisson. We next examine whether the gain in likelihood also compensates for the additional parameter according to the AIC rule.

\subsubsection{Relative model comparison and final selection}
\label{sec:aic-results}

  Among the ON/OFF model combinations reported in \autoref{tab:joint-aic}, Poisson ON--NegBin OFF has the smallest AIC ($998.80$), with an Akaike weight of $0.491$. The combination Poisson--Poisson remains competitive (AIC $=1000.23$; Akaike weight $=0.240$), with a difference of only $1.43$. In the OFF phase, the observed likelihood improvement $T_{\mathrm{obs}}=3.43$ exceeds the penalty of two for the single additional parameter in \eqref{eq:aic-dispersion-rule}. The modest AIC separation reflects the limited gain beyond this penalty. In the ON phase, the NegBin fit approaches the Poisson limit without a comparable likelihood gain, so the additional parameter is not favored.

  \begin{table}[htbp]
    \centering
    \caption{Joint Akaike information criteria for the four $\mathrm{ON/OFF}$ count-model assignments, with all mean templates held fixed. The parameter count $K$ includes freely fitted dispersion parameters; $\Delta$ is relative to the minimum AIC and $w$ is the Akaike weight among these four candidates.}
    \label{tab:joint-aic}
    \begin{tabular}{llrrrr}
      \toprule \toprule
      $\mathrm{ON}$ model & $\mathrm{OFF}$ model & $K$ & AIC & $\Delta$ & $w$ \\
      \midrule
      Poisson & NegBin & $1$ & $998.80$ & $0.00$ & $0.491$ \\
      Poisson & Poisson & $0$ & $1000.23$ & $1.43$ & $0.240$ \\
      NegBin & NegBin & $2$ & $1000.80$ & $2.00$ & $0.180$ \\
      NegBin & Poisson & $1$ & $1002.23$ & $3.43$ & $0.088$ \\
      \bottomrule \bottomrule
    \end{tabular}
  \end{table}

  Taken together, the diagnostics support a Poisson description for the ON phase and a NegBin description for the OFF phase: the former is compatible with the fixed-mean fluctuation model without an extra parameter, while the latter improves the global and distributional diagnostics and is stable under single-bin deletion. The modest AIC separation does not by itself establish a unique count law. The count model adopted for the likelihood analysis is therefore

  \begin{equation}
    \begin{aligned}
      \boldsymbol{(}M,M'\boldsymbol{)}&=
      \boldsymbol{(}\mathrm P,\mathrm{NB}\boldsymbol{)},\\
      N_i^{\mathrm{ON}}&\sim\operatorname{Poisson}
      \boldsymbol{(}\mu_i^{\mathrm{ON}}\boldsymbol{)},\\
      N_i^{\mathrm{OFF}}&\sim\operatorname{NegBin}
      \boldsymbol{(}\mu_i^{\mathrm{OFF}},\kappa^{\mathrm{OFF}}\boldsymbol{)}.
    \end{aligned}
    \label{eq:selected-model}
  \end{equation}

  where $\kappa^{\mathrm{OFF}}$ is fixed at its unrounded fitted value in the subsequent count-based inference. We use the shorthand $L_{\mathrm{NSI}}\equiv L_{\mathrm P,\mathrm{NB}}^{\mathrm{NSI}}$, $L_{\mathrm{sig}}\equiv L_{\mathrm P,\mathrm{NB}}^{\mathrm{sig}}$, $\Delta q\equiv\Delta q_{\mathrm P,\mathrm{NB}}$, and $\Delta q_{\mathrm{ref}}\equiv\Delta q_{\mathrm{ref};\mathrm P,\mathrm{NB}}$. The adopted model describes the reconstructed spectra conditionally; it does not establish the collaboration's underlying event-count law.

\subsection{\texorpdfstring{\CEvNS{}}{CEvNS} signal significance and SM compatibility}
\label{sec:signal-results}

  Using the count law in \eqref{eq:selected-model}, we test two distinct hypotheses through \eqref{eq:signal-profile-statistic}: the absence of the modeled \CEvNS{} signal ($s=0$) and its nominal SM normalization ($s=1$). Both tests use the same signal template and background nuisance parameter.

\subsubsection{Joint background and signal-strength analysis}
\label{sec:normalization-diagnostic}

  Before testing either reference signal strength, we fit the background scale $b$ and the nonnegative SM signal strength $s$ through \eqref{eq:signal-strength-means}. Writing $\ell_{\mathrm{sig}}\boldsymbol{(}b,s\boldsymbol{)}=\log L_{\mathrm{sig}}\boldsymbol{(}s,b\boldsymbol{)}$, their joint estimates are

  \begin{equation}
    \boldsymbol{(}\hat b,\hat s\boldsymbol{)}
    =\underset{b>0,\,s\geq0}{\operatorname{arg\,max}}\,
    \ell_{\mathrm{sig}}\boldsymbol{(}b,s\boldsymbol{)}.
    \label{eq:normalization-estimators}
  \end{equation}

\begin{figure}[htbp]
    \centering
    \IfFileExists{Pictures/Validation/Perfiles_Normalizacion_b_s.pdf}{%
      \includegraphics[width=0.96\linewidth]{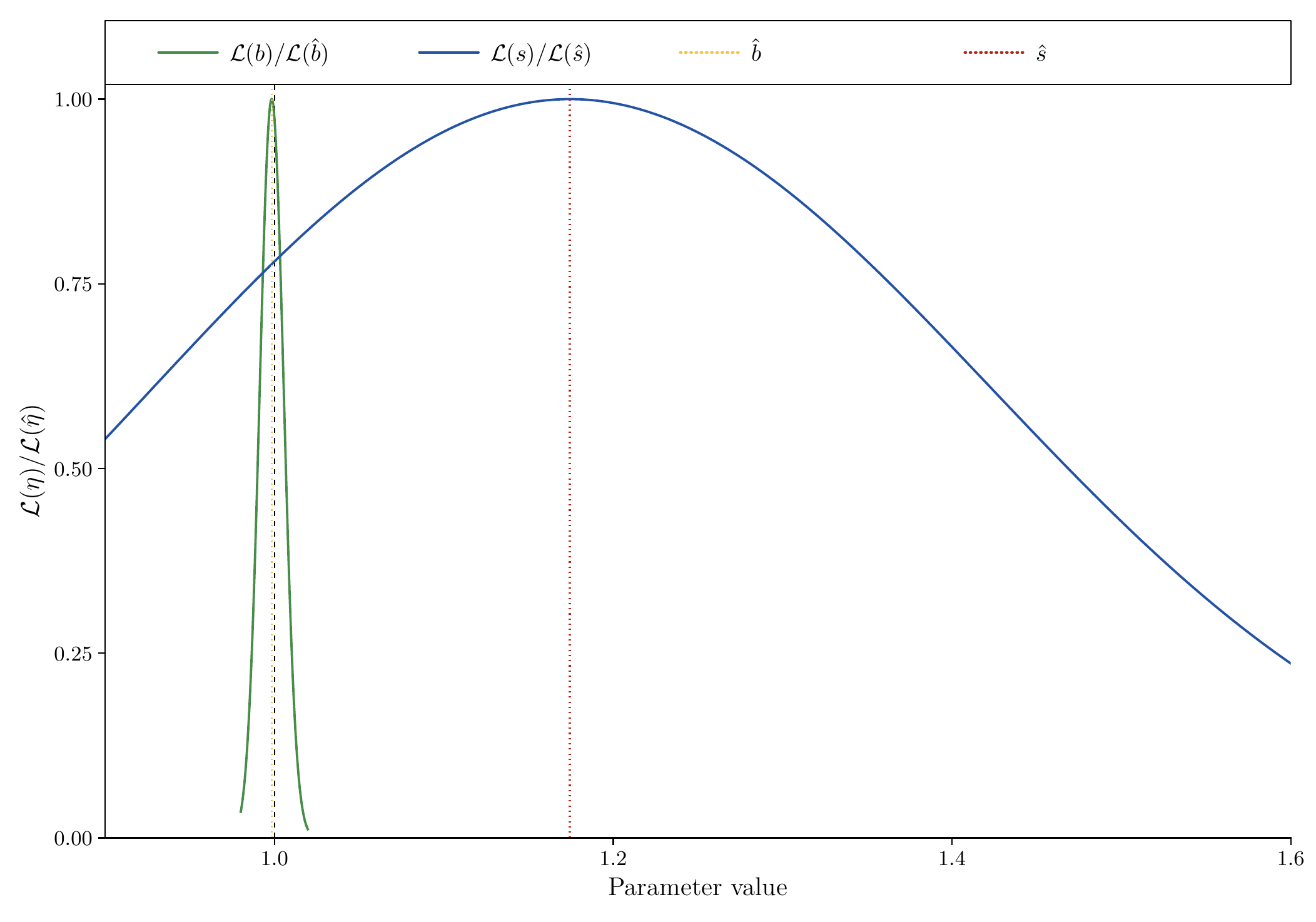}
    }{%
      \fbox{\parbox[c][4cm][c]{0.82\linewidth}{\centering Figura pendiente:\\ \texttt{Perfiles\_Normalizacion\_b\_s.pdf}}}%
    }
    \caption{Profile log-likelihood curves for the common background scale $b$ and SM signal scale $s$, optimizing the other parameter at each point. The maximum occurs near $b=0.998$ and $s=1.17$ for the selected Poisson $\mathrm{ON}$-NegBin $\mathrm{OFF}$ model.}
    \label{fig:normalization-profiles}
  \end{figure}

  The normalization and signal-significance fits use the same independently calculated SM template, whose sum is given by \eqref{eq:sm-template-total}. As discussed after \autoref{tab:conus-parameters}, its $320.1501$ events are obtained with our stated physical and numerical approximations and lie $7.74\%$ below, but remain statistically compatible with, the collaboration's global expectation of $347\pm59$ events~\cite{Ackermann:2025obx}. Using a single template here is essential because the tests of $s=0$ and $s=1$ must refer to the same spectral prediction. The joint maximum is $\hat\ell=-498.15$. Solving the unit-level crossings in \eqref{eq:normalization-interval-profiles} gives nominal intervals $[0.9913,1.0055]$ for $b$ and $[0.927,1.424]$ for $s$. The lower and upper errors are nearly symmetric for both parameters; following the PDG rounding convention~\cite{ParticleDataGroup:2024cfk}, we report
  \begin{equation}
    \hat b=0.998\pm0.007,\qquad
    \hat s=1.17\pm0.25.
    \label{eq:normalization-estimates-uncertainties}
  \end{equation}
  Both nominal intervals in \eqref{eq:normalization-estimates-uncertainties} include unity. The unrounded best-fit background normalization is about $0.2\%$ below unity, whereas the signal strength is $17.4\%$ above its SM reference. The profile-likelihood curves in \autoref{fig:normalization-profiles} show the dependence on the background-normalization and signal-strength parameters. The quoted uncertainties are conditional on the selected count law and the fixed OFF dispersion.

  At the nominal SM and background-normalization point $(b,s)=(1,1)$, the log likelihood is $\ell_{\mathrm{sig}}\boldsymbol{(}1,1\boldsymbol{)}=-498.40$. Relative to the joint maximum, this gives
  \begin{equation}
    \frac{L_{\mathrm{sig}}\boldsymbol{(}1,1\boldsymbol{)}}
    {L_{\mathrm{sig}}\boldsymbol{(}\hat s,\hat b\boldsymbol{)}}
    =\exp\boldsymbol{(}\ell_{\mathrm{sig}}\boldsymbol{(}1,1\boldsymbol{)}-\hat\ell\boldsymbol{)}\simeq0.778.
    \label{eq:nominal-relative-likelihood}
  \end{equation}

  The value of $0.778$ describes the relative likelihood of these two parameter points; it is not a goodness-of-fit probability. Profiling $b$ at fixed $s=1$ instead gives $\hat\ell_{s=1}=-498.40$, the reference used in the SM compatibility test. The proximity of these two log likelihoods shows that allowing the small background adjustment has little effect on that test.

\subsubsection{Signal detection and SM compatibility}
\label{sec:signal-tests-results}

  The joint maximum is $\hat\ell=-498.15$ at the joint estimates in \eqref{eq:normalization-estimates-uncertainties}. Profiling the background at fixed $s=0$ gives $\hat\ell_{s=0}=-509.85$, while profiling at fixed \(s=1\) gives $\hat\ell_{s=1}=-498.40$. The resulting profile-likelihood comparisons are summarized in \autoref{tab:signal-significance}.

  \begin{table*}[htbp]
    \centering
    \caption{Profile-likelihood tests using the reconstructed $64$-bin $\mathrm{ON/OFF}$ spectra under the selected Poisson $\mathrm{ON}$-NegBin $\mathrm{OFF}$ model. Detection tests $s=0$ against $s>0$; the two SM rows report bilateral and one-sided tail probabilities for the same test of $s=1$. The results are conditional on digitized counts and fixed templates and are not collaboration-level significances.}
    \label{tab:signal-significance}
    \begin{tabular}{lccccc}
      \toprule \toprule
      Question & Reference & Alternative & $\Delta q$ & $Z$ & $p$ \\
      \midrule
      Positive \CEvNS{} signal & $s=0$ & $s>0$ & $23.40$ & $4.84$ & $6.60\times10^{-7}$ \\
      Bilateral SM compatibility & $s=1$ & $s\neq1$ & $0.50$ & $0.70$ & $0.481$ \\
      Upper excess over SM & $s=1$ & $s>1$ & $0.50$ & $0.70$ & $0.241$ \\
      \bottomrule \bottomrule
    \end{tabular}
  \end{table*}

  Using the one-sided asymptotic prescription in \autoref{sec:signal-profile-methods}, the no-signal hypothesis gives $\Delta q_0=23.40$, corresponding to a significance $Z_0=4.84\sigma$ and $p_0=6.60\times10^{-7}$. This is strong evidence for a positive signal within the reconstructed-data model. It does not replace the collaboration's $3.7\sigma$ observation~\cite{Ackermann:2025obx}, which accounts for the full experimental likelihood and nuisance parameters that are unavailable here.

  At the SM normalization $s=1$, the profile-likelihood-ratio statistic is $\Delta q_1=0.50$ ($0.70\sigma$), with bilateral probability $p_{\mathrm{bil}}=0.481$. The one-sided upward-excess probability is $p_+=0.241$. Thus the best-fit strength $\hat s=1.17\pm0.25$ remains compatible with the SM rate: evidence for a positive \CEvNS{} signal does not imply evidence for a nonstandard contribution.

\subsection{NSI inference and geometric confidence regions}
\label{sec:nsi-inference-results}

  The one- and two-parameter vector-NSI regions are obtained from the reconstructed $\mathrm{ON/OFF}$ likelihood and the separate $19$-bin excess-spectrum statistic. The nuclear weak charge fixes their interference and cancellation degeneracies, as already analyzed in our earlier \CEvNS{} and E$\nu$eS study~\cite{Benavides:2026jdz}. The present comparison quantifies the changes in allowed boundaries and regional overlap between the two public-data analyses.

  As a check of the $19$-bin quadratic implementation, the SM point $\boldsymbol{\varepsilon}=\mathbf{0}$ gives $\hat\alpha_{\SM}=0.018284$ and $\chi^2_{\SM}=7.398790$ after profiling the signal normalization. At this SM benchmark, the unconstrained solution in \eqref{eq:chi-square-alpha-profile} lies within the NSI scan bounds in \eqref{eq:chi-square-alpha-bounded}. The following intervals and contours use the nominal asymptotic thresholds specified in \autoref{sec:inference-methods}.

\subsubsection{One-dimensional NSI profiles}
\label{sec:nsi-results}

  Each one-dimensional scan varies a single real NSI coupling, with the others fixed at zero. The constrained $\mathrm{ON/OFF}$ likelihood profiles the background scale $b$ and signal-normalization shift $\alpha$ jointly, whereas the excess-spectrum statistic profiles only $\alpha$. The label $e\mu/\tau$ denotes separate fits of $e\mu$ and $e\tau$ couplings, which are identical for this electron-flavor source when only one is varied.

\begin{figure}[htbp]
    \centering
    \IfFileExists{Pictures/NSI/plot_nsi_1d.pdf}{%
      \includegraphics[width=0.98\linewidth]{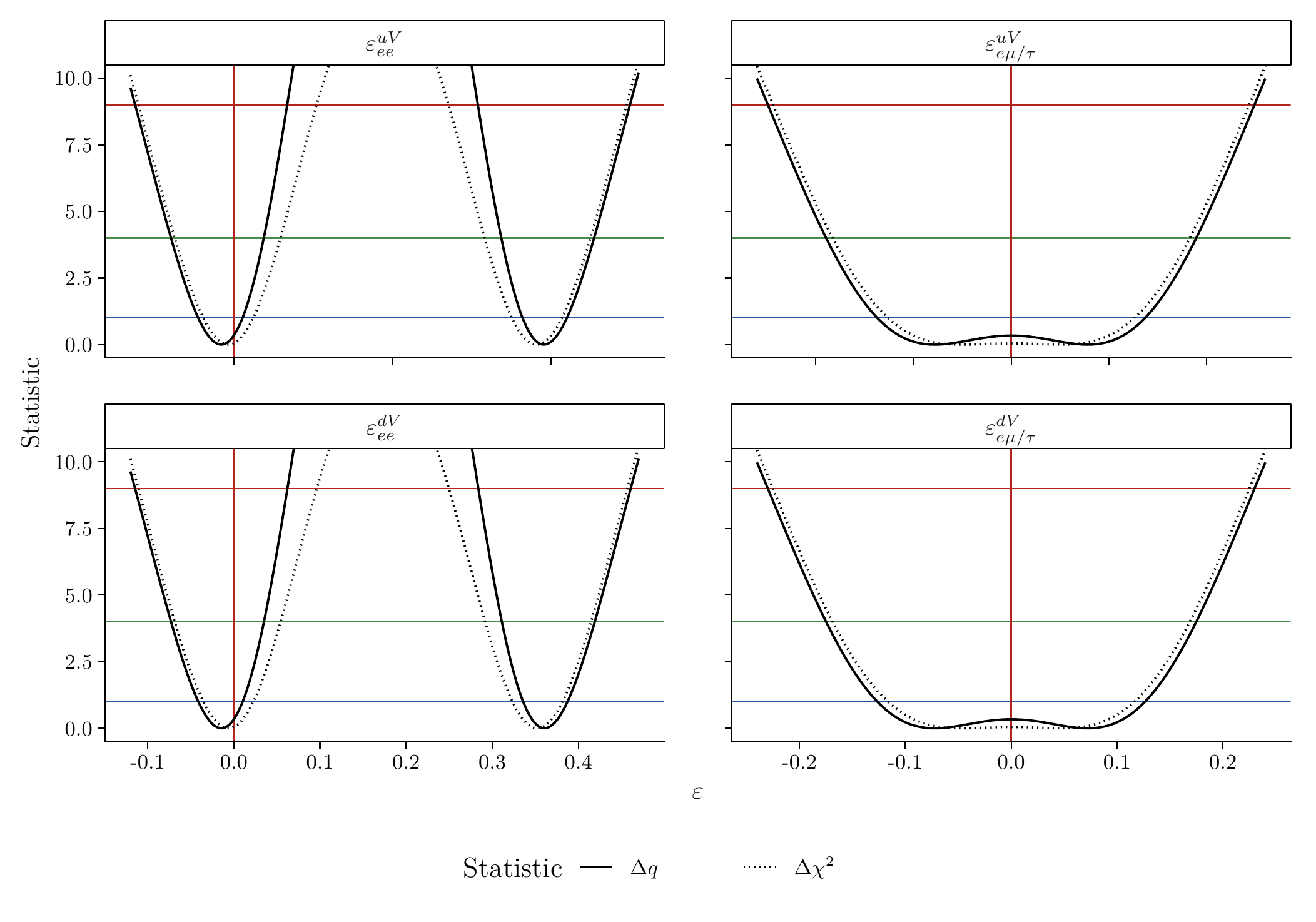}
    }{%
      \fbox{\parbox[c][5cm][c]{0.86\linewidth}{\centering Figura pendiente:\\
      \texttt{Pictures/NSI/plot\_nsi\_1d.pdf}}}%
    }
    \caption{One-dimensional \CEvNS{}-only NSI profiles for the four vector couplings. Solid curves denote $\Delta q$ from the reconstructed $\mathrm{ON/OFF}$ counts and dashed curves denote $\Delta\chi^2$ from the $19$-bin excess spectrum. Horizontal lines mark the nominal $1\sigma$, $2\sigma$, and $3\sigma$ thresholds. Diagonal couplings show charge-reflection solutions, while flavor-changing profiles are symmetric about zero.}
    \label{fig:nsi-1d}
  \end{figure}

  The profiles in \autoref{fig:nsi-1d} retain the charge degeneracies predicted by \eqref{eq:nsi-effective-charge}. A diagonal coupling interferes with the SM weak charge, producing a solution near zero and a second solution with the opposite effective-charge sign. A flavor-changing coupling contributes quadratically, so its profile is symmetric under a sign reversal. In the present fits, the nominal $1\sigma$ flavor-changing intervals remain connected and include zero.

  The nominal $1\sigma$ intervals in \autoref{tab:nsi-comparison} are obtained from the threshold crossings of these profiles. Both constructions reproduce the two disconnected diagonal branches and the connected flavor-changing intervals. The previously published \CONUS{} and COHERENT phenomenological fits~\cite{DeRomeri:2025csu,DeRomeri:2022twg} are included for context; they use different nuclear targets, flavor compositions, and statistical treatments.

  \begin{table*}[htbp]
    \centering
    \caption{Nominal $1\sigma$ vector-NSI intervals obtained by varying one coupling at a time. The $\Delta q$ column uses the reconstructed $64$-bin $\mathrm{ON/OFF}$ likelihood; $\Delta\chi^2$ uses the $19$-bin background-subtracted excess. Endpoints are interpolated at the corresponding unit threshold. The final columns reproduce published phenomenological fits and are not independent reanalyses in this work.}
    \label{tab:nsi-comparison}
    \scriptsize
    \begin{tabular}{@{}ccccc@{}}
      \toprule \toprule
      Parameter & Public-data likelihood $\Delta q$ & Excess-spectrum $\Delta\chi^2$ & \shortstack{Literature \CONUS{} fit\\\cite{DeRomeri:2025csu}} & \shortstack{COHERENT CsI+LAr fit\\\cite{DeRomeri:2022twg}} \\
      \midrule
      $\varepsilon^{uV}_{ee}$
      & $[-0.0447,0.0111]\cup[0.3638,0.4195]$
      & $[-0.0384,0.0245]\cup[0.3503,0.4132]$
      & $[-0.037,0.026]\cup[0.348,0.411]$
      & $[-0.024,0.045]\cup[0.34,0.43]$ \\
      $\varepsilon^{dV}_{ee}$
      & $[-0.0413,0.0102]\cup[0.3358,0.3873]$
      & $[-0.0354,0.0226]\cup[0.3234,0.3815]$
      & $[-0.034,0.024]\cup[0.322,0.380]$
      & $[-0.027,0.048]\cup[0.30,0.39]$ \\
      $\varepsilon^{uV}_{e\mu/\tau}$
      & $[-0.1369,0.1369]$
      & $[-0.1260,0.1260]$
      & $[-0.123,0.123]$
      & \shortstack{$e\mu:\ [-0.081,0.081]$\\$e\tau:\ [-0.13,0.13]$} \\
      $\varepsilon^{dV}_{e\mu/\tau}$
      & $[-0.1264,0.1264]$
      & $[-0.1163,0.1163]$
      & $[-0.114,0.114]$
      & \shortstack{$e\mu:\ [-0.071,0.071]$\\$e\tau:\ [-0.12,0.12]$} \\
      \bottomrule \bottomrule
    \end{tabular}
  \end{table*}

  Using the unrounded endpoints underlying \autoref{tab:nsi-comparison}, the total accepted length for each diagonal coupling is approximately $11.4\%$ smaller with $\Delta q$ than with $\Delta\chi^2$. The flavor-changing $u$ and $d$ intervals are instead about $8.7\%$ wider. Thus the change in the constraints is sector dependent and cannot be attributed solely to the likelihood-ratio statistic, since the two analyses also differ in data reduction, energy range, and background treatment.

  For each coupling, the nominal $1\sigma$ allowed set includes zero in both analyses; the second diagonal branch and flavor-changing symmetry arise from the weak charge and were also present in the earlier analysis~\cite{Benavides:2026jdz}. The differences here are quantitative, not evidence of a new degeneracy or of a nonzero coupling.

\subsubsection{Two-dimensional NSI regions}
\label{sec:nsi-2d-results}

  Two-parameter scans reveal the cancellation directions absent from one-coupling fits. For the six selected parameter pairs, the effective-nucleus charge yields two open-band cases and four closed conic cases. The boundaries use the nominal two-parameter threshold $\Delta q=\Delta\chi^2=4.605$, applied to the respective profiled statistic.

\begin{figure}[htbp]
    \centering
    \IfFileExists{Pictures/NSI/plot_nsi_linear.pdf}{%
      \includegraphics[width=0.98\linewidth]{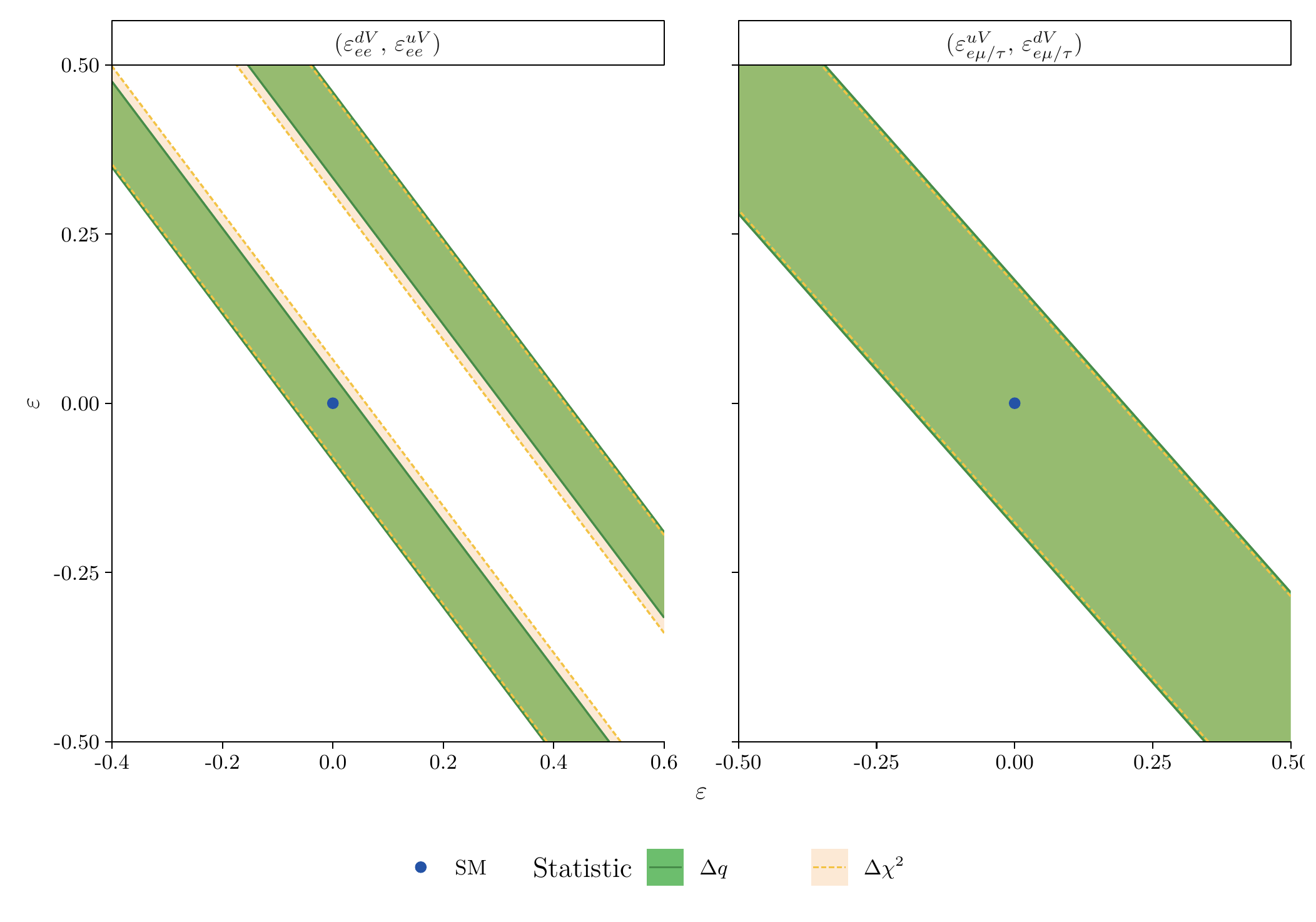}%
    }{%
      \fbox{\parbox[c][4.8cm][c]{0.86\linewidth}{\centering Figura pendiente:\\ \texttt{Pictures/NSI/plot\_nsi\_linear.pdf}}}%
    }
    \caption{Open NSI regions at nominal $90\%$ C.L. Left: diagonal $(\varepsilon_{ee}^{dV},\varepsilon_{ee}^{uV})$ plane. Right: flavor-changing $(\varepsilon_{e\mu/\tau}^{uV},\varepsilon_{e\mu/\tau}^{dV})$ plane. In the common coordinate order $(\varepsilon^{dV},\varepsilon^{uV})$, the theoretical cancellation slope is $-C_d/C_u\simeq-1.0831$; reversing the axes in the right panel gives the plotted slope $-C_u/C_d\simeq-0.9232$. The accepted band widths are determined by the two statistical constructions.}
    \label{fig:nsi-linear-regions}
  \end{figure}

  The linear regions in \autoref{fig:nsi-linear-regions} illustrate the cancellation between up- and down-quark contributions. In the diagonal plane, interference with the SM produces two parallel bands corresponding to opposite signs of the effective weak charge. The pure flavor-changing plane has a connected band at the displayed threshold. Both leave the orthogonal NSI combination unconstrained within the effective single-target model, and the plotted boundaries are limited by the chosen scan domain.

\begin{figure}[htbp]
    \centering
    \IfFileExists{Pictures/NSI/plot_nsi_circular.pdf}{%
      \includegraphics[width=0.98\linewidth]{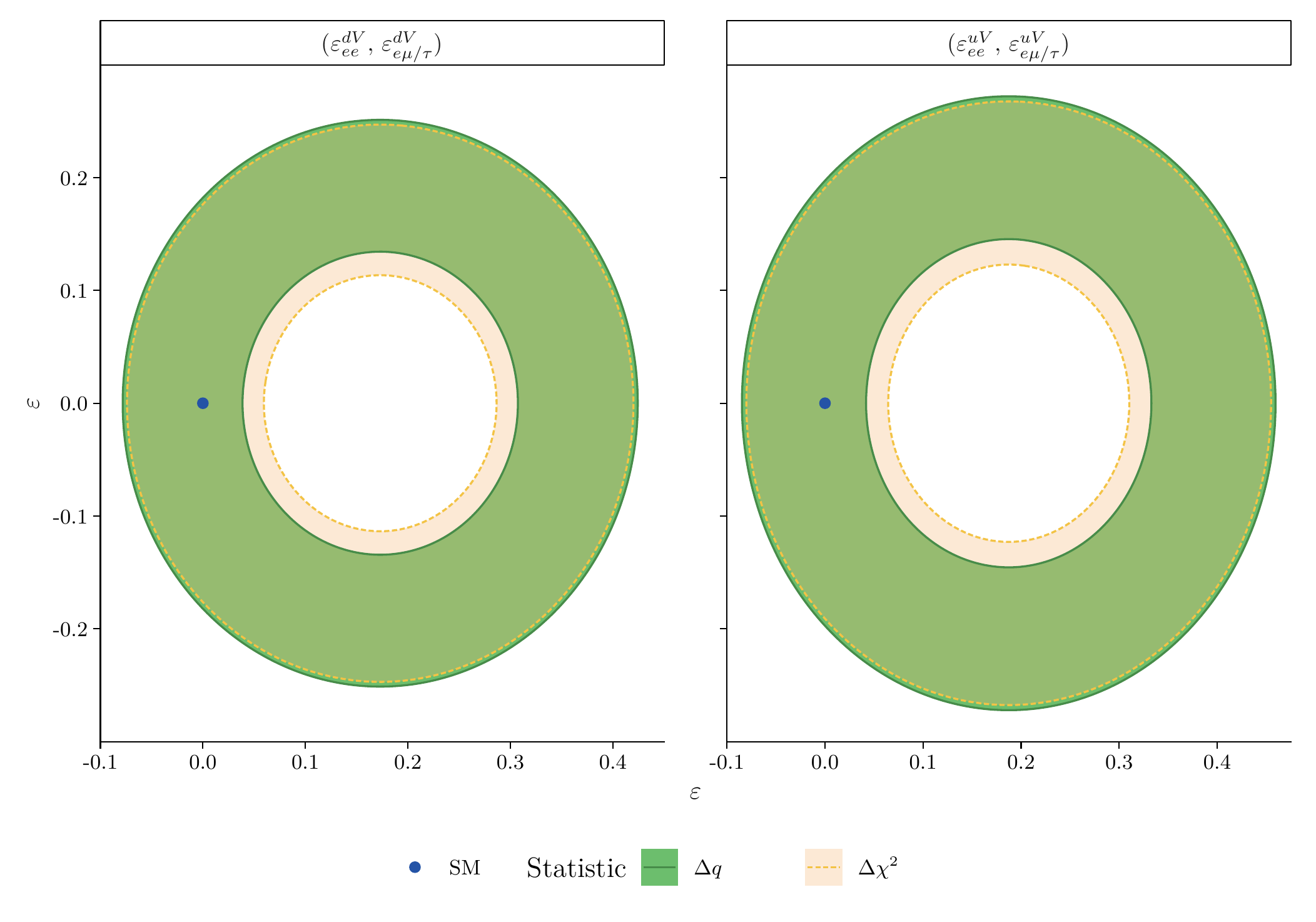}%
    }{%
      \fbox{\parbox[c][4.8cm][c]{0.86\linewidth}{\centering Figura pendiente:\\ \texttt{Pictures/NSI/plot\_nsi\_circular.pdf}}}%
    }
    \caption{Circular NSI regions at nominal $90\%$ C.L. The left and right panels show $(\varepsilon_{ee}^{dV},\varepsilon_{e\mu/\tau}^{dV})$ and $(\varepsilon_{ee}^{uV},\varepsilon_{e\mu/\tau}^{uV})$, respectively. The effective weak charge fixes their centers, while the two statistics admit different inner and outer radii.}
    \label{fig:nsi-circular-regions}
  \end{figure}

  For the mixed diagonal-flavor-changing planes, equal quark coefficients yield the circular regions in \autoref{fig:nsi-circular-regions}. The centers $c_d$ and $c_u$ are fixed by cancellation between the diagonal NSI contribution and the SM charge. The inner and outer boundaries correspond to the charge levels admitted by each statistic, rather than to distinct physical signal components.

\begin{figure}[htbp]
    \centering
    \IfFileExists{Pictures/NSI/plot_nsi_elliptical.pdf}{%
      \includegraphics[width=0.98\linewidth]{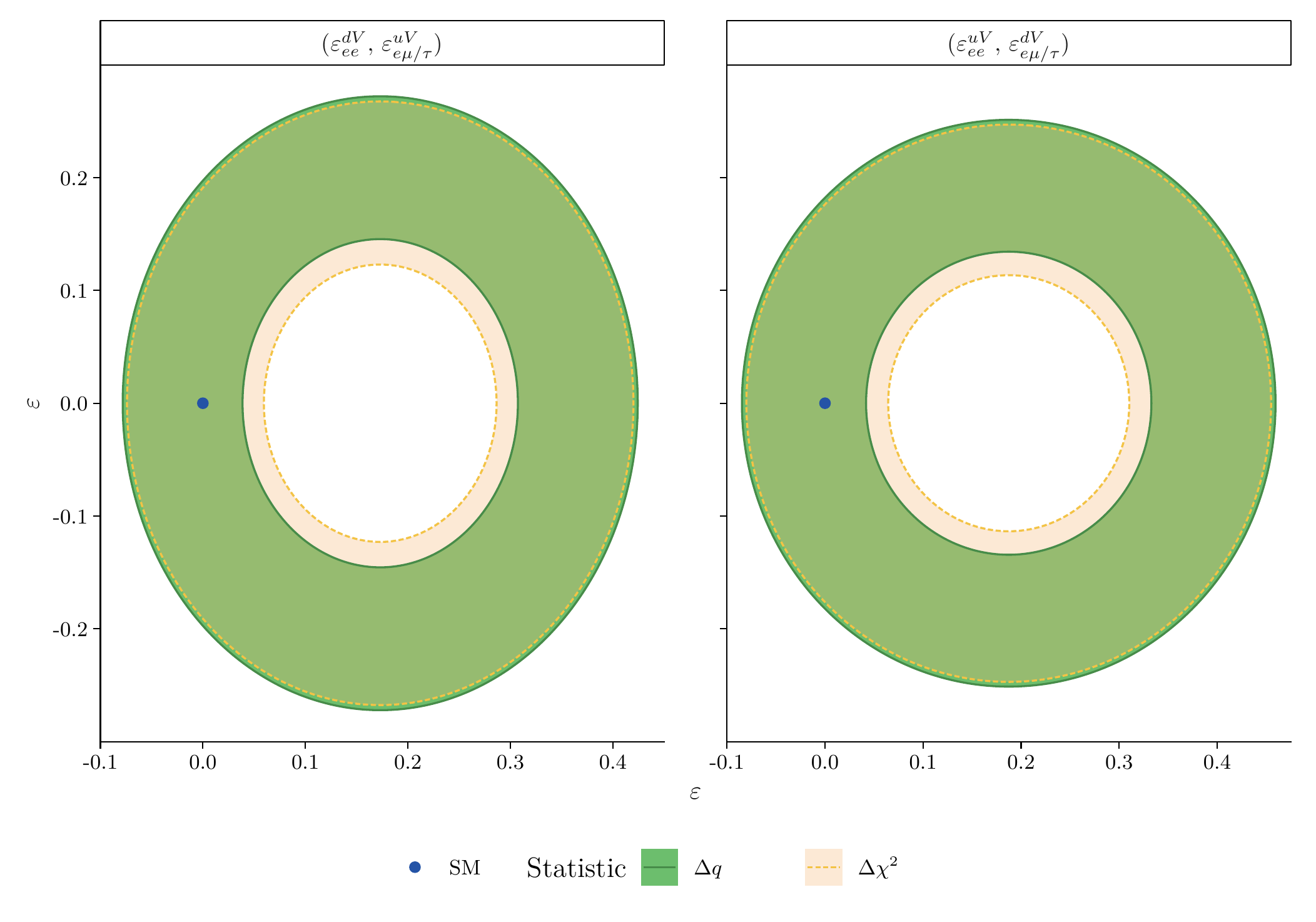}%
    }{%
      \fbox{\parbox[c][4.8cm][c]{0.86\linewidth}{\centering Figura pendiente:\\ \texttt{Pictures/NSI/plot\_nsi\_elliptical.pdf}}}%
    }
    \caption{Elliptical NSI regions at nominal $90\%$ C.L. The left panel shows $(\varepsilon_{ee}^{dV},\varepsilon_{e\mu/\tau}^{uV})$ and the right panel $(\varepsilon_{ee}^{uV},\varepsilon_{e\mu/\tau}^{dV})$. The effective nuclear coefficients fix the centers, orientations, and semiaxis ratio, while $\Delta q$ and $\Delta\chi^2$ give different accepted contours.}
    \label{fig:nsi-elliptical-regions}
  \end{figure}

  When the diagonal and flavor-changing interactions involve different quark species, the coefficients differ and the allowed contours become ellipses, as shown in \autoref{fig:nsi-elliptical-regions}. Interchanging the quark assignments rotates the major axis by $90^{\circ}$; the predicted semiaxis ratio remains $C_d/C_u\simeq1.0831$. Consequently, the comparison between the two statistics concerns the admitted semiaxis lengths and areas, rather than the orientation imposed by the nuclear model.

  The SM point lies within all six nominal $90\%$ regions. This is consistent with, but distinct from, the conditional signal-strength result in \autoref{sec:signal-results}. Along an exactly degenerate direction, the fitted coordinates are not unique; the effective charge and full allowed set are the relevant quantities.

\subsubsection{Geometric checks and regional overlap}
\label{sec:nsi-geometric-results}

  The charge-level equations in \autoref{sec:nsi-geometry} predict the contour slopes, centers, and semiaxis ratios. These predictions are checked numerically using centerlines interpolated from the open bands and conics fitted to the closed boundaries. \autoref{tab:nsi-geometry} compares the two methods with the analytic values.

  \begin{table*}[htbp]
    \centering
    \caption{Analytic and numerically fitted NSI contour properties at nominal $90\%$ C.L. The linear slopes are expressed in the common coordinate order $(\varepsilon^{dV},\varepsilon^{uV})$ for both flavor channels, irrespective of the displayed plot axes. Fitted values check contour extraction within the effective-nucleus approximation; they are not independent measurements of the nuclear coefficients.}
    \label{tab:nsi-geometry}
    \begin{tabular}{llccc}
      \toprule \toprule
      Geometry & Quantity & Theory & $\Delta q$ & $\Delta\chi^2$ \\
      \midrule
      Linear, diagonal near SM & Slope
      & $-1.0831$ & $-1.0831$ & $-1.0831$ \\
      Linear, flavor changing & Slope
      & $-1.0831$ & $-1.0831$ & $-1.0831$ \\
      Circle, $d$-$d$ & Center $c_x$
      & $0.1730$ & $0.1730$ & $0.1730$ \\
      Circle, $u$-$u$ & Center $c_x$
      & $0.1874$ & $0.1874$ & $0.1874$ \\
      Ellipse, $d$-$u$ & $a_{\mathrm{maj}}/a_{\mathrm{min}}$
      & $1.0831$ & $1.0831$ & $1.0831$ \\
      Ellipse, $u$-$d$ & $a_{\mathrm{maj}}/a_{\mathrm{min}}$
      & $1.0831$ & $1.0831$ & $1.0831$ \\
      Ellipse, $d$-$u$ & Orientation
      & $90^{\circ}$ & $90^{\circ}$ & $90^{\circ}$ \\
      Ellipse, $u$-$d$ & Orientation
      & $0^{\circ}$ & $0^{\circ}$ & $0^{\circ}$ \\
      \bottomrule \bottomrule
    \end{tabular}
  \end{table*}

  The fitted quantities agree with the analytic values at the reported precision. Within the single-effective-nucleus model, this tests the contour extraction rather than measuring the germanium isotope composition. The radii, thicknesses, and accepted areas, by contrast, depend on the allowed charge levels and distinguish the statistical constructions.

  To quantify differences in accepted area, let $\mathcal R_q$ and $\mathcal R_{\chi^2}$ be the regions defined by the two nominal thresholds. Following the regional comparison used in Ref.~\cite{Benavides:2026jdz}, we express their Jaccard overlap as a percentage:
  \begin{equation}
    J\boldsymbol{(}\mathcal R_q,\mathcal R_{\chi^2}\boldsymbol{)}
    =100\,
    \frac{A\boldsymbol{(}\mathcal R_q\cap\mathcal R_{\chi^2}\boldsymbol{)}}
    {A\boldsymbol{(}\mathcal R_q\cup\mathcal R_{\chi^2}\boldsymbol{)}},
    \label{eq:nsi-jaccard}
  \end{equation}
  where $A$ is the area estimated by counting accepted grid cells. The values in \autoref{tab:nsi-overlap} therefore refer to the numerical scan, and for open bands also depend on its finite boundaries.

  \begin{table}[htbp]
    \centering
    \caption{Percentage Jaccard overlap between the nominal $90\%$ NSI regions obtained from $\Delta q$ and $\Delta\chi^2$, evaluated over the same finite parameter grids.}
    \label{tab:nsi-overlap}
    \begin{tabular}{lc}
      \toprule \toprule
      Parameter plane & $J\,[\%]$ \\
      \midrule
      $\boldsymbol{(}\varepsilon_{ee}^{dV},\varepsilon_{ee}^{uV}\boldsymbol{)}$
      & $81.0$ \\
      $\boldsymbol{(}\varepsilon_{ee}^{dV},\varepsilon_{e\mu/\tau}^{dV}\boldsymbol{)}$
      & $85.6$ \\
      $\boldsymbol{(}\varepsilon_{ee}^{dV},\varepsilon_{e\mu/\tau}^{uV}\boldsymbol{)}$
      & $85.5$ \\
      $\boldsymbol{(}\varepsilon_{e\mu/\tau}^{dV},\varepsilon_{e\mu/\tau}^{uV}\boldsymbol{)}$
      & $97.1$ \\
      $\boldsymbol{(}\varepsilon_{ee}^{uV},\varepsilon_{e\mu/\tau}^{uV}\boldsymbol{)}$
      & $85.5$ \\
      $\boldsymbol{(}\varepsilon_{ee}^{uV},\varepsilon_{e\mu/\tau}^{dV}\boldsymbol{)}$
      & $85.6$ \\
      \bottomrule \bottomrule
    \end{tabular}
  \end{table}

  The overlap ranges from $81.0\%$ in the diagonal up-down plane to $97.1\%$ in the pure flavor-changing plane. On the sampled grids, five $\Delta\chi^2$ regions have larger areas than their $\Delta q$ counterparts, although their shifted boundaries prevent a simple containment relation. In the pure flavor-changing plane the $\Delta\chi^2$ region is instead contained in the larger $\Delta q$ region at the grid resolution. Thus the statistical construction changes the accepted fraction of parameter space without changing the analytic type of the contours.

  The earlier study compared \CEvNS{}-only and \CEvNS{}+E$\nu$eS predictions within the excess-spectrum framework~\cite{Benavides:2026jdz}. Here both the data representation and statistical model change, so the different Jaccard overlaps cannot be interpreted as the isolated effect of replacing $\chi^2$ by a count likelihood.

\section{Discussion and limitations}
\label{sec:discussion}

  The validation results in \autoref{sec:validation} support a phase-dependent effective count law conditional on the nominal templates. Poisson adequately describes the ON counts (deviance $p=0.833$), whereas OFF Poisson is rejected at the nominal $5\%$ level ($p=0.0300$). NegBin OFF with $\hat\kappa^{\mathrm{OFF}}=150^{+220}_{-70}$ improves compatibility ($p=0.469$), and the bootstrap dispersion test gives $p_{\mathrm{MC}}=0.0291$. The calibrated OFF dispersion set excludes Poisson at the $2\sigma$ level, consistent with the direct bootstrap test. However, the AIC gain over Poisson--Poisson is modest ($1.43$), and the OFF-Poisson rejection depends partly on a few bins. The combined evidence motivates the selected model without establishing a unique underlying count law.

  The smaller OFF exposure increases relative fluctuations, an effect already accounted for by Poisson sampling. The fitted additional dispersion instead measures variation relative to the fixed reconstructed background template. Background uncertainty, time dependence, or missing correlations may contribute to that variation, but its origin cannot be determined from the public spectra. Consequently, the finite OFF dispersion should be interpreted as an effective description rather than as evidence of an intrinsic non-Poisson counting mechanism.

  The reference signal-strength analysis (\autoref{sec:signal-results}) gives a conditional significance of $4.84\sigma$ and an SM-compatible normalization ($\hat s=1.17\pm0.25$, $p_{\mathrm{bil}}=0.481$). This significance is conditional on the reconstructed counts, the fixed signal and background shapes, the fitted background scale, and the selected OFF dispersion. It is not a remeasurement of the collaboration's $3.7\sigma$ observation~\cite{Ackermann:2025obx}. The upward best-fit signal strength does not constitute evidence of NSI.

  To assess sensitivity to the graphical tolerance in \eqref{eq:rounding-and-count-tolerance}, we assigned each bin its lowest or highest compatible integer, coherently across bins or with alternating choices in energy, and also repeated the analysis using an earlier digitization of the ON background. The OFF bootstrap probabilities ranged from $0.0197$ to $0.0308$, and Poisson ON with NegBin OFF minimized AIC in every scenario. The coherent endpoint assignments gave the largest changes in the signal fit: the conditional significance ranged from $4.30\sigma$ to $5.39\sigma$, the strength $\hat s$ from $1.052$ to $1.300$, and the bilateral SM-compatibility probabilities from $0.223$ to $0.839$. Thus the model choice and SM signal-strength compatibility are stable within the scenarios examined, but the magnitude of the conditional significance is reconstruction sensitive, including its position relative to the conventional $5\sigma$ threshold~\cite{Cowan:2010js}.

  The vector-NSI comparison in \autoref{sec:nsi-inference-results} shows a sector-dependent change rather than a uniform tightening: relative to the excess-spectrum analysis, the count likelihood reduces the total length of the diagonal nominal $1\sigma$ allowed sets by $11.4\%$ and increases that of flavor-changing intervals by about $8.7\%$. The nominal $90\%$ two-dimensional Jaccard overlaps range from $81.0\%$ to $97.1\%$ on the sampled grids. The weak-charge degeneracies found in the earlier study~\cite{Benavides:2026jdz} persist within the effective-nucleus approximation; it is the accepted charge range, not the underlying geometric form, that differs.

  The reconstruction scenarios also affect the NSI constraints. The coherent lower-endpoint scenario moves the SM point outside the nominal unit-level allowed sets in all four one-parameter scans and splits the flavor-changing intervals at zero. Comparing the perturbed likelihood regions with the same excess-spectrum regions gives two-dimensional Jaccard overlaps of $67.4\%$ to $97.5\%$ across scenarios and planes. These changes in nominal NSI sets are distinct from the separate bilateral test of SM signal strength, which remains compatible in every scenario. Neither exclusion at a nominal unit threshold nor these overlap ranges establishes evidence for nonstandard interactions.

  Several limitations qualify these findings. First, $\Delta q$ uses $64$ reconstructed bins per phase over $160$-$800\,\mathrm{eVee}$, whereas $\Delta\chi^2$ uses $19$ excess bins over $160$-$350\,\mathrm{eVee}$. Although both NSI analyses include the same Gaussian signal-normalization constraint, their fluctuation models and explicit background nuisance treatments differ. The global constraint does not reproduce the full energy-dependent or correlated experimental systematics. Second, the original continuous digitization readings are unavailable; the tested integer assignments probe a specified graphical tolerance around archived rates rather than a calibrated distribution of reconstruction errors. The resulting ranges are scenario-dependent sensitivity summaries, not confidence intervals for the reported statistics.

  Finally, the OFF dispersion is fixed during inference within each scenario, so its estimation uncertainty and count-model selection uncertainty are not propagated. The effective-germanium approximation omits isotope-specific recoil spectra. In particular, the reduction to a single squared charge in \eqref{eq:nsi-statistic-charge-reduction} can make a two-coupling fit nonidentifiable along charge-level curves. As discussed in \autoref{sec:nsi-profile-methods}, the nominal one- and two-parameter thresholds should therefore not be interpreted as coverage-calibrated confidence levels. A common detector-level dataset with a validated uncertainty model and dedicated coverage studies would be needed to isolate the influence of the statistical procedure and quantify these limitations.

\section{Conclusions}
\label{sec:conclusions}

  The validation of reconstructed public \CONUS{} spectra supports Poisson ON and negative-binomial OFF fluctuations at fixed nominal means. The selected effective count law is supported by complementary discrepancy and residual diagnostics, the bootstrap dispersion test, and the AIC comparison, and remains preferred throughout the reconstruction scenarios examined. The Monte Carlo calibrated OFF dispersion set excludes the Poisson endpoint at the $2\sigma$ level. The modest AIC separation and sensitivity of the OFF-Poisson rejection to individual bins limit conclusions about the underlying experimental count process.

  Under the selected count law, the reference conditional significance for a positive \CEvNS{} signal is $4.84\sigma$, and the fitted strength is compatible with the SM ($\hat s=1.17\pm0.25$, $p_{\mathrm{bil}}=0.481$). The reconstruction scenarios preserve SM signal-strength compatibility but vary the significance from $4.30\sigma$ to $5.39\sigma$. These results support a positive signal within the reconstructed-data model without constituting an independent determination of the collaboration's discovery significance.

  For vector quark NSI, the reference count analysis yields nominal $1\sigma$ diagonal allowed sets with $11.4\%$ smaller total length and flavor-changing intervals about $8.7\%$ wider than the excess-spectrum analysis, with nominal $90\%$ two-dimensional Jaccard overlaps of $81.0$-$97.1\%$. Both analyses profile an effective Gaussian signal-normalization nuisance and preserve the weak-charge degeneracies. The reconstructed-count constraints are nevertheless sensitive to the tested count assignments, for which regional overlaps span $67.4\%$-$97.5\%$. The SM belongs to the reference allowed sets at the reported thresholds; the unit-level inclusion is not stable under every reconstruction scenario.

  The principal contribution is therefore the validation and selection of an effective count likelihood and the assessment of the stability of the resulting physical inferences. The comparison does not establish a universal improvement over $\chi^2$, and the nominal regions are not calibrated for frequentist coverage. More complete detector-level data and experimental nuisance information are required to distinguish reconstruction effects from statistical-model effects and to establish calibrated NSI constraints.

  \begin{acknowledgments}
    This article is derived from a master's thesis in Applied Statistics at the Universidad de Nariño, Colombia, entitled ``Estimación estadística de regiones de confianza para interacciones no estándar y bosones $Z'$ ligeros a partir de datos CE$\nu \mathcal{N}$S del experimento \textsc{Conus}$+$''. Ch.M.B., C.S.M. and E.R. acknowledge the Universidad de Nariño for its academic support during the development of this work, as well as support from the Vicerrectoría de Investigaciones e Interacción Social (VIIS) of the Universidad de Nariño under project numbers 3595 and 3899. B.C. and E.R. acknowledge support from the Minciencias Grant CD 82315 CT ICETEX 2021-1080.
  \end{acknowledgments}

  \bibliographystyle{apsrev4-2}
  \bibliography{References}

@article{ParticleDataGroup:2024cfk,
    author = "Navas, S. and others",
    collaboration = "Particle Data Group",
    title = "{Review of particle physics}",
    doi = "10.1103/PhysRevD.110.030001",
    journal = "Phys. Rev. D",
    volume = "110",
    number = "3",
    pages = "030001",
    year = "2024"
}

@article{Freedman:1973yd,
    author = "Freedman, Daniel Z.",
    title = "{Coherent Neutrino Nucleus Scattering as a Probe of the Weak Neutral Current}",
    reportNumber = "NAL-PUB-73-76-THY, FERMILAB-PUB-73-076-T",
    doi = "10.1103/PhysRevD.9.1389",
    journal = "Phys. Rev. D",
    volume = "9",
    pages = "1389--1392",
    year = "1974"
}

@article{Papoulias:2015vxa,
    author = "Papoulias, D. K. and Kosmas, T. S.",
    title = "{Standard and Nonstandard Neutrino-Nucleus Reactions Cross Sections and Event Rates to Neutrino Detection Experiments}",
    eprint = "1502.02928",
    archivePrefix = "arXiv",
    primaryClass = "nucl-th",
    doi = "10.1155/2015/763648",
    journal = "Adv. High Energy Phys.",
    volume = "2015",
    pages = "763648",
    year = "2015"
}

@article{Ohlsson:2012kf,
    author = "Ohlsson, Tommy",
    title = "{Status of non-standard neutrino interactions}",
    eprint = "1209.2710",
    archivePrefix = "arXiv",
    primaryClass = "hep-ph",
    doi = "10.1088/0034-4885/76/4/044201",
    journal = "Rept. Prog. Phys.",
    volume = "76",
    pages = "044201",
    year = "2013"
}

@article{Farzan:2017xzy,
    author = "Farzan, Y. and Tortola, M.",
    title = "{Neutrino oscillations and Non-Standard Interactions}",
    eprint = "1710.09360",
    archivePrefix = "arXiv",
    primaryClass = "hep-ph",
    doi = "10.3389/fphy.2018.00010",
    journal = "Front. in Phys.",
    volume = "6",
    pages = "10",
    year = "2018"
}

@article{Barranco:2005yy,
    author = "Barranco, J. and Miranda, O. G. and Rashba, T. I.",
    title = "{Probing new physics with coherent neutrino scattering off nuclei}",
    eprint = "hep-ph/0508299",
    archivePrefix = "arXiv",
    reportNumber = "MPP-2005-85",
    doi = "10.1088/1126-6708/2005/12/021",
    journal = "JHEP",
    volume = "12",
    number= {12},
    pages = "021",
    year = "2005"
}

@article{Canas:2019fjw,
    author = "Canas, B. C. and Garces, E. A. and Miranda, O. G. and Parada, A. and Sanchez Garcia, G.",
    title = "{Interplay between nonstandard and nuclear constraints in coherent elastic neutrino-nucleus scattering experiments}",
    eprint = "1911.09831",
    archivePrefix = "arXiv",
    primaryClass = "hep-ph",
    doi = "10.1103/PhysRevD.101.035012",
    journal = "Phys. Rev. D",
    volume = "101",
    number = "3",
    pages = "035012",
    year = "2020"
}

@article{Giunti:2019xpr,
    author = "Giunti, C.",
    title = "{General COHERENT constraints on neutrino nonstandard interactions}",
    eprint = "1909.00466",
    archivePrefix = "arXiv",
    primaryClass = "hep-ph",
    doi = "10.1103/PhysRevD.101.035039",
    journal = "Phys. Rev. D",
    volume = "101",
    number = "3",
    pages = "035039",
    year = "2020"
}

@article{DeRomeri:2022twg,
    author = "De Romeri, V. and Miranda, O. G. and Papoulias, D. K. and Sanchez Garcia, G. and T{\'o}rtola, M. and Valle, J. W. F.",
    title = "{Physics implications of a combined analysis of COHERENT CsI and LAr data}",
    eprint = "2211.11905",
    archivePrefix = "arXiv",
    primaryClass = "hep-ph",
    doi = "10.1007/JHEP04(2023)035",
    journal = "JHEP",
    volume = "04",
    number= {4},
    pages = "035",
    year = "2023"
}

@article{Baldoncini:2014vda,
    author = "Baldoncini, Marica and Callegari, Ivan and Fiorentini, Giovanni and Mantovani, Fabio and Ricci, Barbara and Strati, Virginia and Xhixha, Gerti",
    title = "{Reference worldwide model for antineutrinos from reactors}",
    eprint = "1411.6475",
    archivePrefix = "arXiv",
    primaryClass = "physics.ins-det",
    doi = "10.1103/PhysRevD.91.065002",
    journal = "Phys. Rev. D",
    volume = "91",
    number = "6",
    pages = "065002",
    year = "2015"
}

@article{Zhang:2023zif,
    author = "Zhang, Chao and Qian, Xin and Fallot, Muriel",
    title = "{Reactor antineutrino flux and anomaly}",
    eprint = "2310.13070",
    archivePrefix = "arXiv",
    primaryClass = "hep-ph",
    doi = "10.1016/j.ppnp.2024.104106",
    journal = "Prog. Part. Nucl. Phys.",
    volume = "136",
    pages = "104106",
    year = "2024"
}

@article{Mueller:2011nm,
    author = "Mueller, Th. A. and others",
    title = "{Improved Predictions of Reactor Antineutrino Spectra}",
    eprint = "1101.2663",
    archivePrefix = "arXiv",
    primaryClass = "hep-ex",
    reportNumber = "IRFU-10-280",
    doi = "10.1103/PhysRevC.83.054615",
    journal = "Phys. Rev. C",
    volume = "83",
    pages = "054615",
    year = "2011"
}

@article{Kopeikin:2012zz,
    author = "Kopeikin, V. I.",
    title = "{Flux and spectrum of reactor antineutrinos}",
    doi = "10.1134/S1063778812020123",
    journal = "Phys. Atom. Nucl.",
    volume = "75",
    pages = "143--152",
    year = "2012"
}

@article{CONUS:2024lnu,
    author = "Ackermann, N. and others",
    collaboration = "CONUS+",
    title = "{CONUS+~Experiment}",
    eprint = "2407.11912",
    archivePrefix = "arXiv",
    primaryClass = "hep-ex",
    doi = "10.1140/epjc/s10052-024-13551-6",
    journal = "Eur. Phys. J. C",
    volume = "84",
    number = "12",
    pages = "1265",
    year = "2024",
    note = "[Erratum: Eur.Phys.J.C 85, 19 (2025)]"
}

@article{Ackermann:2025obx,
    author = "Ackermann, N. and others",
    title = "{Direct observation of coherent elastic antineutrino{\textendash}nucleus scattering}",
    eprint = "2501.05206",
    archivePrefix = "arXiv",
    primaryClass = "hep-ex",
    doi = "10.1038/s41586-025-09322-2",
    journal = "Nature",
    volume = "643",
    number = "8074",
    pages = "1229--1233",
    year = "2025",
    note = "Published online 30 July 2025"
}

@phdthesis{Rink:2022rsx,
    author = "Rink, Thomas",
    title = "{Investigating Neutrino Physics within and beyond the Standard Model using CONUS Experimental Data.}",
    doi = "10.11588/heidok.00031274",
    school = "U. Heidelberg (main)",
    year = "2022"
}

@article{Bonhomme:2022lcz,
    author = "Bonhomme, A. and others",
    title = "{Direct measurement of the ionization quenching factor of nuclear recoils in germanium in the keV energy range}",
    eprint = "2202.03754",
    archivePrefix = "arXiv",
    primaryClass = "physics.ins-det",
    doi = "10.1140/epjc/s10052-022-10768-1",
    journal = "Eur. Phys. J. C",
    volume = "82",
    number = "9",
    pages = "815",
    year = "2022"
}

@article{Lindner:2024eng,
    author = "Lindner, Manfred and Rink, Thomas and Sen, Manibrata",
    title = "{Light vector bosons and the weak mixing angle in the light of future germanium-based reactor CE{\ensuremath{\nu}}NS experiments}",
    eprint = "2401.13025",
    archivePrefix = "arXiv",
    primaryClass = "hep-ph",
    doi = "10.1007/JHEP08(2024)171",
    journal = "JHEP",
    volume = "08",
    number= {8},
    pages = "171",
    year = "2024"
}

@article{DeRomeri:2025csu,
    author = "De Romeri, Valentina and Papoulias, Dimitrios K. and Sanchez Garcia, Gonzalo",
    title = "{Implications of the first CONUS+ measurement of coherent elastic neutrino-nucleus scattering}",
    eprint = "2501.17843",
    archivePrefix = "arXiv",
    primaryClass = "hep-ph",
    doi = "10.1103/PhysRevD.111.075025",
    journal = "Phys. Rev. D",
    volume = "111",
    number = "7",
    pages = "075025",
    year = "2025"
}

@article{AtzoriCorona:2025ygn,
    author = "Atzori Corona, M. and Cadeddu, M. and Cargioli, N. and Dordei, F. and Giunti, C.",
    title = "{Reactor antineutrinos CE{\ensuremath{\nu}}NS on germanium: CONUS+ and TEXONO as a new gateway to SM and BSM physics}",
    eprint = "2501.18550",
    archivePrefix = "arXiv",
    primaryClass = "hep-ph",
    doi = "10.1103/n563-8v8d",
    journal = "Phys. Rev. D",
    volume = "112",
    number = "1",
    pages = "015007",
    year = "2025"
}

@article{Benavides:2026jdz,
    author = "Benavides, Ch. M. and Mu{\~n}oz, C. S. and Canas, B. C. and Rojas, Eduardo",
    title = "{Nonstandard interactions and light $Z'$ bosons from CE{\ensuremath{\nu}}NS data at CONUS+}",
    eprint = "2606.24005",
    archivePrefix = "arXiv",
    primaryClass = "hep-ph",
    doi = "10.1103/mwvn-zgkr",
    journal = "Phys. Rev. D",
    month = "6",
    year = "2026"
}

@article{Chattaraj:2025fvx,
    author = "Chattaraj, Ayan and Majumdar, Anirban and Srivastava, Rahul",
    title = "{Probing standard model and beyond with reactor CE{\ensuremath{\nu}}NS data of CONUS+ experiment}",
    eprint = "2501.12441",
    archivePrefix = "arXiv",
    primaryClass = "hep-ph",
    doi = "10.1016/j.physletb.2025.139438",
    journal = "Phys. Lett. B",
    volume = "864",
    pages = "139438",
    year = "2025"
}

@article{COHERENT:2017ipa,
    author = "Akimov, D. and others",
    collaboration = "COHERENT",
    title = "{Observation of Coherent Elastic Neutrino-Nucleus Scattering}",
    eprint = "1708.01294",
    archivePrefix = "arXiv",
    primaryClass = "nucl-ex",
    doi = "10.1126/science.aao0990",
    journal = "Science",
    volume = "357",
    number = "6356",
    pages = "1123--1126",
    year = "2017"
}

@article{Baker:1983tu,
    author = "Baker, Steve and Cousins, Robert D.",
    title = "{Clarification of the Use of Chi Square and Likelihood Functions in Fits to Histograms}",
    reportNumber = "UCLA-HEP-83-3",
    doi = "10.1016/0167-5087(84)90016-4",
    journal = "Nucl. Instrum. Meth.",
    volume = "221",
    pages = "437--442",
    year = "1984"
}

@article{Bohme:2025nvu,
    author = {B{\"o}hme, Lukas and Schwarz, Dominik J. and Tiwari, Prabhakar and Pashapour-Ahmadabadi, Morteza and Bahr-Kalus, Benedict and Bilicki, Maciej and Hale, Catherine L. and Heneka, Caroline S. and Siewert, Thilo M.},
    title = "{Overdispersed Radio Source Counts and Excess Radio Dipole Detection}",
    eprint = "2509.16732",
    archivePrefix = "arXiv",
    primaryClass = "astro-ph.CO",
    doi = "10.1103/6z32-3zf4",
    journal = "Phys. Rev. Lett.",
    volume = "135",
    number = "20",
    pages = "201001",
    year = "2025"
}

@article{Grosso:2023scl,
    author = "Grosso, Gaia and Letizia, Marco and Pierini, Maurizio and Wulzer, Andrea",
    title = "{Goodness of fit by Neyman-Pearson testing}",
    eprint = "2305.14137",
    archivePrefix = "arXiv",
    primaryClass = "hep-ph",
    doi = "10.21468/SciPostPhys.16.5.123",
    journal = "SciPost Phys.",
    volume = "16",
    number = "5",
    pages = "123",
    year = "2024"
}

@article{Beaujean:2011zza,
    author = "Beaujean, F. and Caldwell, A. and Kollar, D. and Kroninger, K.",
    title = "{p-values for model evaluation}",
    eprint = "1011.1674",
    archivePrefix = "arXiv",
    primaryClass = "physics.data-an",
    doi = "10.1103/PhysRevD.83.012004",
    journal = "Phys. Rev. D",
    volume = "83",
    pages = "012004",
    year = "2011"
}

@article{Christ:2024nxz,
    author = "Christ, Norman and Eranki, Rajiv and Kelly, Christopher",
    collaboration = "RBC, UKQCD",
    title = "{Bootstrap-determined p values in lattice QCD}",
    eprint = "2409.11379",
    archivePrefix = "arXiv",
    primaryClass = "hep-lat",
    doi = "10.1103/PhysRevD.111.074514",
    journal = "Phys. Rev. D",
    volume = "111",
    number = "7",
    pages = "074514",
    year = "2025"
}

@article{Wilks:1938dza,
    author = "Wilks, S. S.",
    title = "{The Large-Sample Distribution of the Likelihood Ratio for Testing Composite Hypotheses}",
    doi = "10.1214/aoms/1177732360",
    journal = "Annals Math. Statist.",
    volume = "9",
    number = "1",
    pages = "60--62",
    year = "1938"
}

@article{Cowan:2010js,
    author = "Cowan, Glen and Cranmer, Kyle and Gross, Eilam and Vitells, Ofer",
    title = "{Asymptotic Formulae for Likelihood-Based Tests of New Physics}",
    eprint = "1007.1727",
    archivePrefix = "arXiv",
    primaryClass = "physics.data-an",
    doi = "10.1140/epjc/s10052-011-1554-0",
    journal = "Eur. Phys. J. C",
    volume = "71",
    pages = "1554",
    year = "2011",
    note = "Erratum: Eur. Phys. J. C 73, 2501 (2013)"
}

@article{Feldman:1997qc,
    author = "Feldman, Gary J. and Cousins, Robert D.",
    title = "{A Unified approach to the classical statistical analysis of small signals}",
    eprint = "physics/9711021",
    archivePrefix = "arXiv",
    reportNumber = "HUTP-97-A096",
    doi = "10.1103/PhysRevD.57.3873",
    journal = "Phys. Rev. D",
    volume = "57",
    pages = "3873--3889",
    year = "1998"
}

@article{Berns:2023dmi,
    author = "Berns, Lukas",
    title = "{Importance sampling method for Feldman-Cousins confidence intervals}",
    eprint = "2303.11290",
    archivePrefix = "arXiv",
    primaryClass = "hep-ex",
    doi = "10.1103/PhysRevD.109.092002",
    journal = "Phys. Rev. D",
    volume = "109",
    number = "9",
    pages = "092002",
    year = "2024"
}

@article{Akaike:1974vps,
    author = "Akaike, H.",
    title = "{A new look at the statistical model identification}",
    doi = "10.1109/TAC.1974.1100705",
    journal = "IEEE Trans. Automatic Control",
    volume = "19",
    number = "6",
    pages = "716--723",
    year = "1974"
}

@article{Liddle:2007fy,
    author = "Liddle, Andrew R",
    title = "{Information criteria for astrophysical model selection}",
    eprint = "astro-ph/0701113",
    archivePrefix = "arXiv",
    doi = "10.1111/j.1745-3933.2007.00306.x",
    journal = "Mon. Not. Roy. Astron. Soc.",
    volume = "377",
    pages = "L74--L78",
    year = "2007"
}

@article{Self:1987liang,
    author = "Self, Steven G. and Liang, Kung-Yee",
    title = "{Asymptotic Properties of Maximum Likelihood Estimators and Likelihood Ratio Tests under Nonstandard Conditions}",
    doi = "10.1080/01621459.1987.10478472",
    journal = "J. Am. Stat. Assoc.",
    volume = "82",
    number = "398",
    pages = "605--610",
    year = "1987"
}

@article{Williams:1987deletion,
    author = "Williams, D. A.",
    title = "{Generalized Linear Model Diagnostics Using the Deviance and Single Case Deletions}",
    doi = "10.2307/2347550",
    journal = "J. Roy. Stat. Soc. C",
    volume = "36",
    number = "2",
    pages = "181--191",
    year = "1987"
}

@article{Pena:2006validation,
    author = "Pe{\~n}a, Edsel A. and Slate, Elizabeth H.",
    title = "{Global Validation of Linear Model Assumptions}",
    doi = "10.1198/016214505000000637",
    journal = "J. Am. Stat. Assoc.",
    volume = "101",
    number = "473",
    pages = "341--354",
    year = "2006"
}

@article{Dunn:1996rqr,
    author = "Dunn, Peter K. and Smyth, Gordon K.",
    title = "{Randomized Quantile Residuals}",
    doi = "10.1080/10618600.1996.10474708",
    journal = "J. Comput. Graph. Stat.",
    volume = "5",
    number = "3",
    pages = "236--244",
    year = "1996"
}

@article{vanBuuren:2001worm,
    author = "van Buuren, Stef and Fredriks, Miranda",
    title = "{Worm Plot: A Simple Diagnostic Device for Modelling Growth Reference Curves}",
    doi = "10.1002/sim.746",
    journal = "Stat. Med.",
    volume = "20",
    number = "8",
    pages = "1259--1277",
    year = "2001"
}

@article{Warton:2023envelopes,
    author = "Warton, David I.",
    title = "{Global Simulation Envelopes for Diagnostic Plots in Regression Models}",
    eprint = "2208.01811",
    archivePrefix = "arXiv",
    primaryClass = "stat.ME",
    doi = "10.1080/00031305.2022.2139294",
    journal = "Am. Stat.",
    volume = "77",
    number = "4",
    pages = "425--431",
    year = "2023"
}

@article{Phipson:2010pvalues,
    author = "Phipson, Belinda and Smyth, Gordon K.",
    title = "{Permutation P-values Should Never Be Zero: Calculating Exact P-values When Permutations Are Randomly Drawn}",
    eprint = "1603.05766",
    archivePrefix = "arXiv",
    primaryClass = "stat.AP",
    doi = "10.2202/1544-6115.1585",
    journal = "Stat. Appl. Genet. Mol. Biol.",
    volume = "9",
    number = "1",
    pages = "39",
    year = "2010"
}

\end{document}